\documentclass[english,aps,preprint]{revtex4-1}
\usepackage[T1]{fontenc}
\usepackage[utf8]{inputenc}
\usepackage{color}
\usepackage{amsmath}
\usepackage{graphicx}

\makeatletter

\newcommand{\lyxdot}{.}

\usepackage{babel}

\usepackage{babel}

\usepackage{babel}

\usepackage{babel}

\usepackage{babel}

\usepackage{babel}

\usepackage{babel}

\makeatother

\usepackage{babel}
\begin{document}

\preprint{This line only printed with preprint option}

\title{Black and gray solitons in holographic superfluids at zero temperature}

\author{Meng Gao}
\email{2015210214@student.cup.edu.cn}

\affiliation{Department of Physics, College of Science, China University of Petroleum,
Beijing 102249, China~~~~~~~~\\
 School of Physical Sciences, University of Chinese Academy of Sciences,
Beijing 100049, China}

\author{Yuqiu Jiao}
\email{napoleonest@aliyun.com}

\affiliation{Department of Physics, College of Science, China University of Petroleum,
Beijing 102249, China}

\author{Xin Li}
\email{lixin615@mails.ucas.edu.cn}

\affiliation{School of Physical Sciences, University of Chinese Academy of Sciences,
Beijing 100049, China}

\author{Yu Tian}
\email{ytian@ucas.ac.cn}

\affiliation{School of Physical Sciences, University of Chinese Academy of Sciences,
Beijing 100049, China~~~~~~~~\\
 Institute of Theoretical Physics, Chinese Academy of Sciences, Beijing
100190, China}

\author{Hongbao Zhang}
\email{hzhang@vub.ac.be}

\affiliation{Department of Physics, Beijing Normal University, Beijing 100875,
China~~~~~~~~\\
 Theoretische Natuurkunde, Vrije Universiteit Brussel, and The International
Solvay Institutes, Pleinlaan 2, B-1050 Brussels, Belgium}
\begin{abstract}
We construct gray soliton configurations, which move at constant speeds,
in holographic superfluids for the first time. Since there should
be no dissipation for a moving soliton to exist, we use the simplest
holographic superfluid model at zero temperature, considering both
the standard and alternative quantizations. For comparison purpose,
we first investigate black solitons in the zero temperature holographic
superfluids, which are static configurations. Then we focus on the
numerical construction of gray solitons under both quantizations,
which interpolate between the (static) black solitons and sound waves
(moving at the speed of sound). Interestingly, under the standard
quantization, a peculiar oscillation of the soliton configurations
is observed, very much resembling the Friedel oscillation in fermionic
superfluids at the BCS regime. Some implications and other aspects
of the soliton configurations are also discussed. 
\end{abstract}

\keywords{holographic superfluid, dark soliton, gray soliton, Friedel oscillation,
AdS soliton spacetime.}
\maketitle

\section{Introduction and motivation}

AdS/CFT correspondence states that a gravitational system in the anti-de
Sitter space is dual to a conformal field theory on its conformal
boundary, realizing the so-called holographic principle\cite{Susskind,Maldacena,Witten}.
The main application of AdS/CFT, or holography, is that it can be
used to explore strongly coupled quantum systems, which are outside
the reach of conventional perturbative techniques. There are many
papers that have applied this correspondence to various condensed
matter systems with strong interaction, for example \cite{Sachdev,Hartnoll,Herzog}.
There are also some papers\cite{Chesler,Ewerz,Yiqiang,Y. Tian} that
have investigated the features of superfluid, such as turbulence,
by means of this correspondence. In applied AdS/CFT, an Abelian Higgs
model in the background of an AdS black hole\cite{S.A,C.P} that exhibits
spontaneous breaking of the bulk U(1) symmetry has been extensively
studied. Due to the principle of AdS/CFT, this broken U(1) symmetry
is global on the boundary, so it is more appropriate that one identifies
these models as superfluids\cite{Kovtun,Montull,Salvio,Alberto,Salvio-1}.
There are some interesting objects in superfluids, such as vortices
and solitons. In \cite{Albash,Tameem,Silva,Maeda,V.Keranen}, vortex
solutions are studied using the holographic model in \cite{S.A} (the
HHH model). A dark (bright) soliton is an interface of reduced (increased)
particle number density between two superfluid phases (for a review,
see \cite{Kivshara}). Black solitons, which are static dark solitons,
in holographic superfluids have been studied in \cite{Keranen,Lan}.
However, the more general dark solitons, which move at constant speeds
and are called gray solitons, as well as bright solitons have not
been constructed in holographic models.

In this paper, we are interested in holographic dark soliton solutions
at zero temperature, including the static black solitons and the gray
solitons moving at constant speeds. This zero temperature holographic
setup is realized by an Abelian Higgs model in the AdS soliton background\cite{Nishioka},
instead of the AdS black hole in \cite{S.A,C.P} for the finite temperature
case. Our major motivations to consider the zero temperature holographic
superfluids are two-fold. First, it is well known that in fermionic
cold atom systems there are two types (or two mechanisms) of superfluidity,
i.e. the BEC superfluidity and the BCS superfluidity, with a crossover
(called the BEC-BCS crossover) in between\cite{M. Randeria,Timmermans}.
With the Bogoliubov-de Gennes (BdG) equations as a mean field description
of the fermionic superfluidity, black soliton configurations in these
two types of superfluids are shown to have a salient difference in
the (particle number) density depletion at their centers\cite{Antezza}.
In the finite temperature holographic superfluids, the authors of
\cite{Keranen} have investigated the density depletion at the centers
of the black solitons under two types of quantizations, i.e. the standard
quantization and alternative quantization, of the bulk scalar field
and argued that these two types of quantizations correspond to the
BCS-like and BEC-like superfluids, respectively (see also \cite{Chaolun}).
However, the BdG analysis in \cite{Antezza} is at zero temperature,
and it is numerically difficult to make the temperature of the HHH
model used in \cite{Keranen} low enough, which renders the argument
of the correspondence between different quantizations and different
types of superfluids limited. In order to avoid this limitation, we
directly use the zero temperature holographic setup.

Next, since there is dissipation when local structures are traveling
in finite temperature (holographic) superfluids\cite{Chesler}, moving
``solitons'' are not expected to preserve their shape under such
backgrounds, i.e. we cannot obtain the constantly traveling gray soliton
configurations, which is more interesting than black solitons in many
aspects. But just like the gray soliton solution to the (zero temperature)
Gross-Pitaevskii (GP) equation\cite{Gross,Pitaevskii,Gross2}, as
will be shown in this paper, we can construct gray soliton solutions
in the zero temperature holographic superfluids instead. Having these
holographic gray solitons at hand, one can extend the study of the
holographic BEC-like and BCS-like behaviors mentioned above and the
stability of holographic black solitons to the more general cases\cite{GKLTZ}.
One can even investigate the collision of two or more holographic
gray solitons and other dynamical properties of them.

The paper is organized as follows. In the next section, we introduce
the zero temperature holographic superfluid model. In Sec.III, we
first present the black soliton solutions under both quantizations
at different chemical potentials. Then we study the density depletion
of the black solitons and show the holographic BEC-like and BCS-like
behaviors at different quantizations. In Sec.IV, with the help of
a ``comoving'' frame, we construct the gray soliton solutions at
different speeds of traveling and investigate their (particle number)
density profiles. As one of the key features of gray solitons, we
show that the (gauge invariant) phase differences between the condensates
on their two sides are not equal to $\pi$, in contrast to the case
of black solitons, which have phase differences exactly equal to $\pi$.
Actually, the obtained gray soliton configurations perfectly interpolate
between the (static) black solitons and sound waves (moving at the
speed of sound). Unexpectedly, for both black solitons and gray solitons
under the standard quantization, we can observe peculiar oscillatory
behaviors of their configurations, which very much resemble the Friedel
oscillation in fermionic superfluids at the BCS limit, further supporting
the correspondence proposed by \cite{Keranen}. In the end, we draw
our conclusions and have some discussions.

\section{Holographic setup}

In this section, we present the holographic model for zero temperature
superfluids. The action is based on the Abelian Higgs model\cite{S.A,C.P}
that is given by 
\begin{equation}
S=\frac{1}{16\pi G}\int_{\mathcal{M}}d^{4}x\sqrt{-g}\left(R+\frac{6}{L^{2}}+\frac{1}{e^{2}}\mathcal{L}_{matter}\right).\label{eq:HHH model}
\end{equation}
Here $G$ is Newton's constant, and the Lagrangian for matter fields
reads 
\begin{equation}
\mathcal{L}_{matter}=-\frac{1}{4}F_{ab}F^{ab}-\left|D\Psi\right|^{2}-m^{2}\left|\Psi\right|^{2},\label{eq:Lagrangian}
\end{equation}
where $D=\nabla-iA$, $F=dA$, with $\nabla$ the covariant derivative
compatible to the metric. $e$ and $m$ are the charge and mass of
the complex scalar field $\Psi$, respectively. To make the model
easier, we will work in the probe limit, namely the limit that the
backreaction of the matter fields onto the bulk spacetime is neglected.
In what follows, we will take the units in which $L=1$, $16\pi Ge^{2}=1$.
We just focus on the action of the matter fields which reads 
\begin{equation}
S=\int_{\mathcal{M}}d^{4}x\sqrt{-g}\left(-\frac{1}{4}F_{ab}F^{ab}-\left|D\Psi\right|^{2}-m^{2}\left|\Psi\right|^{2}\right).\label{eq:action}
\end{equation}

In AdS/CFT, a bulk spacetime with a black hole corresponds to a boundary
system at finite temperature, with the Hawking temperature of the
black hole corresponding to the temperature of the boundary system.
But here we would like to investigate the superfluid model under the
AdS soliton background\cite{Guo,Guo2,Nishioka}, which implies the
temperature of the holographic superfluids is zero. The metric of
the (1+3)D AdS soliton spacetime in the Schwarzschild coordinates,
which is just a double Wick rotation of the Schwarzschild-AdS black
brane, reads 
\begin{equation}
ds^{2}=\frac{1}{z^{2}}\left[-dt^{2}+\frac{dz^{2}}{f(z)}+dx^{2}+f(z)d\xi^{2}\right].\label{eq:metric}
\end{equation}
Here $f(z)=1-(\frac{z}{z_{0}})^{3}$ with $z=z_{0}$ the tip where
our geometry caps off and $z=0$ the AdS boundary. In order to make
the geometry smooth at the tip, we are required to impose the periodicity
$\frac{4\pi z_{0}}{3}$ onto the $\xi$ coordinate. With the help
of the scale invariance\cite{C.P} and for numerical simplicity, we
shall set $z_{0}=1$.

Variation of the action gives rise to the equations of motion for
the matter fields, which can be written as 
\begin{align}
\nabla_{a}F^{ab}= & i[\Psi^{*}D^{b}\Psi-\Psi(D^{b}\Psi)^{*}],\label{eq:Maxwell equation}\\
0= & D_{a}D^{a}\Psi-m^{2}\Psi.\label{eq:Klein-Gordon equation}
\end{align}
The asymptotical behavior for the bulk fields near the AdS boundary
goes as\cite{C.P,Kovtun} 
\begin{equation}
A_{\mu}=a_{\mu}+b_{\mu}z+\cdots,\label{eq:A_mu asymptotic solution}
\end{equation}
\begin{equation}
\Psi=\Psi_{-}z^{\Delta_{-}}+\Psi_{+}z^{\Delta_{+}}+\cdots.\label{eq:psi asymptotic solution}
\end{equation}

The mass of scalar field is related to the conformal dimension $\Delta$
of the condensate as $\Delta_{\pm}=3/2\pm\sqrt{9/4+m^{2}}$ \cite{Klebanov}
in four dimensional bulk spacetime. The condition $m^{2}\geq-9/4$
for $\Delta$ to be real is known as the Breitenlohner-Freedman (BF)
bound\cite{Freedman}. We set $m^{2}=-2$ for numerical simplicity
in this paper,\textcolor{red}{{} }namely $\Delta_{-}=1$ and $\Delta_{+}=2$.
In this case there are two choices for the source\cite{C.P}, which
are called the standard quantization and the alternative quantization,
respectively.\textcolor{red}{{} }According to the holographic dictionary,
we can obtain the vacuum expectation of the operators on the boundary
as 
\begin{align}
\left\langle j^{\mu}\right\rangle  & =\frac{\delta S_{onshell}}{\delta a_{\mu}}=b^{\mu},\label{eq:dictionary}\\
\left\langle O_{\pm}\right\rangle  & =\frac{\delta S_{onshell}}{\delta\Psi_{\mp}}=\pm\Psi_{\pm}^{*}.\label{eq:order parameter dictionary}
\end{align}
Here $j^{\mu}$ is the conserved current on the boundary, corresponding
to the conserved particle number of the boundary system. Particularly,
$j^{t}$ is the particle number density and $a_{t}=A_{t}|_{z=0}$
the chemical potential conjugate to the conserved particle number.
The vacuum expectation value of the scalar operator $\left\langle O_{\pm}\right\rangle $
is interpreted as the condensate in the holographic superfluid model.
The choice of $\Psi_{-}$ or $\Psi_{+}$ as the source corresponds
to the standard or alternative quantization, respectively. If the
condensate (or order parameter in the language of Ginzburg-Landau
phase transition) is not zero when the source is turned off, the U(1)
symmetry is spontaneously broken and the boundary system is in a superfluid
phase, otherwise it is in a normal phase. For an equilibrium holographic
superfluid, where the configuration is independent of $x$ and $\xi$,
investigation of this phase transition leads to a critical chemical
potential $\mu_{c}$ (see \cite{Guo2} for the specific case considered
here) for each of the quantizations, beyond which the system will
be in the superfluid phase.

In the following, we will focus on the superfluid phase, i.e. the
chemical potential $\mu$ is above $\mu_{c}$, and consider the configuration
that is inhomogeneous in the $x$ direction. For the sake of solving
the equations of motion above, we should first choose the axial gauge
$A_{z}=0$ for the U(1) gauge fields. For simplicity, we assume that
the non-vanishing bulk fields are $\Psi:=z\psi$, $A_{t}$ and $A_{x}$,
which do not depend on the coordinate $\xi$. Therefore, the equations
of motion become 
\begin{align}
0= & \partial_{t}^{2}\psi+\left(z+A_{x}^{2}-A_{t}^{2}+i\partial_{x}A_{x}-i\partial_{t}A_{t}\right)\psi+2iA_{x}\partial_{x}\psi-2iA_{t}\partial_{t}\psi-\partial_{x}^{2}\psi\nonumber \\
 & +3z^{2}\partial_{z}\psi+\left(z^{3}-1\right)\partial_{z}^{2}\psi,\label{eq:psi equation}\\
0= & \partial_{t}^{2}A_{x}-\partial_{t}\partial_{x}A_{t}-i\left(\psi\partial_{x}\psi^{*}-\psi^{*}\partial_{x}\psi\right)+2A_{x}\psi\psi^{*}+3z^{2}\partial_{z}A_{x}\nonumber \\
 & +\left(z^{3}-1\right)\partial_{z}^{2}A_{x},\label{eq:Ax equation}\\
0= & \left(z^{3}-1\right)\partial_{z}^{2}A_{t}+3z^{2}\partial_{z}A_{t}-\partial_{x}^{2}A_{t}+\partial_{t}\partial_{x}A_{x}+2A_{t}\psi\psi^{*}\nonumber \\
 & +i\left(\psi^{*}\partial_{t}\psi-\psi\partial_{t}\psi^{*}\right),\label{eq:At equation}\\
0= & \partial_{t}\partial_{z}A_{t}+i\left(\psi\partial_{z}\psi^{*}-\psi^{*}\partial_{z}\psi\right)-\partial_{z}\partial_{x}A_{x},\label{eq:boundary current}
\end{align}
where the third one can be taken as the constraint equation.

\section{Black solitons}

Now we study the black soliton structure in the zero temperature holographic
model. Without loss of generality, we assume that the inhomogeneous
direction of the soliton configuration is the $x$ direction. There
is a density depletion at some positions between two superfluid phases,
where it produces the black soliton structure with the order parameter
changing sign across the interface. Therefore, the matter fields are
functions of both $x$ and $z$ for (static) black solitons. Namely,
$\psi=\psi\left(z,x\right)$, $A_{t}=A_{t}\left(z,x\right)$ and $A_{x}=A_{x}\left(z,x\right)$.
We further recast the complex scalar field $\psi$ in the form $\psi\left(z,x\right)=\phi\left(z,x\right)\exp\left[i\varphi\left(z,x\right)\right]$.
Then we substitute $\psi$, $A_{t}$ and $A_{x}$ into equations (\ref{eq:psi equation})
to (\ref{eq:boundary current}) and obtain the following equations:
\begin{align}
0= & \left(z+A_{x}^{2}-A_{t}^{2}\right)\phi-2A_{x}\phi\partial_{x}\varphi-\partial_{x}^{2}\phi-\phi\left(\partial_{x}\varphi\right)^{2}+3z^{2}\partial_{z}\phi\nonumber \\
 & +\left(z^{3}-1\right)\left[\partial_{z}^{2}\phi-\phi\left(\partial_{z}\varphi\right)^{2}\right],\label{eq:real part}\\
0= & \phi\partial_{x}A_{x}+2A_{x}\partial_{x}\phi+2\left(\partial_{x}\phi\right)\partial_{x}\varphi+\phi\partial_{x}^{2}\varphi+3z^{2}\phi\partial_{z}\varphi\nonumber \\
 & +\left(z^{3}-1\right)\left[2\left(\partial_{z}\phi\right)\partial_{z}\varphi+\phi\partial_{z}^{2}\varphi\right],\label{eq:imaginary part}\\
0= & -2\phi^{2}\partial_{x}\varphi+2A_{x}\phi^{2}+3z^{2}\partial_{z}A_{x}+\left(z^{3}-1\right)\partial_{z}^{2}A_{x},\label{eq:Ax}\\
0= & \left(z^{3}-1\right)\partial_{z}^{2}A_{t}+3z^{2}\partial_{z}A_{t}-\partial_{x}^{2}A_{t}+2\phi^{2}A_{t},\label{eq:At}\\
0= & 2\phi^{2}\partial_{z}\varphi-\partial_{z}\partial_{x}A_{x}.\label{eq:Ax-1}
\end{align}

Because we are looking for static soliton solutions, the currents
in the holographic system should be zero. The RHS of (\ref{eq:Maxwell equation})
is zero, namely $j^{x}=0=j^{z}$, so we have 
\begin{align}
\partial_{z}\varphi & =0\label{eq:cphi condition}\\
\partial_{x}\varphi & =A_{x}\label{eq:cphi Ax}
\end{align}
With the constraint conditions above, (\ref{eq:Ax-1}) and (\ref{eq:Ax})
are satisfied automatically. Moreover, we can choose the gauge $A_{x}=0$,
to fix the local U(1) symmetry. And then we only need to choose $\varphi=0$
to fix the global U(1) symmetry. As a result, the equations of motion
(\ref{eq:real part}) to (\ref{eq:Ax-1}) will be simplified as 
\begin{align}
0 & =(z-A_{t}^{2})\phi-\partial_{x}^{2}\phi+3z^{2}\partial_{z}\phi+(z^{3}-1)\partial_{z}^{2}\phi\label{eq:phi}\\
0 & =(z^{3}-1)\partial_{z}^{2}A_{t}+3z^{2}\partial_{z}A_{t}-\partial_{x}^{2}A_{t}+2\phi^{2}A_{t}\label{eq:At-1}
\end{align}
In the following, we shall numerically solve these nonlinear differential
equations by means of the pseudospectral method\cite{Guo2} and Newton-Raphson
iteration.

Ideally, the size of the system in the $x$ direction should be infinite.
However, a cutoff is needed to impose for numerical treatments. Consider
a box of size $1\times2L$ ($L$\textcolor{blue}{{} }is much larger
than the healing length of black solitons, so that further increasing
$L$ will not affect the structure of black solitons) in the $z$
and $x$ directions. Moreover, there are two kinds of boundary conditions
that can be imposed to these equations, corresponding to standard
and alternative quantizations, respectively. For the standard quantization,
we choose $\Psi_{-}$ to be the source. Therefore, the boundary conditions
are $\psi=0$ and $A_{t}=\mu$ at $z=0$. As a result, the order parameter
of superfluid is $\Psi_{+}=\partial_{z}\psi$ at $z=0$. For the alternative
quantization, the source is taken as $\Psi_{+}$. The boundary condition
at $z=0$ is then $\partial_{z}\psi=0$ and $A_{t}=\mu$. As a result,
the order parameter of superfluid is $\left\langle O\right\rangle |_{z=0}=\psi$.
Taking into account these boundary conditions, we apply the Chebyshev
and Fourier pseudospectral methods\cite{Guo2} and Newton iteration
to solve the equations. In addition to these boundary conditions in
the $z$ direction, we still need the Neumann boundary conditions
in the $x$ direction, which are written as $\partial_{x}\psi=0$
and $\partial_{x}A_{t}=0$ at both $x=L$ and $x=-L$.

\begin{figure}
\includegraphics[scale=0.4]{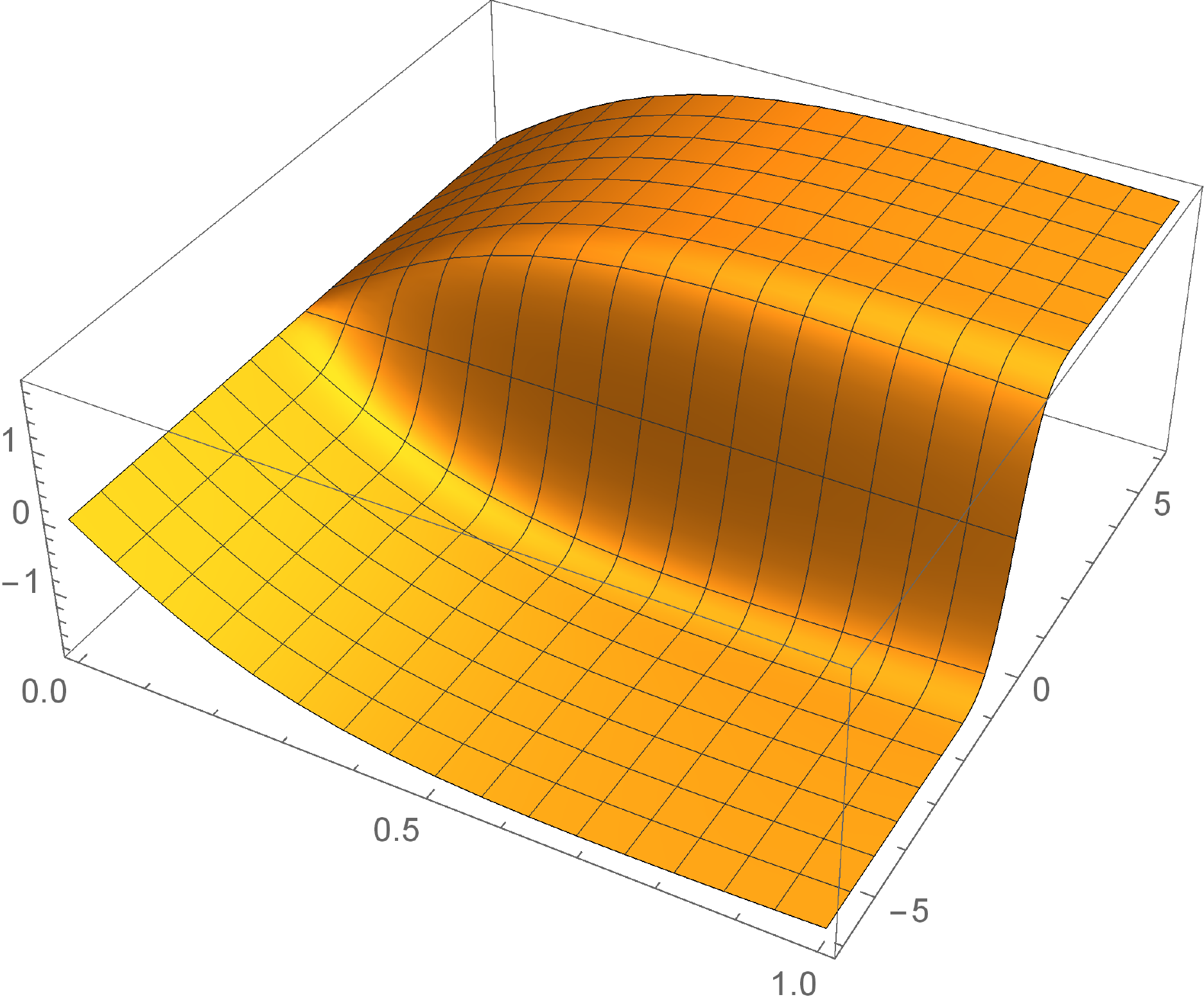}\qquad{}\includegraphics[scale=0.4]{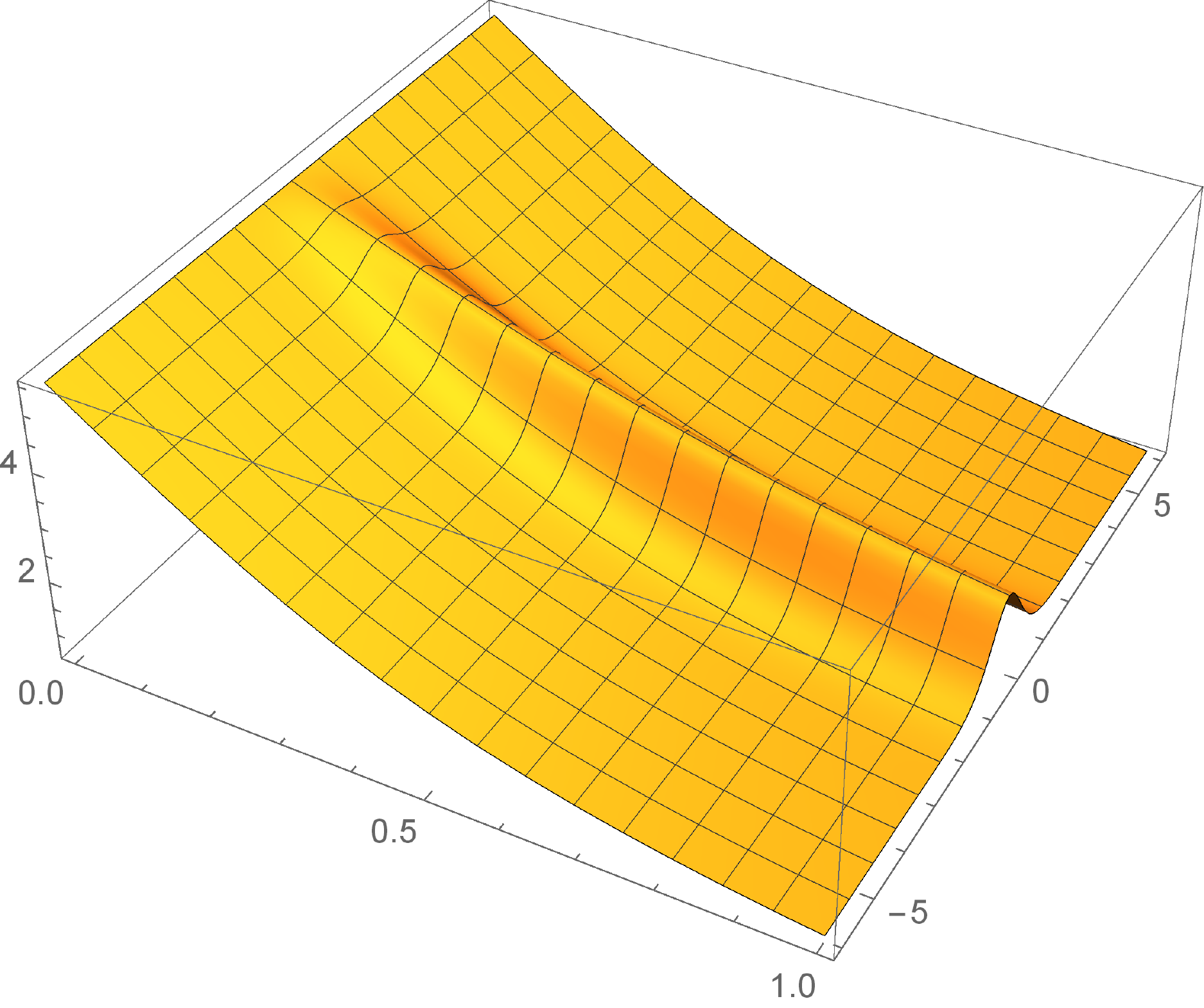}

\caption{The bulk configurations of the fields $\phi\left(z,x\right)$ (left
panel) and $A_{t}\left(z,x\right)$ (right panel) for $\mu=5.5$\label{standard field for black soliton}}
\end{figure}

\begin{figure}
\includegraphics[scale=0.43]{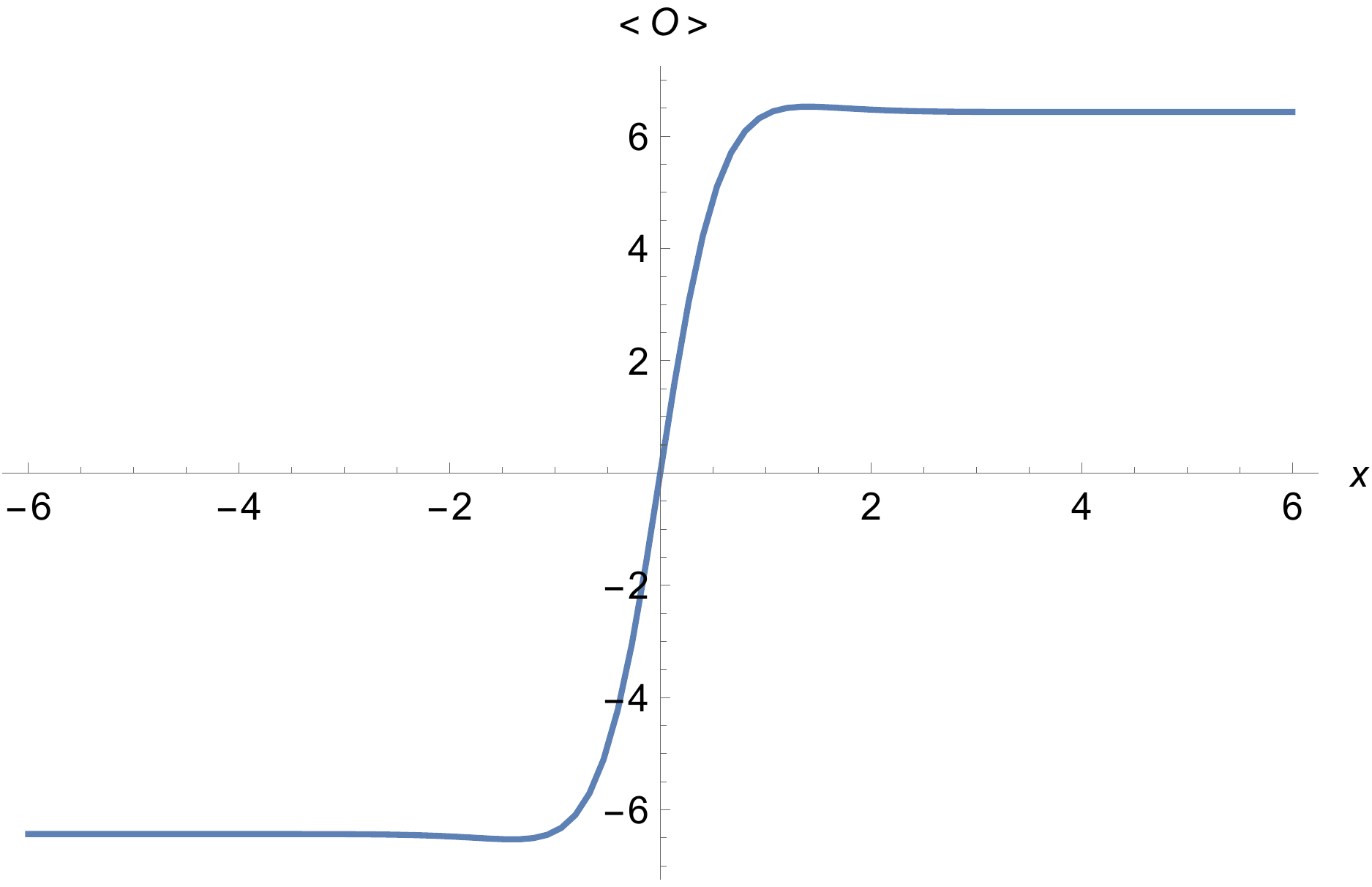}\quad{}\includegraphics[scale=0.4]{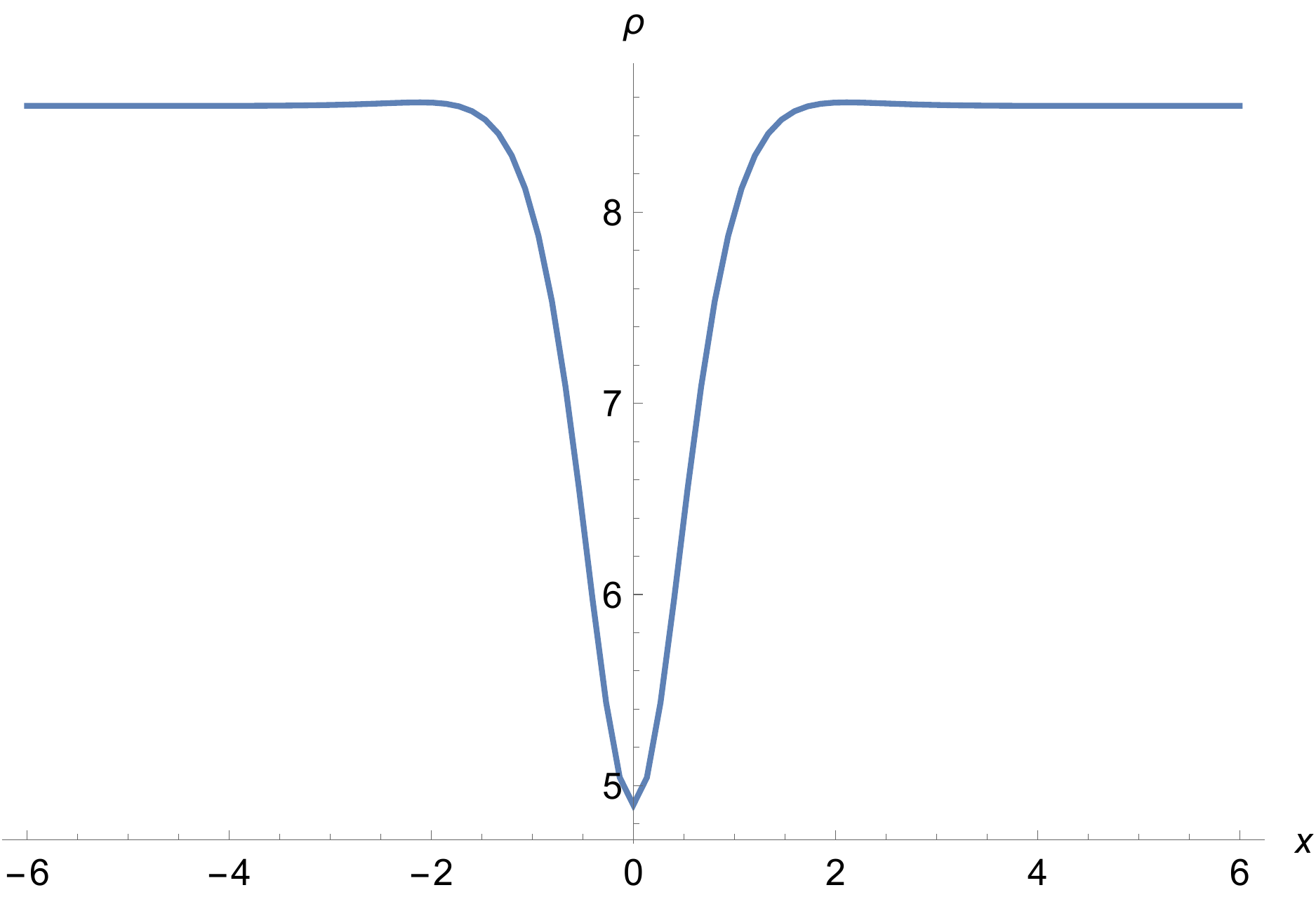}

\caption{The order parameter (left panel) and particle number density (right
panel) with respect to $x$ for $\mu=5.5$\label{order and density in standard black}}
\end{figure}

For standard quantization and the chemical potential $\mu=5.5$, we
present the numerical results of the bulk field configurations in
Fig.\ref{standard field for black soliton}. The order parameter and
density as functions of $x$ are shown in Fig.\ref{order and density in standard black}.
The asymptotical behavior for $A_{t}$ in the AdS boundary is written
as $A_{t}=\mu-\rho z$. Thus we can read off the charge density as
$\rho=-\partial_{z}A_{t}|_{z=0}$.

Fig.\ref{order and density in standard black} (right panel) shows
that there is a density depletion around the centers of the black
solitons. As well, the order parameter is zero at the center and dramatically
changes around the center. Far away enough (much larger than the characteristic
scale set by the so-called healing length) from the cores of the black
solitons, the order parameter is homogeneous.

\begin{figure}
\includegraphics[scale=0.4]{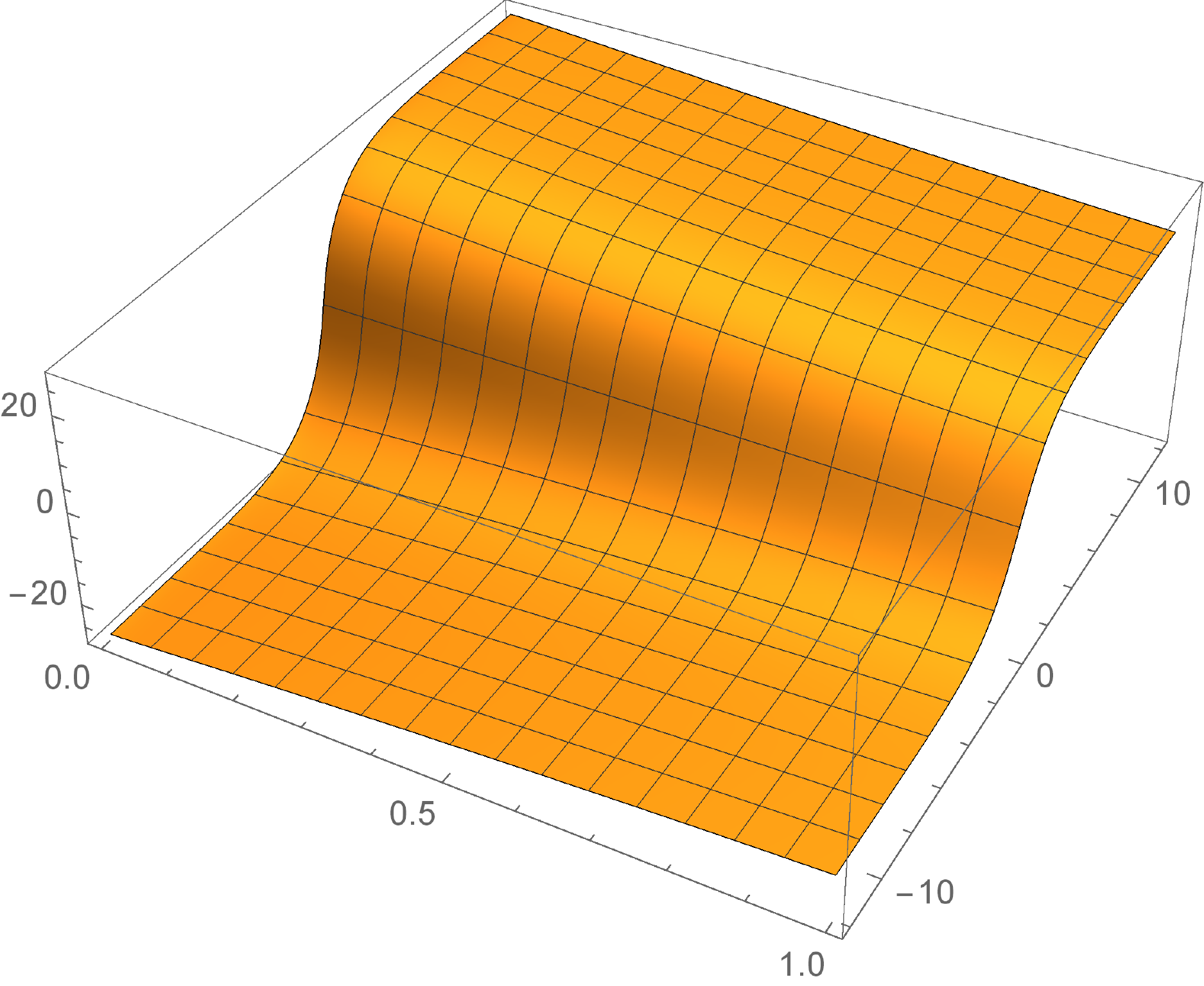}\qquad{}\includegraphics[scale=0.4]{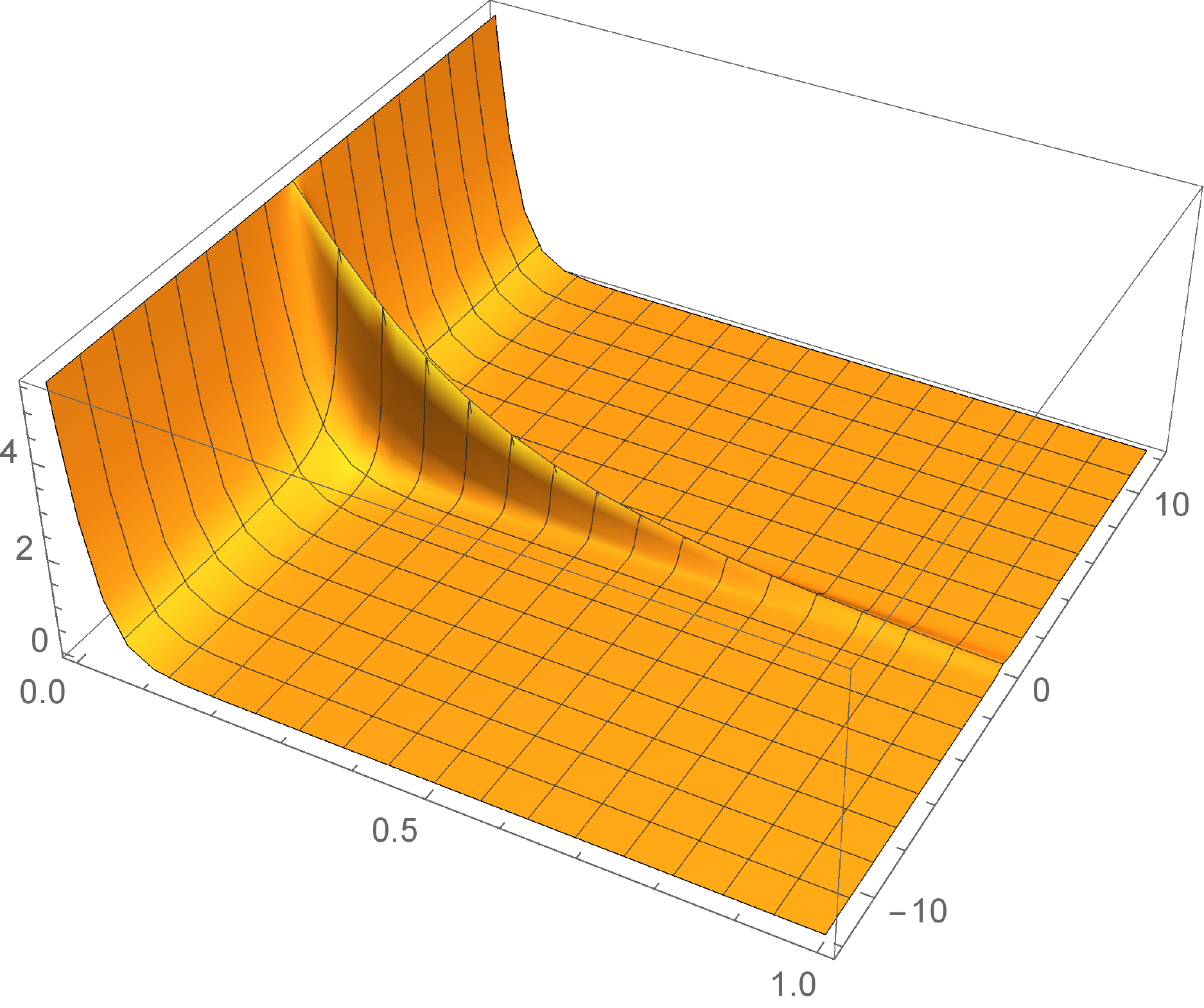}

\caption{The bulk configurations of the fields $\phi\left(z,x\right)$ (left
panel) and $A_{t}\left(z,x\right)$ (right panel) for $\mu=5.5$\label{alternative field for black soliton}}
\end{figure}

\begin{figure}
\includegraphics[scale=0.4]{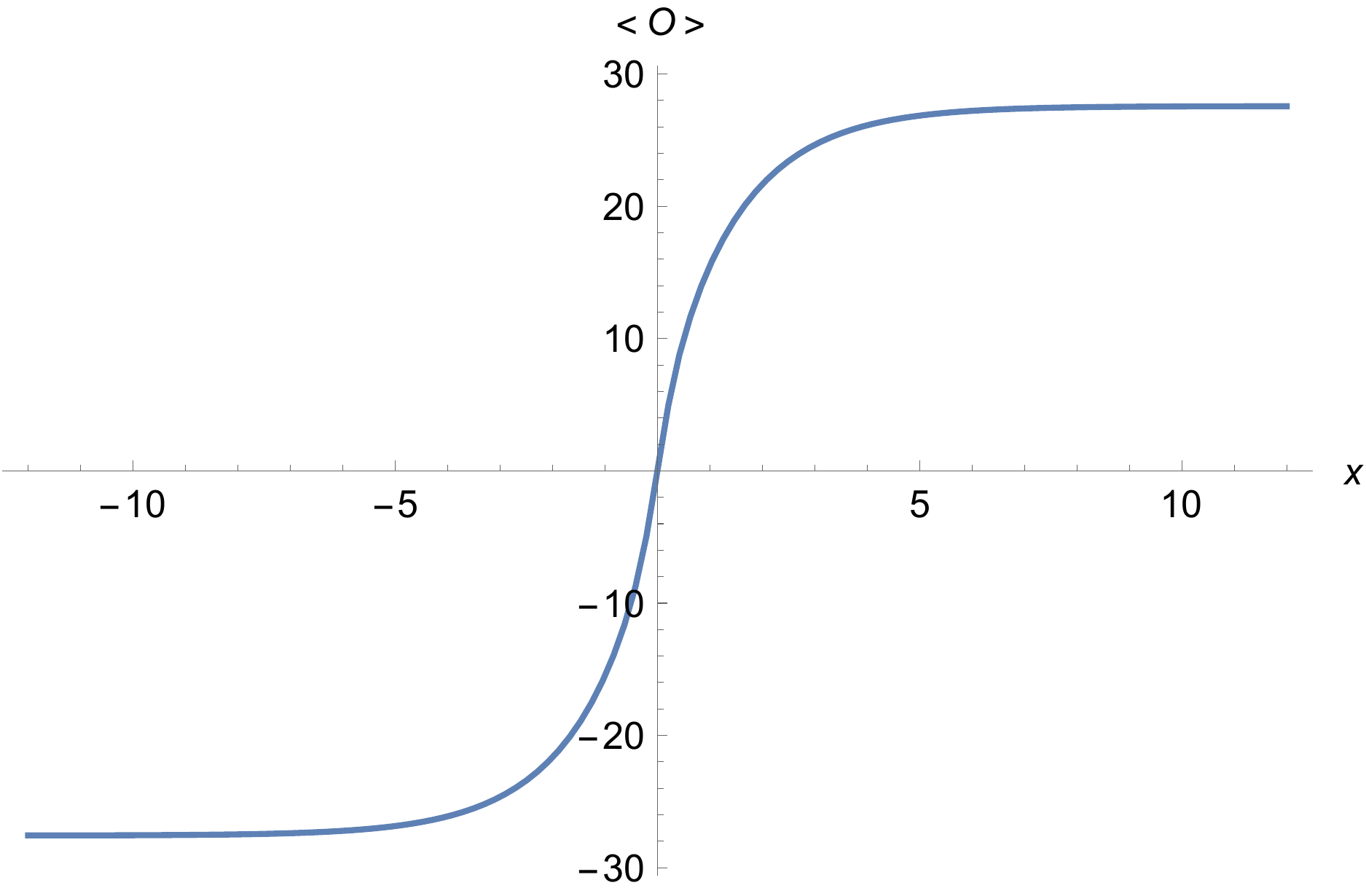}\includegraphics[scale=0.4]{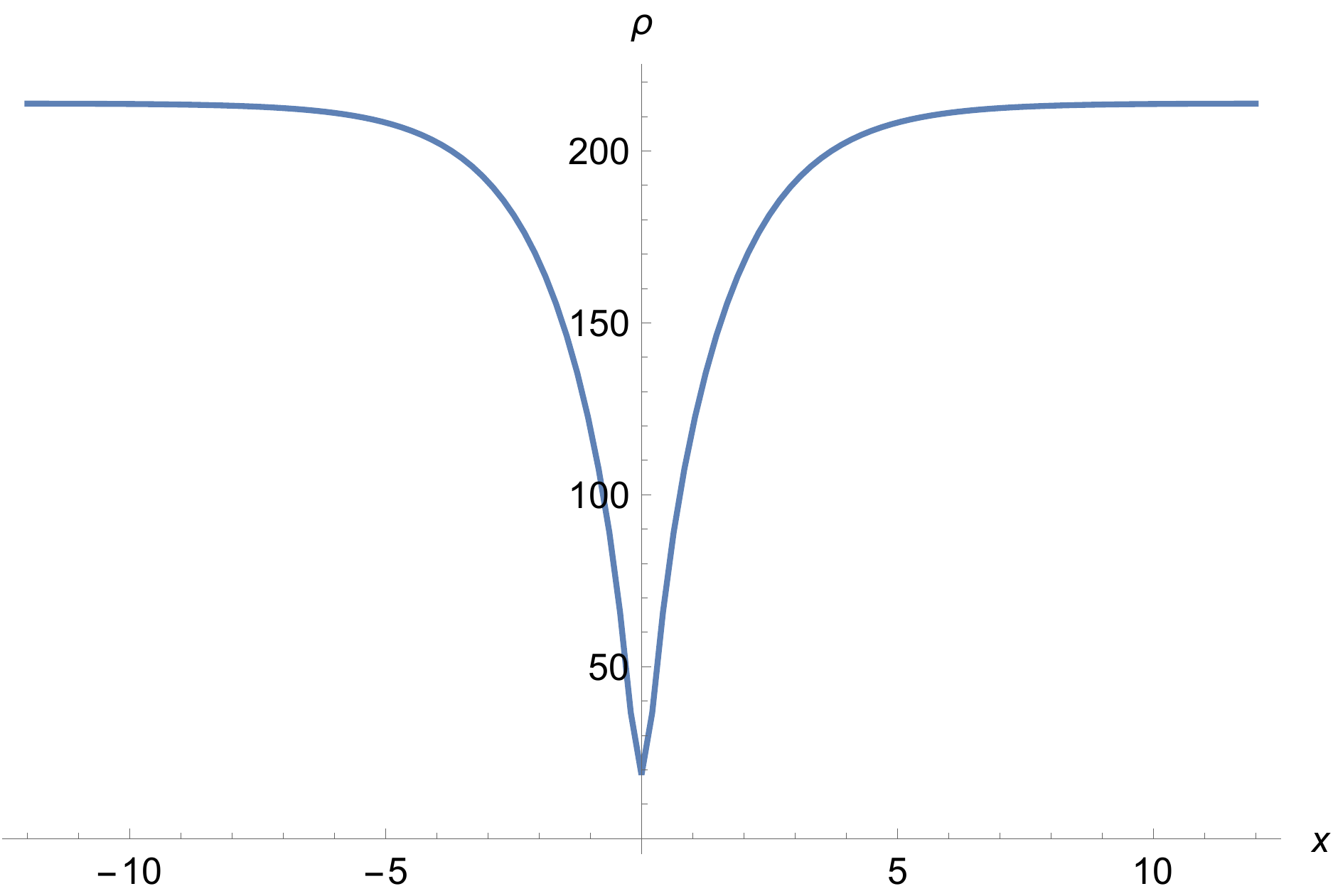}

\caption{The order parameter (left panel) and particle number density (right
panel) with respect to $x$ for $\mu=5.5$\label{order and density in alternative black}}
\end{figure}

For the alternative quantization, we show the numerical results of
the bulk field configurations in Fig.\ref{alternative field for black soliton}.
The order parameter and particle number density as functions of $x$
are shown in Fig.\ref{order and density in alternative black}.

\begin{figure}
\includegraphics[scale=0.45]{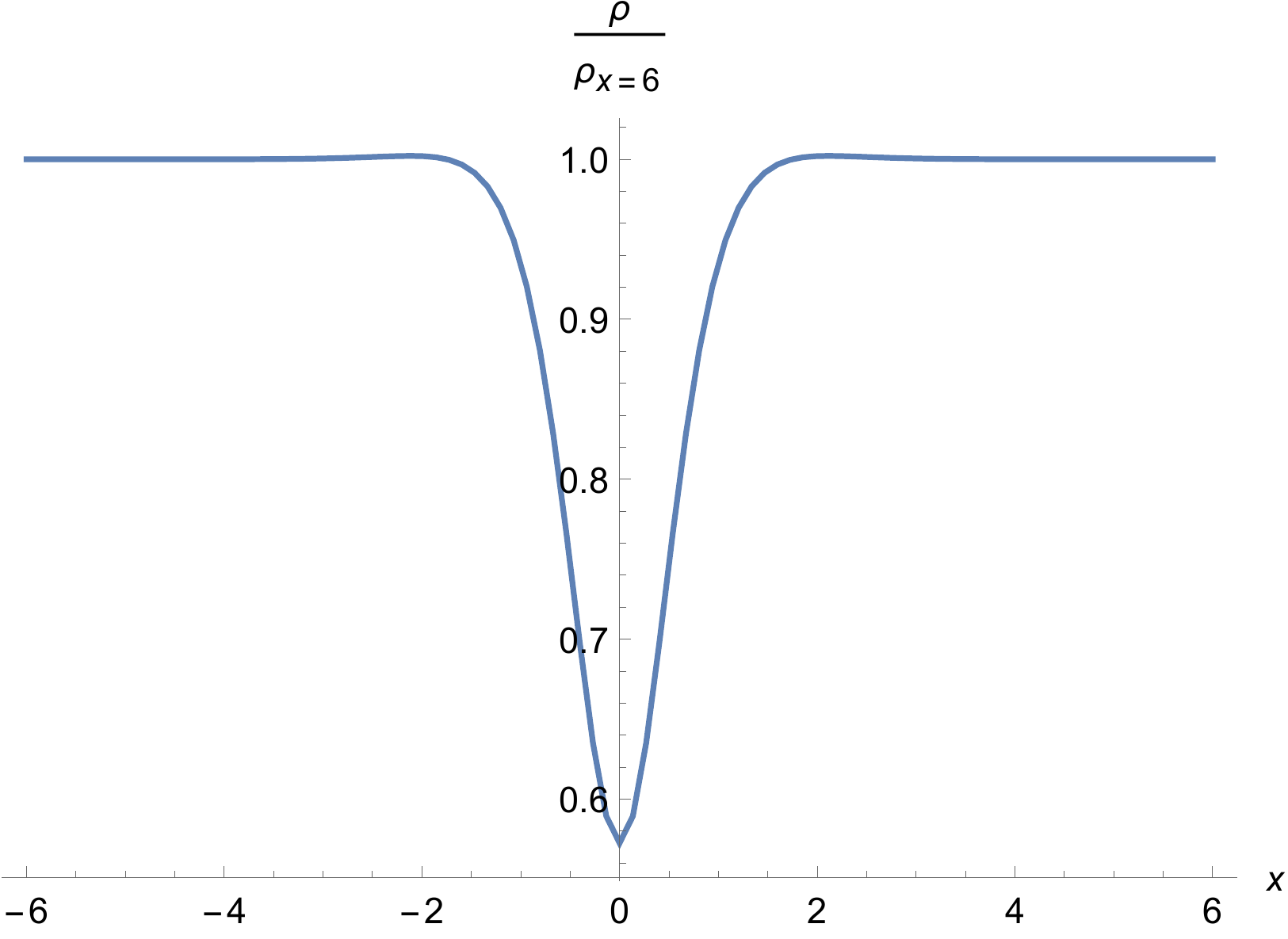}\qquad{}\includegraphics[scale=0.4]{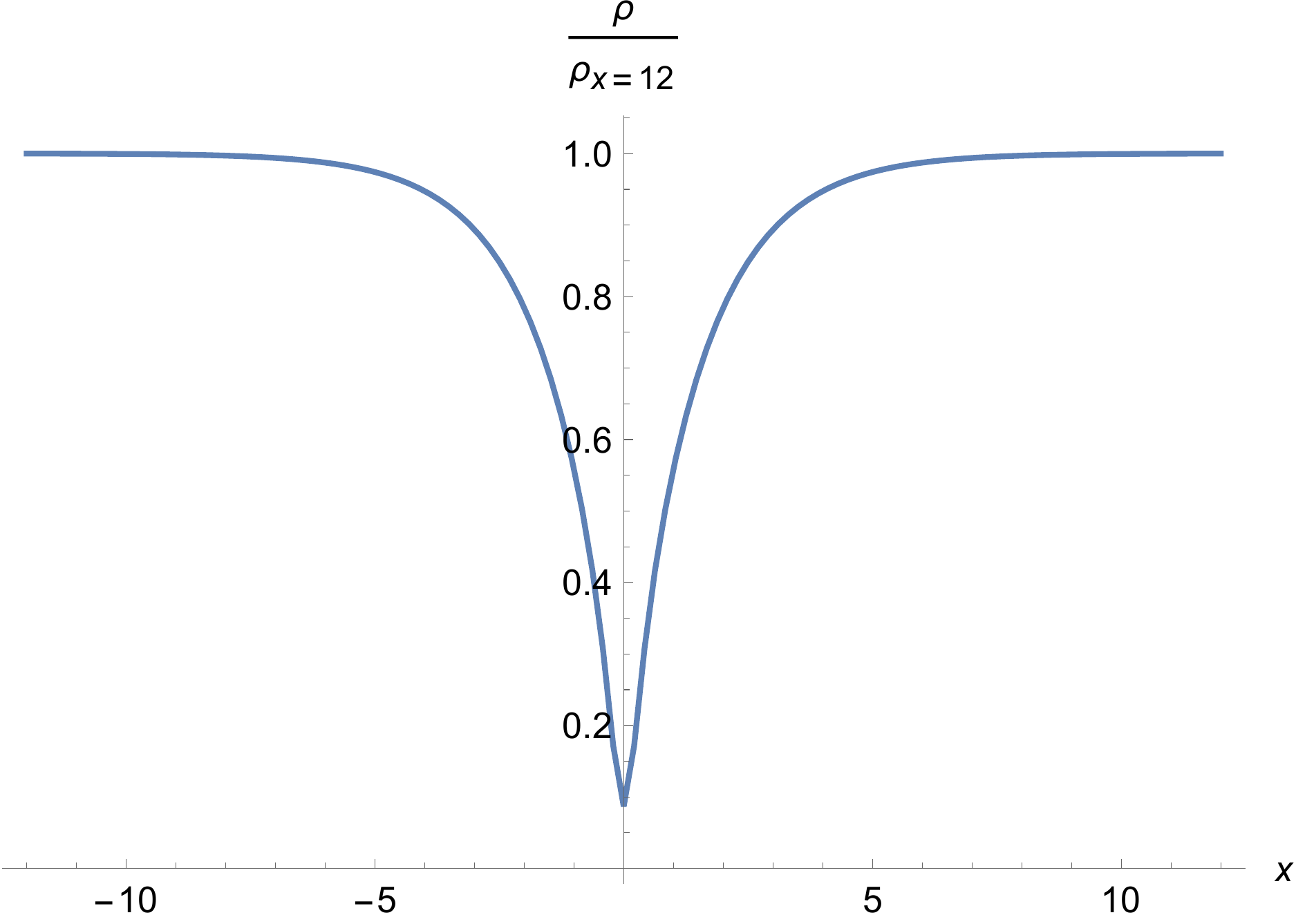}

\caption{The normalized density profile ($\mu=5.5$) for the standard quantization
(left panel) and the alternative quantization (right panel). The values
at the soliton cores are 0.572606 and 0.0866772 for the standard and
alternative quantizations, respectively.\label{normalized density in black}}
\end{figure}

\begin{figure}
\includegraphics[scale=0.5]{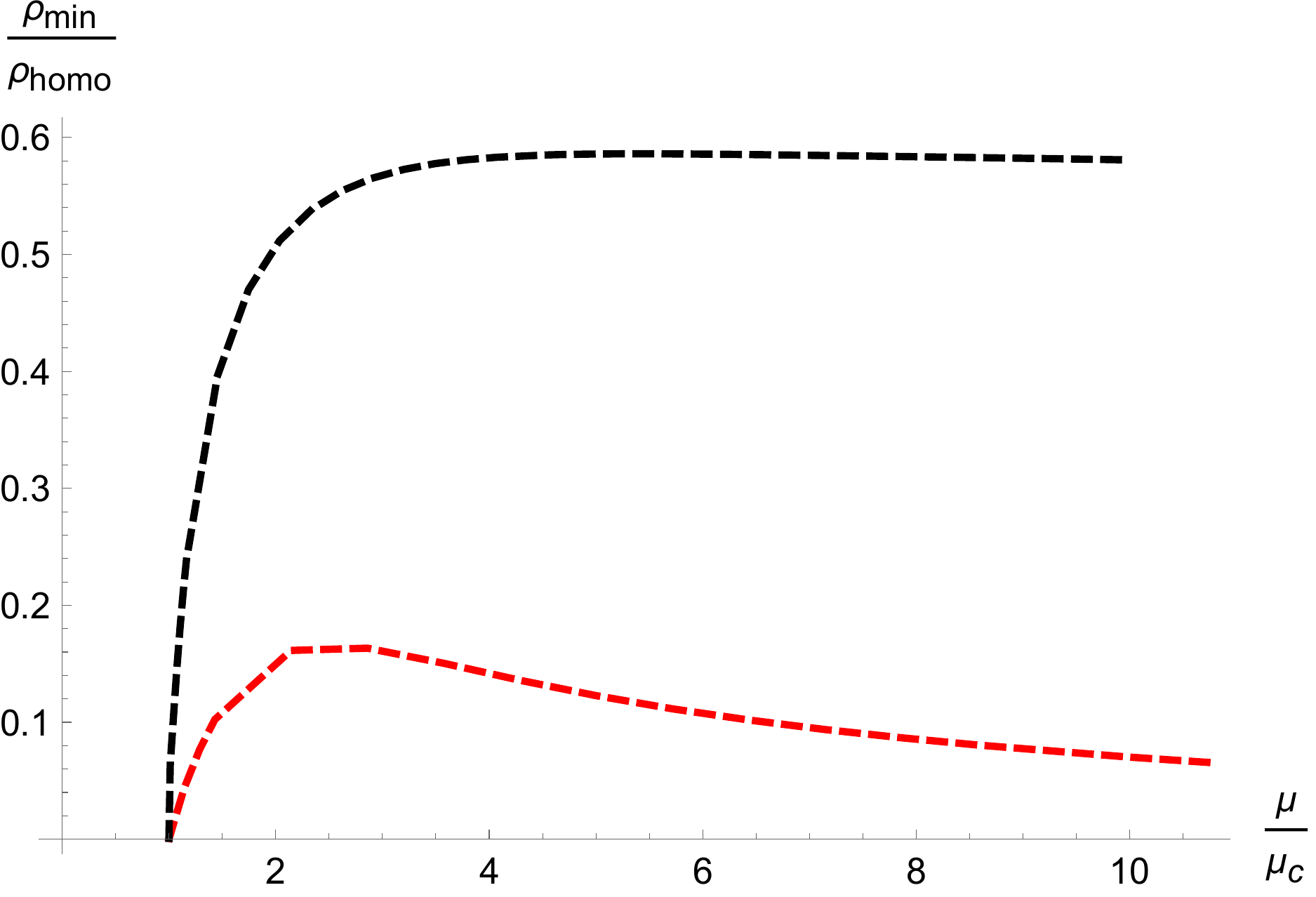}

\caption{The specific value of density $\rho_{min}/\rho_{homo}$ with respect
to the relative chemical potential $\mu/\mu_{c}$ for the standard
(black dashed) and alternative (red dashed) quantizations.\label{depletion difference in black}}
\end{figure}

We note that the density depletion at the soliton core is much deeper
in Fig.\ref{order and density in alternative black} than that in
Fig.\ref{order and density in standard black}. In order to see it
more clearly, we present the normalized density profiles in Fig.\ref{normalized density in black}.
We also show the density depletion fraction with respect to the chemical
potential (in units of its critical value) in Fig.\ref{depletion difference in black}
to clarify the difference between the two quantizations. And also
$\rho_{homo}$ represents the particle number density in the homogeneous
superfluid.

It is interesting to note that the density depletion for holographic
solitons is strongly dependent on which type of quantization we choose.
In \cite{Antezza}, it is shown that the density depletion fraction
for solitons from the BdG equations (at zero temperature) is related
to whether the system is BEC-type (large depletion) or BCS-type (small
depletion). In Fig.\ref{depletion difference in black}, we note that
the density depletion fraction is larger for the alternative quantization
than the standard one. In particular, it can be seen that under the
large $\mu/\mu_{c}$ limit the fraction $\rho_{min}/\rho_{homo}$
tends to a constant finite value ($\approx0.6$) for the standard
quantization while it seems to approach zero or some value smaller
than 0.1 for the alternative case. Here the relative chemical potential
$\mu/\mu_{c}$ comes into play because this holographic superfluid
model has an extra scale (fixed as $z_{0}=1$ in our discussion) instead
of a temperature and the large $\mu/\mu_{c}$ limit just means that
this extra scale becomes unimportant enough to be neglected, in order
not to spoil our comparison with the discussion in \cite{Antezza}.
Therefore, we see that the holographic superfluids at the standard
and alternative quantizations resemble the BCS-type and BEC-type fermionic
superfluids, respectively, consistent with the proposal in\cite{Chaolun}.

\begin{figure}
\includegraphics[scale=0.4]{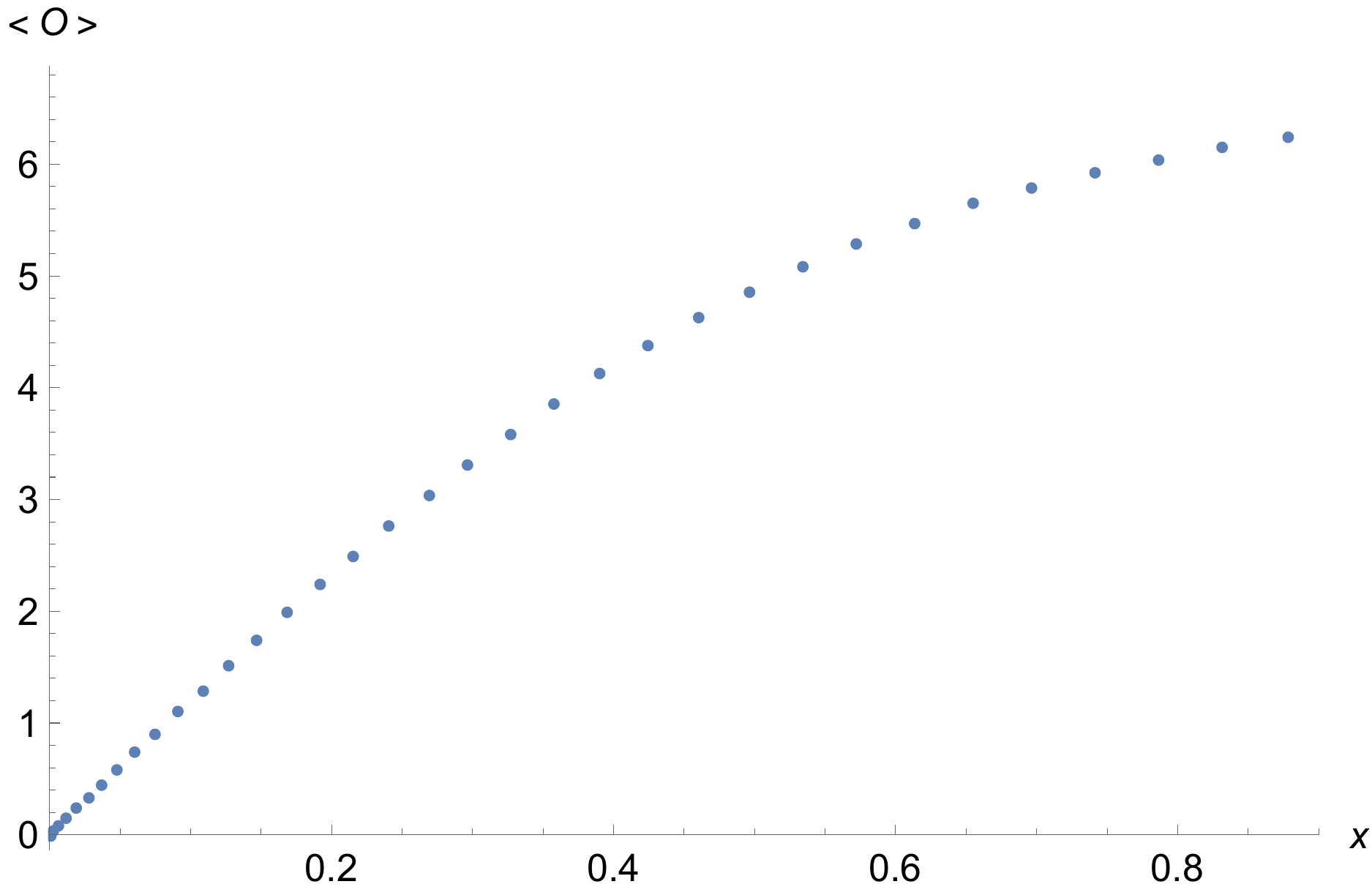} \qquad{}\includegraphics[scale=0.4]{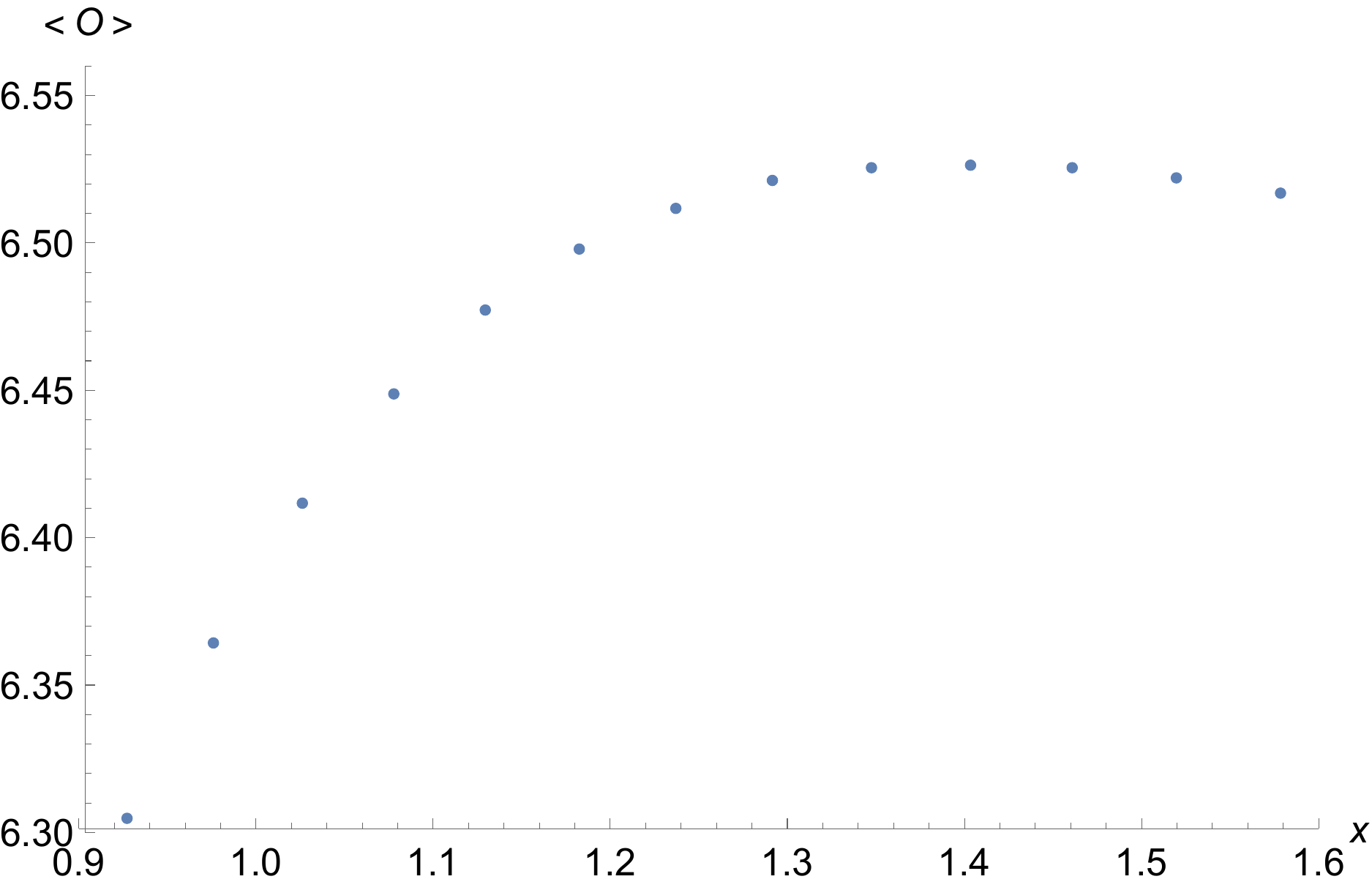}

\includegraphics[scale=0.4]{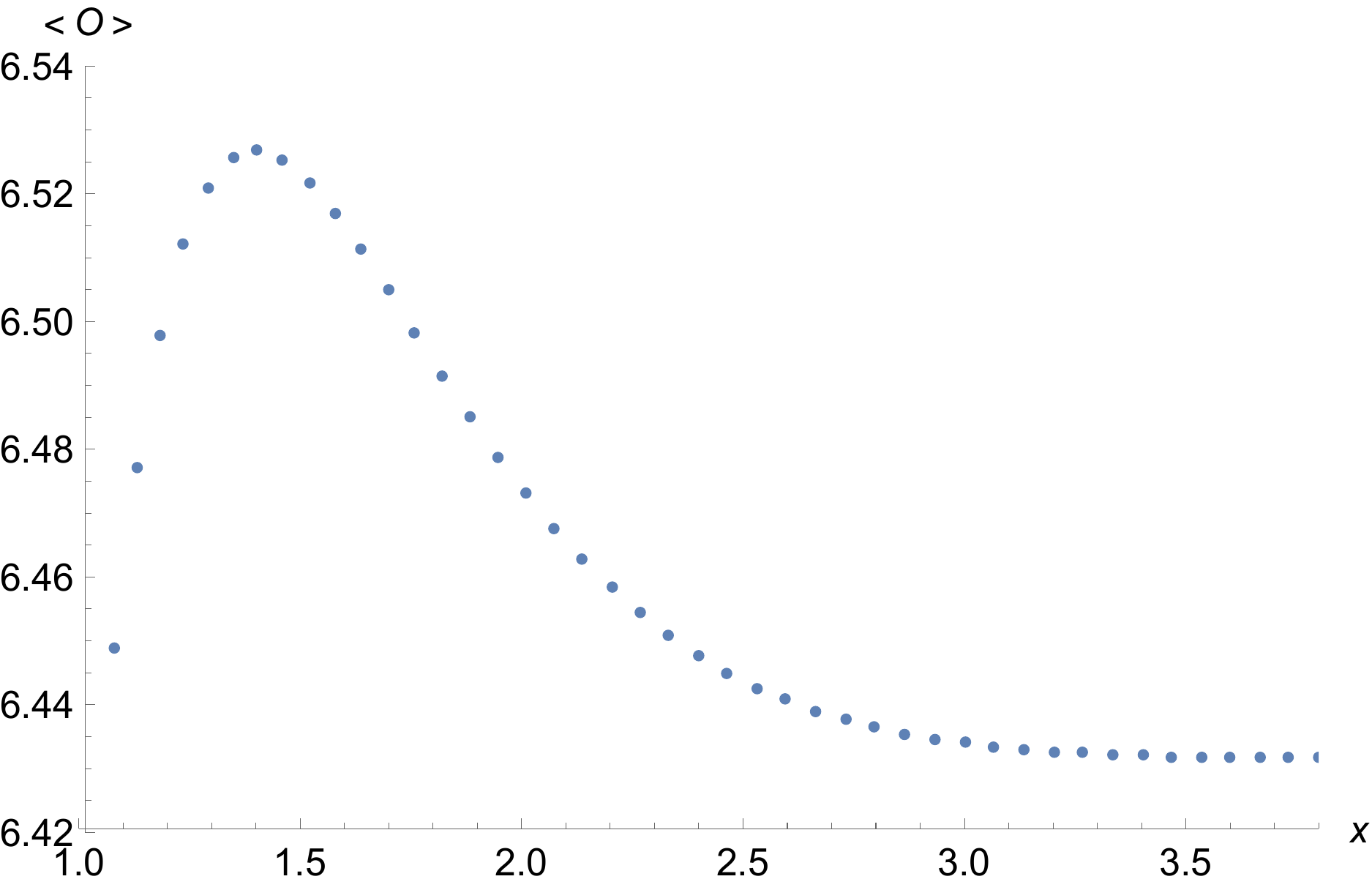} \qquad{}\includegraphics[scale=0.4]{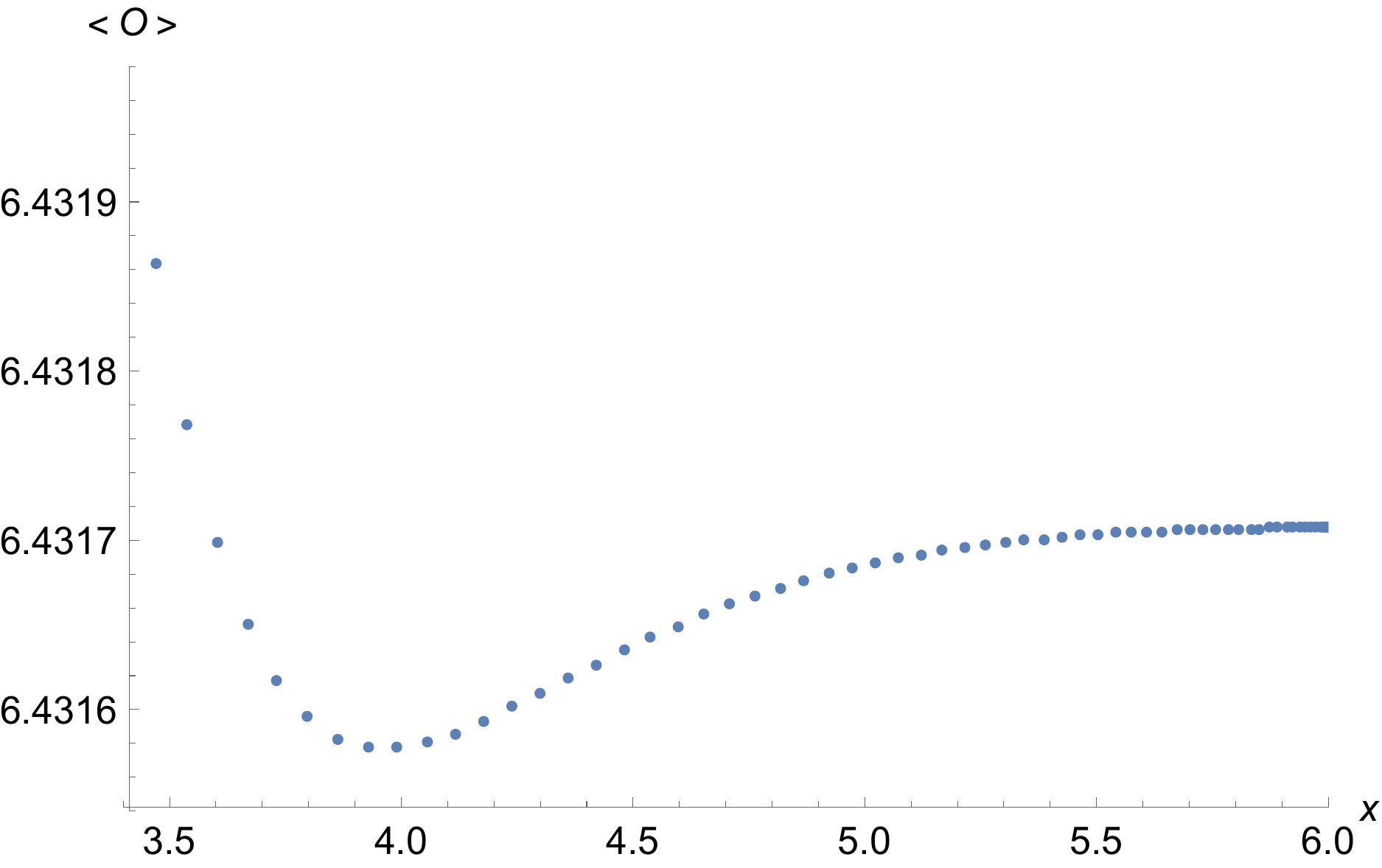}

\caption{The magnified order parameter, at $\mu=5.5$, as a function of $x$
from 0 to 6 (half plot of the box) under the standard quantization.
It is clearly shown that they have fluctuations.\label{standard magnified order black}}
\end{figure}

\begin{figure}
\includegraphics[scale=0.4]{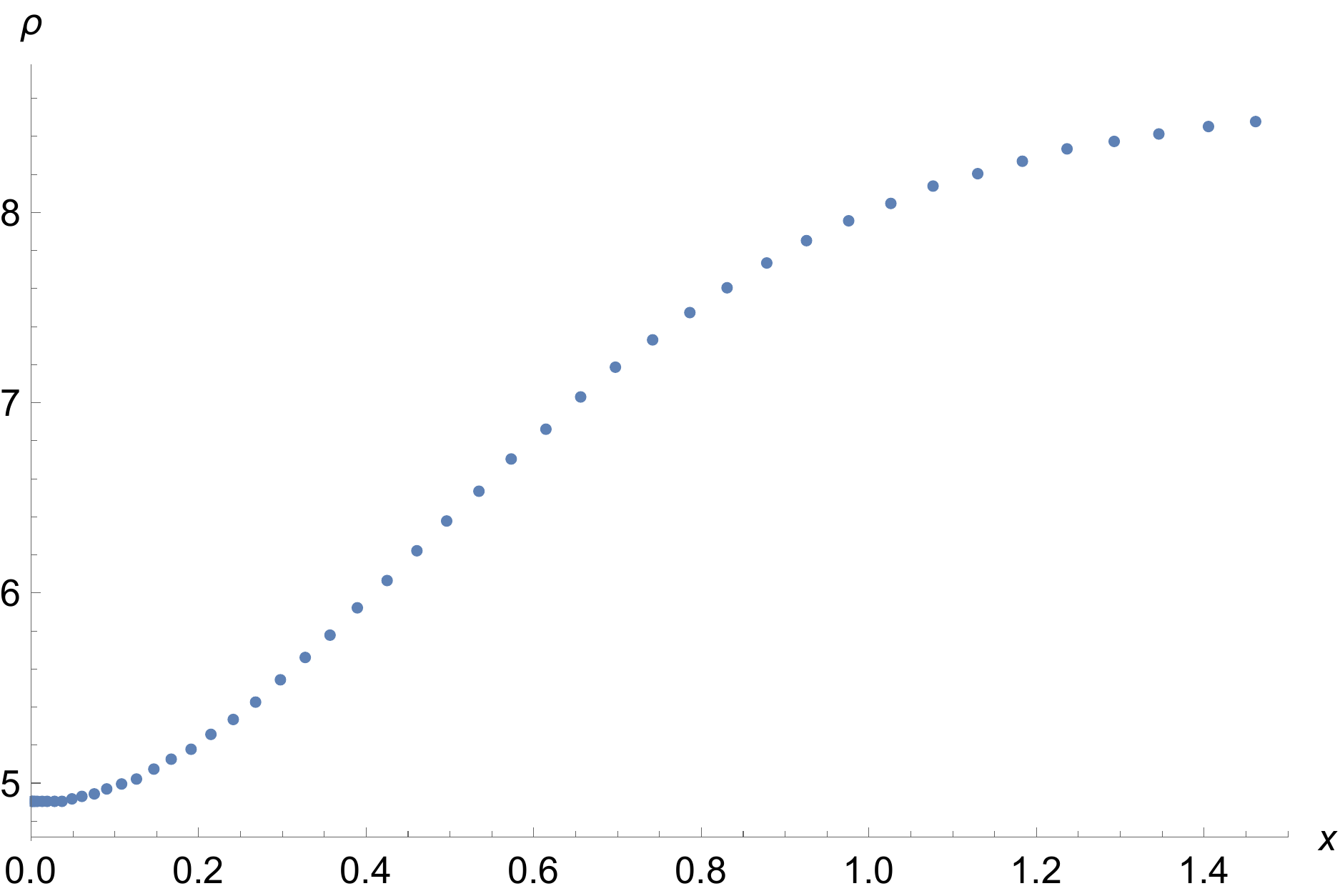} \qquad{}\includegraphics[scale=0.4]{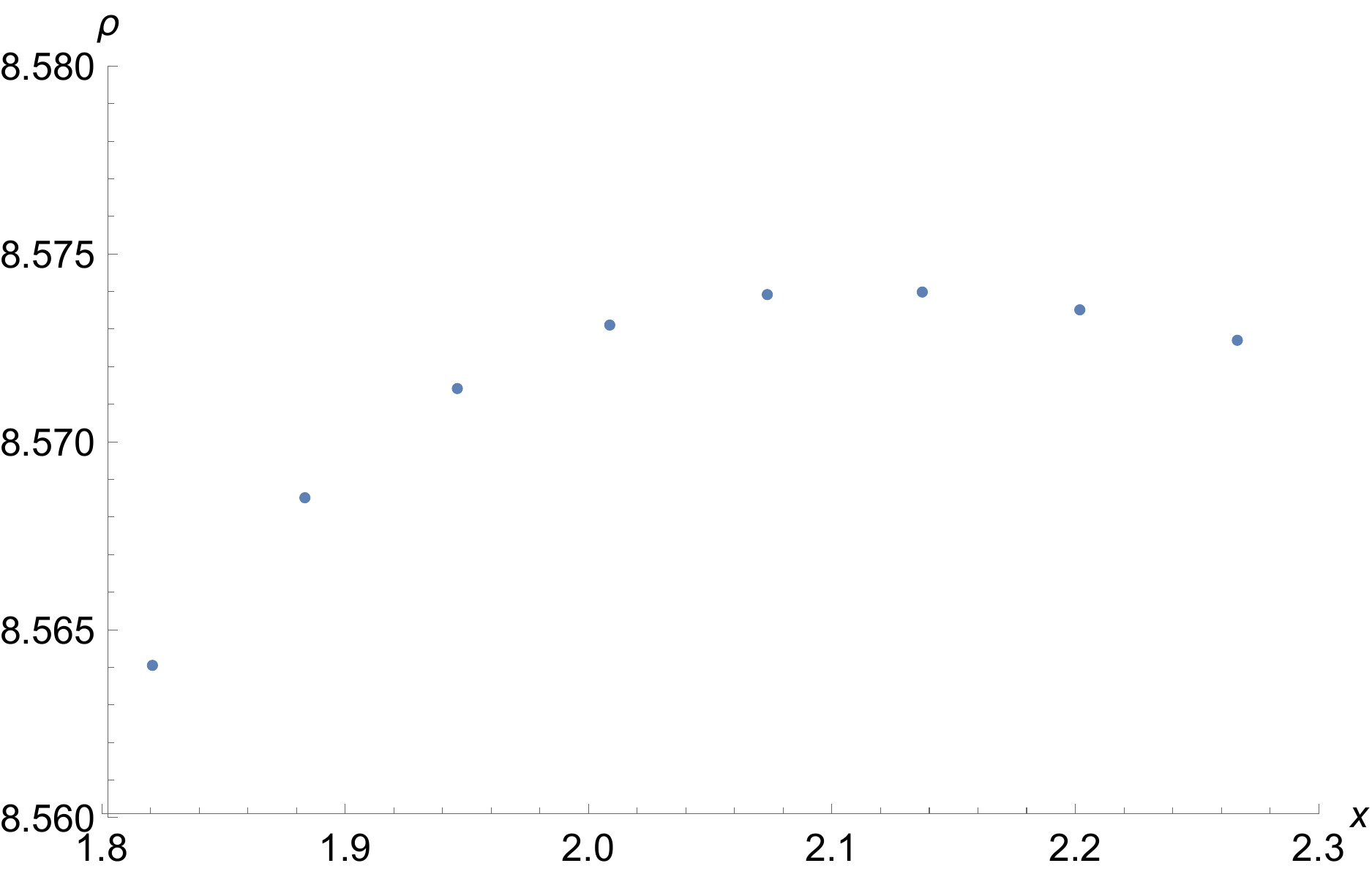}

\includegraphics[scale=0.4]{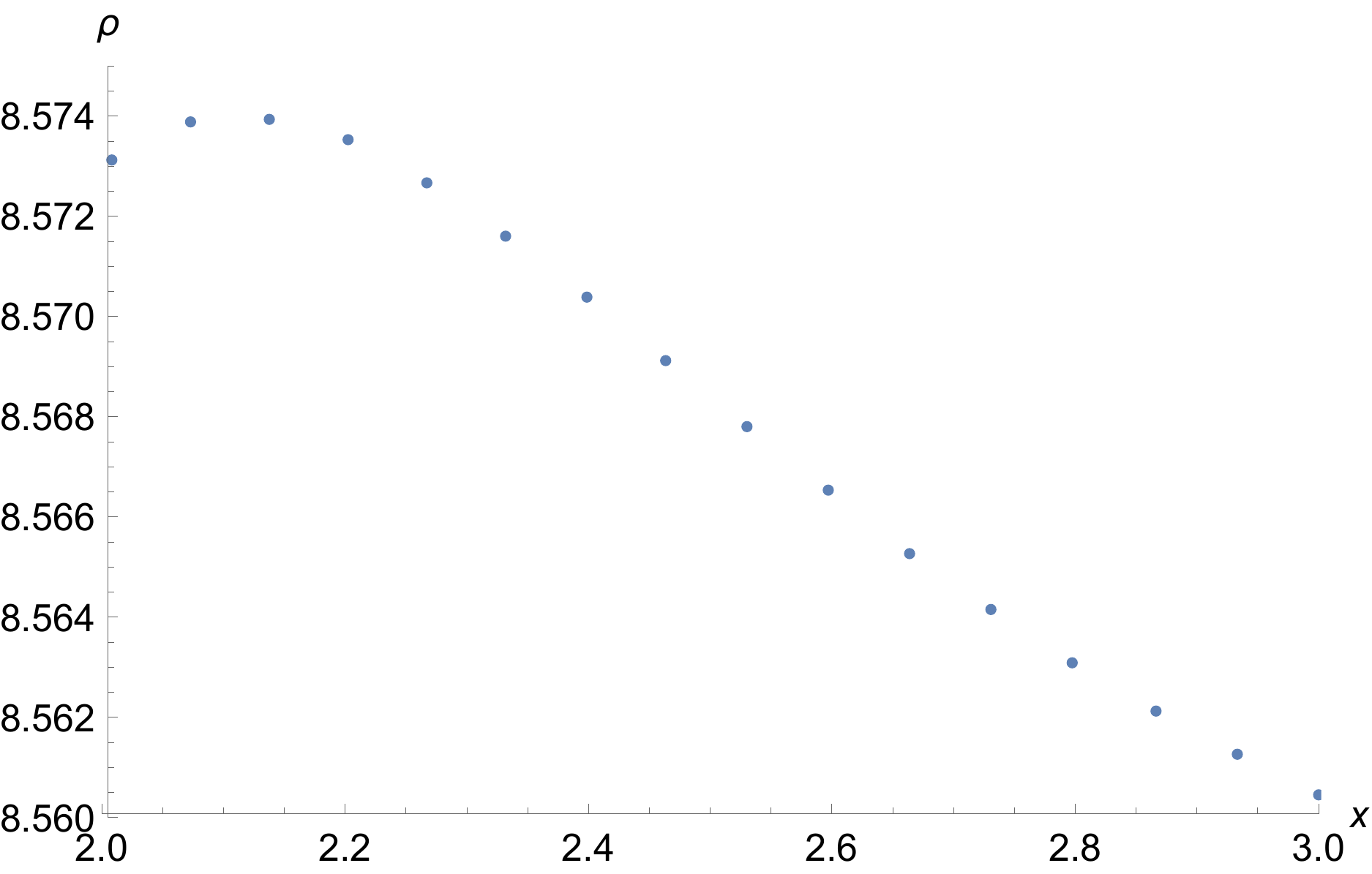} \qquad{}\includegraphics[scale=0.4]{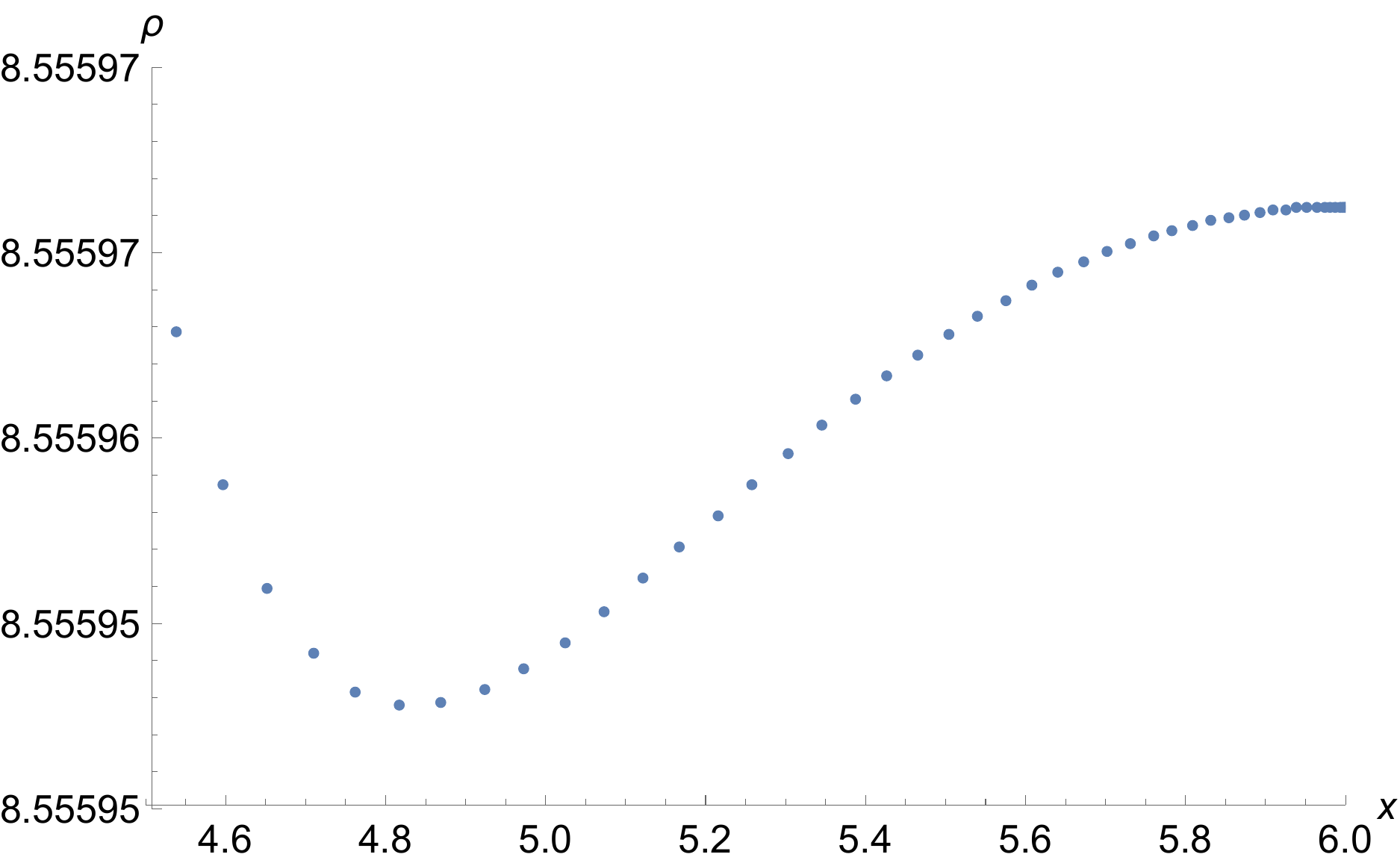}

\caption{The magnified particle number density, at $\mu=5.5$, as a function
of $x$ from 0 to 6 (half plot of the box) under the standard quantization.
It is clearly shown that they have fluctuations.\label{standard magnified charge black}}
\end{figure}

\begin{figure}
\includegraphics[scale=0.4]{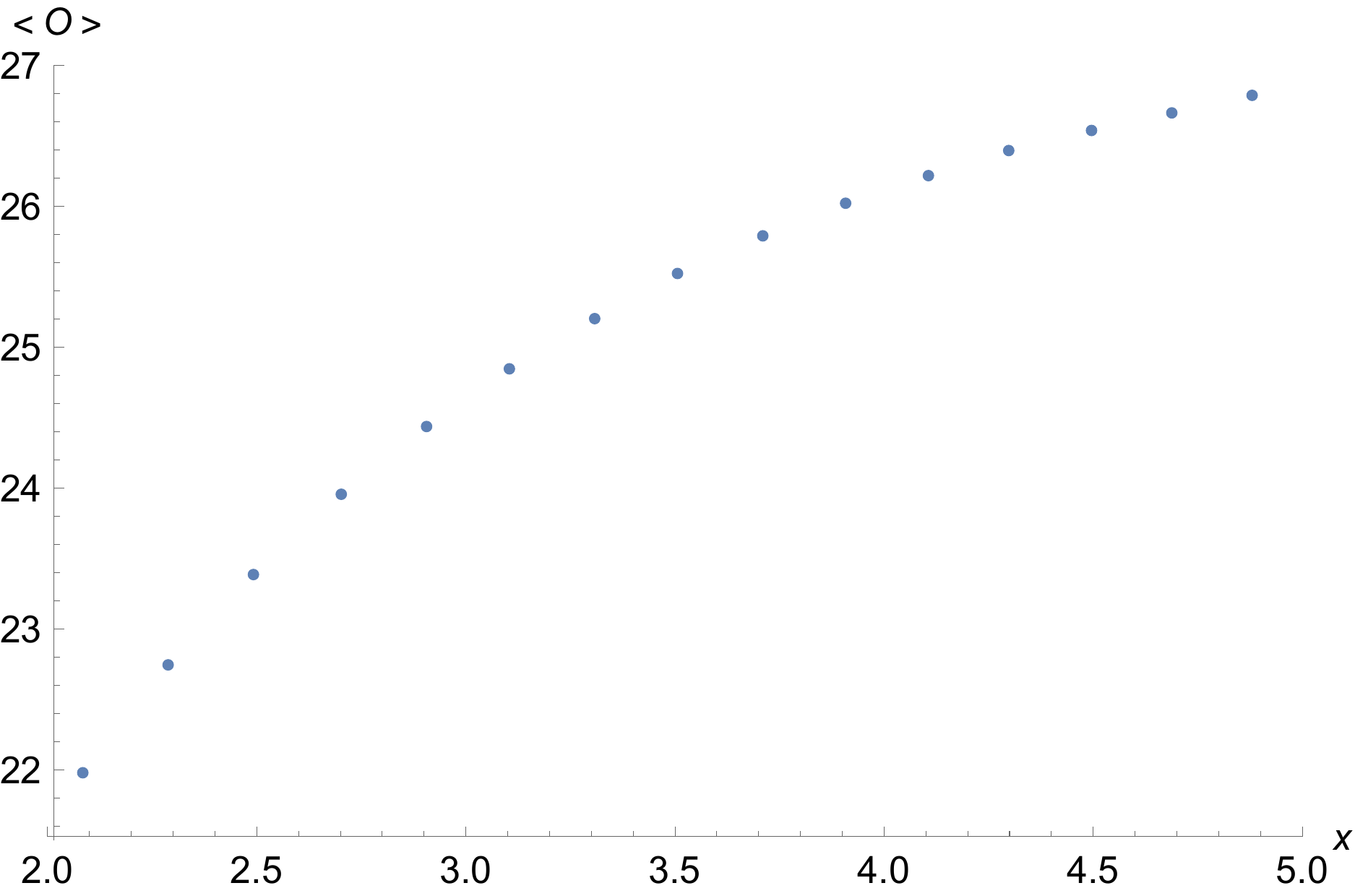} \qquad{}\includegraphics[scale=0.4]{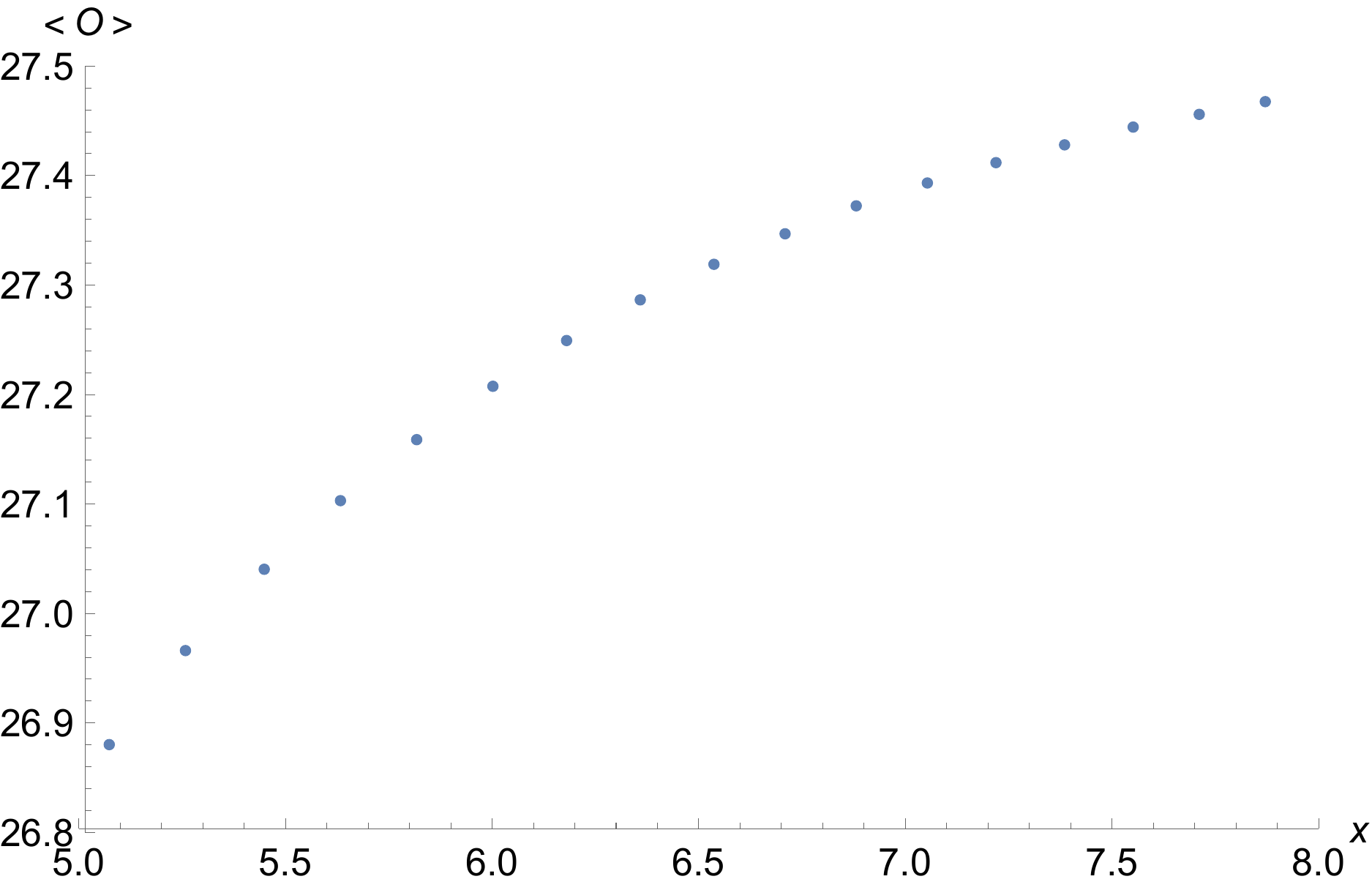}

\includegraphics[scale=0.4]{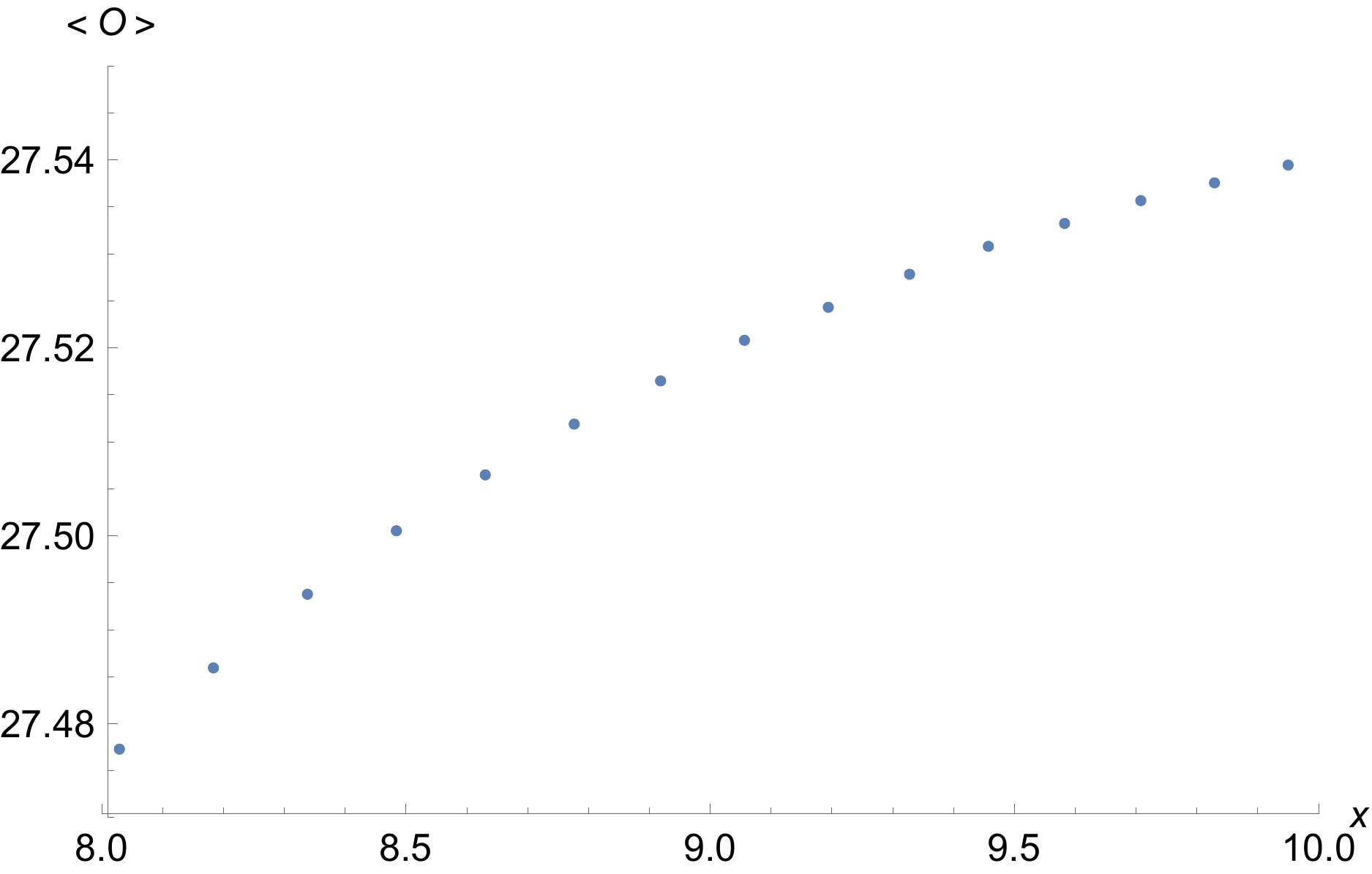} \qquad{}\includegraphics[scale=0.4]{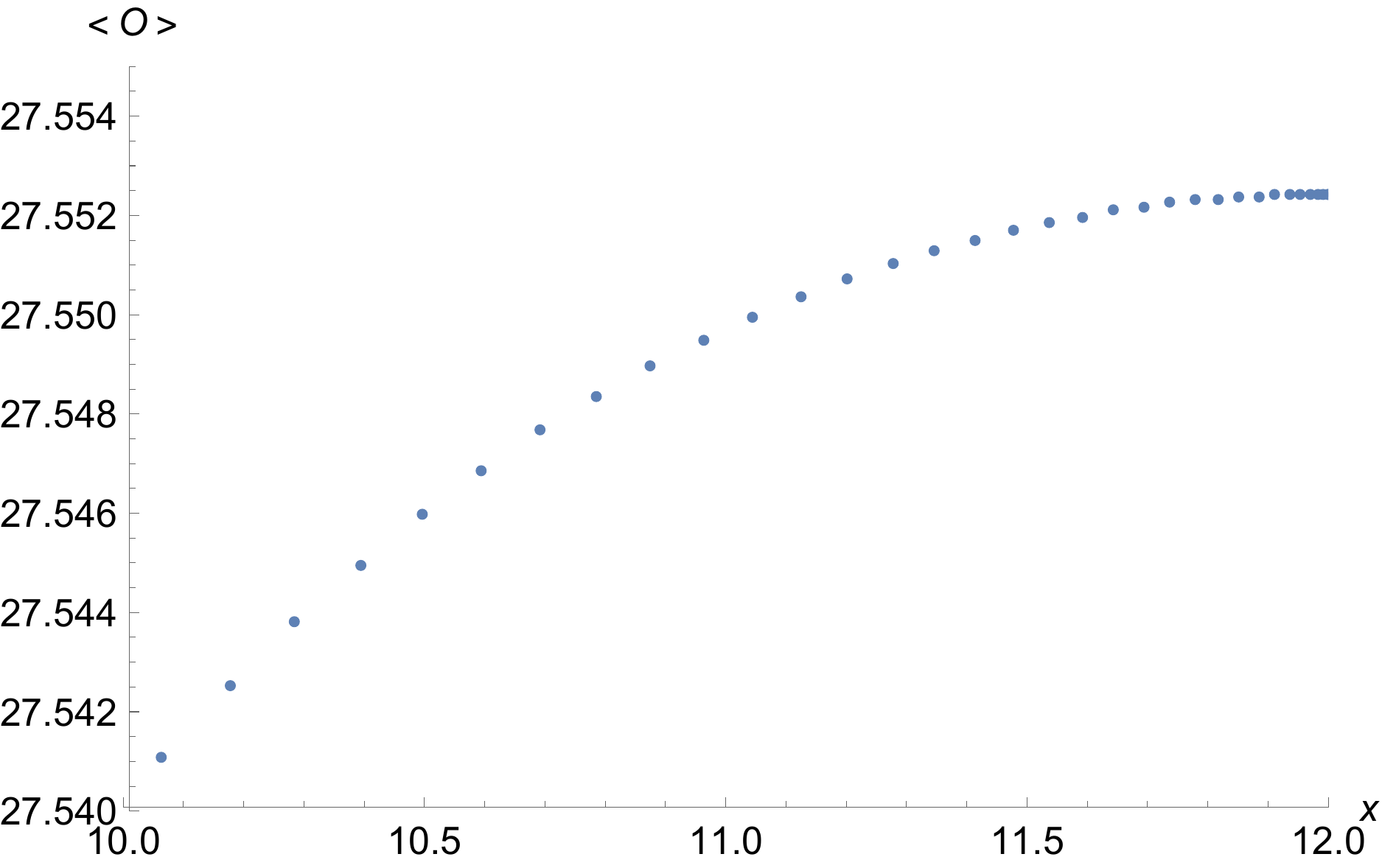}

\caption{The magnified order parameter under the alternative quantization with
the chemical potential $\mu=5.5$. It can be seen that the order parameter
monotonously increases in the $x$ direction.\label{alternative magnified order}}
\end{figure}

Under the standard quantization, there is another interesting feature
of the black soliton solutions. Peculiar fluctuations, weak but identifiable,
appear around the edges of cliffs of the order parameter and density
profiles. In order to further explore the fluctuations, we zoom in
on the order parameter and density profiles and present the results
in Fig.\ref{standard magnified order black} and Fig.\ref{standard magnified charge black},
respectively. These fluctuation behaviors reminds us the Friedel oscillation
in the black soliton configurations in the BCS regime of fermionic
superfluids\cite{Antezza}, which stems from the fact that there are
two length scales, the length scale $k_{F}^{-1}$ corresponding to
the Fermi momentum and the coherence length of Cooper pairs, in the
BCS-type superfluids. In our case, it is also likely that the fluctuation
behaviors result from the interference of two length scales. Moreover,
no such fluctuation can be observed under the alternative quantization,
even after we zoom in on the order parameter and density profiles,
as shown in Fig.\ref{alternative magnified order}. All these facts
appear to be consistent with the proposal in \cite{Keranen}. Additional
discussions on the holographic BCS-like configurations can be found
in Appendix \ref{sec:fit}.

\section{Gray solitons}

Then we will investigate non-static local structures in superfluids,
namely gray solitons, which are really new in the context of holography.
In the zero temperature (holographic) superfluids, gray solitons will
keep traveling at constant speeds with their shapes exactly preserved.

In order to obtain such kinds of solutions, we take the following
useful trick: transforming the original coordinates into the comoving
coordinates with respect to a gray soliton. As a result, the gray
soliton is static for us in these comoving coordinates. Specifically,
\begin{align}
x' & =x-vt,\nonumber \\
\psi\left(z,t,x\right) & \rightarrow\psi\left(z,x'\right),\nonumber \\
A_{t}\left(z,t,x\right) & \rightarrow A_{t}\left(z,x'\right),\label{eq:coordinate transformation}\\
A_{x}\left(z,t,x\right) & \rightarrow A_{x}\left(z,x'\right),\nonumber 
\end{align}
where $v$ is the velocity of the gray soliton, $x'$ is the comoving
coordinate and $x$ is the original coordinate (reference frame of
the superfluid background).

Then we decompose the complex scalar field $\psi$ into its real part
and imaginary part, namely $\psi\left(x-vt,z\right)=a\left(x-vt,z\right)+ib\left(x-vt,z\right)$.
Finally the equations of gray solitons are written as 
\begin{align}
0= & \left(v^{2}-1\right)\partial_{x}^{2}a+a\left(z+A_{x}^{2}-A_{t}^{2}\right)-b\left(v\partial_{x}A_{t}+\partial_{x}A_{x}\right)-2A_{x}\partial_{x}b-2vA_{t}\partial_{x}b\nonumber \\
 & +3z^{2}\partial_{z}a+\left(z^{3}-1\right)\partial_{z}^{2}a,\label{eq:gray real}\\
0= & \left(v^{2}-1\right)\partial_{x}^{2}b+a\left(v\partial_{x}A_{t}+\partial_{x}A_{x}\right)+b\left(z+A_{x}^{2}-A_{t}^{2}\right)+2A_{x}\partial_{x}a+2vA_{t}\partial_{x}a\nonumber \\
 & +3z^{2}\partial_{z}b+\left(z^{3}-1\right)\partial_{z}^{2}b,\label{eq:gray imaginary}\\
0= & v^{2}\partial_{x}^{2}A_{x}+v\partial_{x}^{2}A_{t}+2b\partial_{x}a-2a\partial_{x}b+2A_{x}\left(a^{2}+b^{2}\right)+3z^{2}\partial_{z}A_{x}\nonumber \\
 & +\left(z^{3}-1\right)\partial_{z}^{2}A_{x},\label{eq:gray Ax-1}\\
0= & \left(z^{3}-1\right)\partial_{z}^{2}A_{t}+3z^{2}\partial_{z}A_{t}-\partial_{x}^{2}A_{t}-v\partial_{x}^{2}A_{x}+2A_{t}\left(a^{2}+b^{2}\right)+2va\partial_{x}b\nonumber \\
 & -2vb\partial_{x}a,\label{eq:gray At-1}\\
0= & -v\partial_{z}\partial_{x}A_{t}+2a\partial_{z}b-2b\partial_{z}a-\partial_{z}\partial_{x}A_{x},\label{eq:gray boundary-1}
\end{align}
where the last one can be taken as the constraint equation.

We still impose the same boundary conditions on the equations above
as in the black soliton case for $\psi$ and $A_{t}$. And we set
$A_{x}|_{x=\pm L}=0$. The most technical step is to fix the U(1)
gauge. First, we still use the gauge $A_{z}=0$. In addition, we will
set $A_{x}\left(z=0\right)=0$, which makes the local U(1) gauge fixed.
Next, we need to fix the global U(1) gauge by imposing the condition
$a=0$(where $z=1,x=0$). Here we also investigate the gray soliton
solutions under the standard and alternative quantizations separately.

\begin{figure}
\includegraphics[scale=0.23]{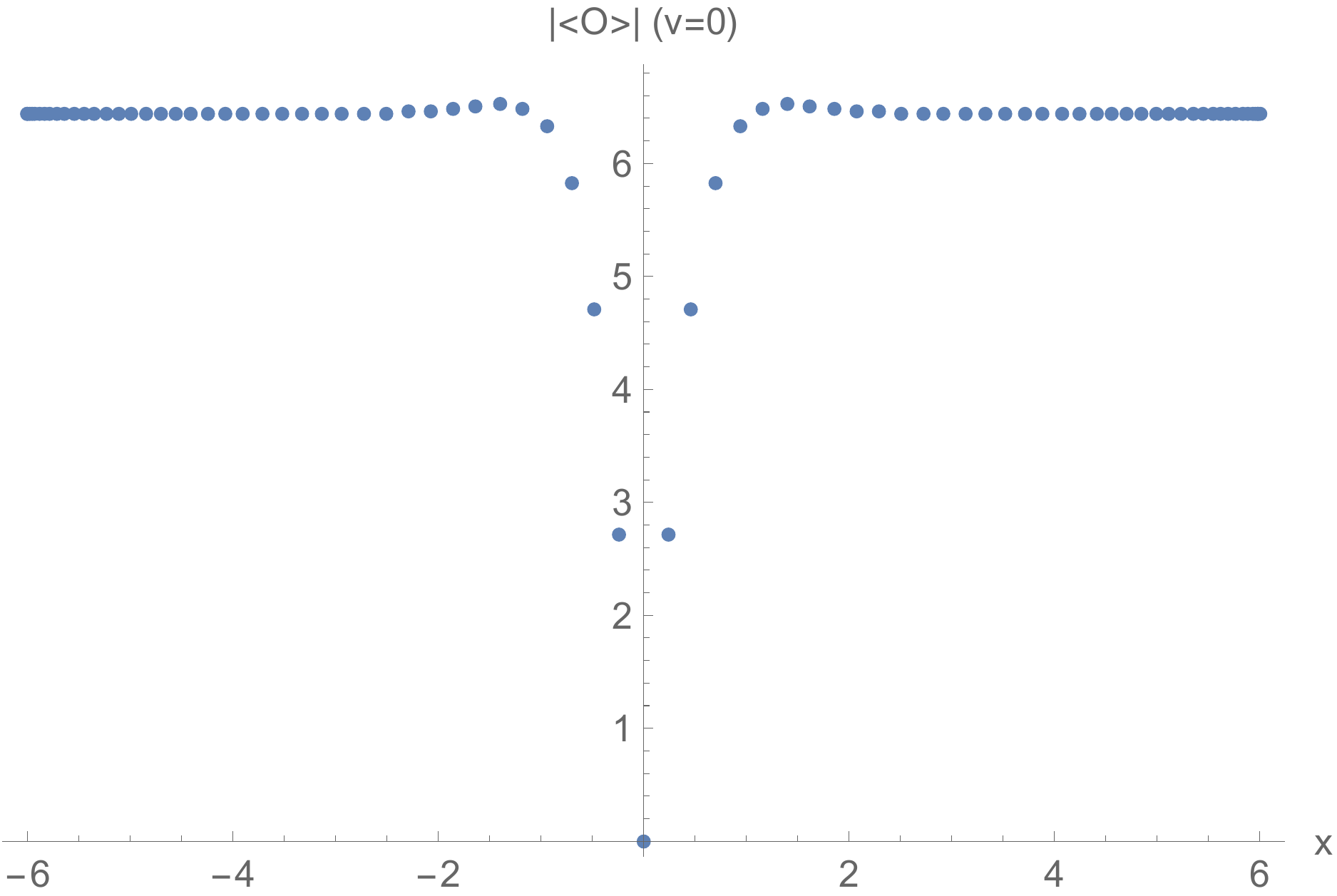}\qquad{}\includegraphics[scale=0.23]{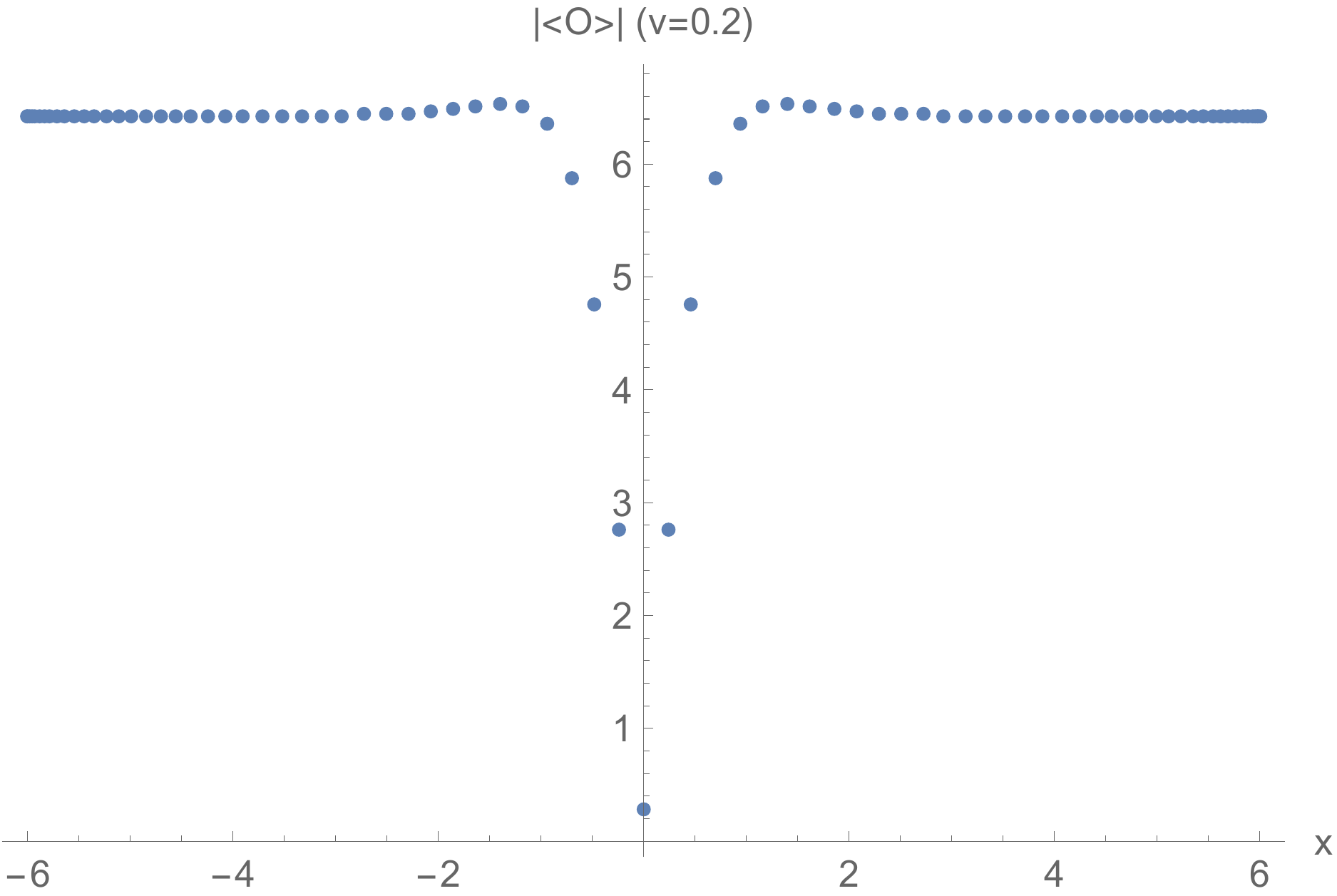}\qquad{}\includegraphics[scale=0.23]{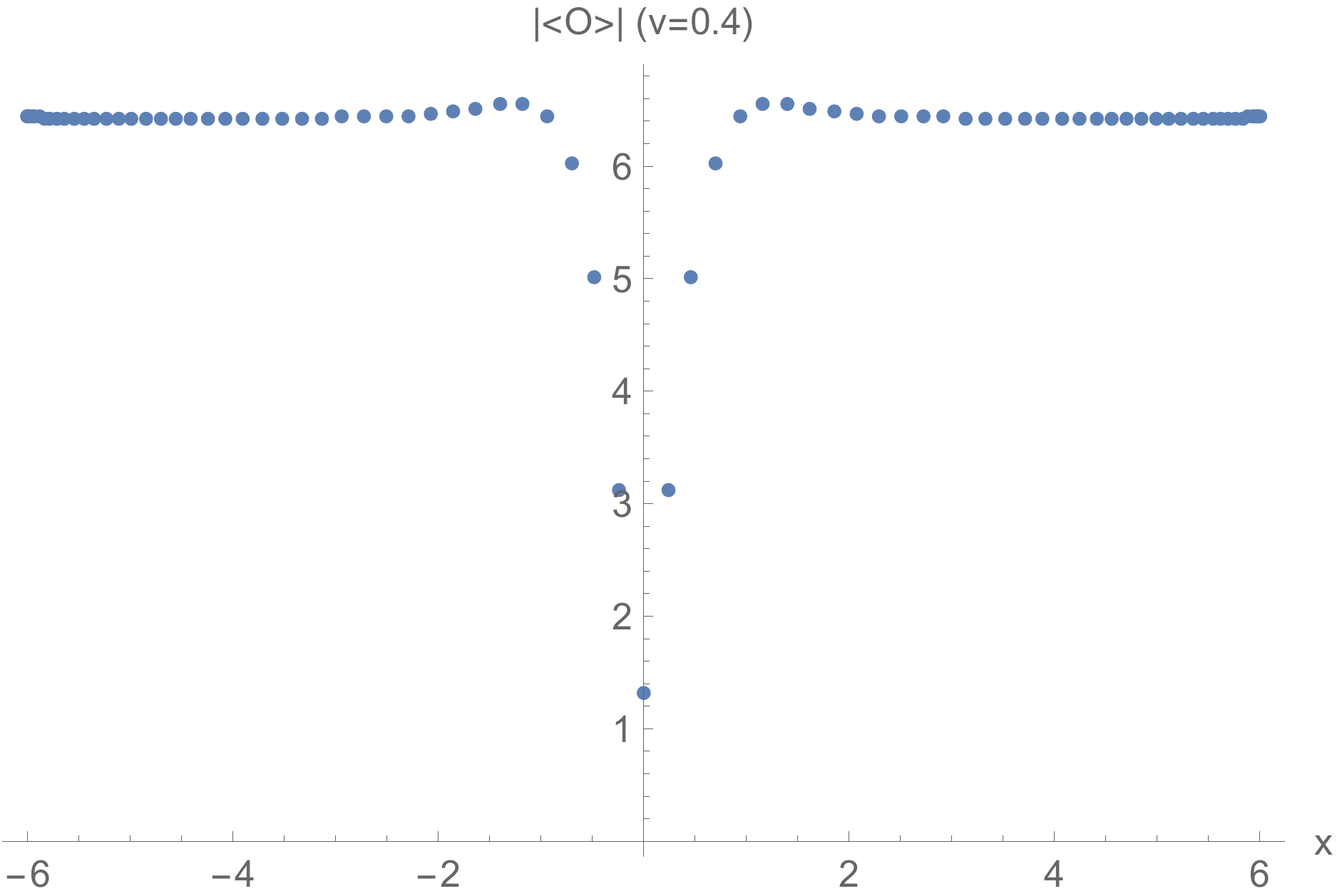}

\includegraphics[scale=0.23]{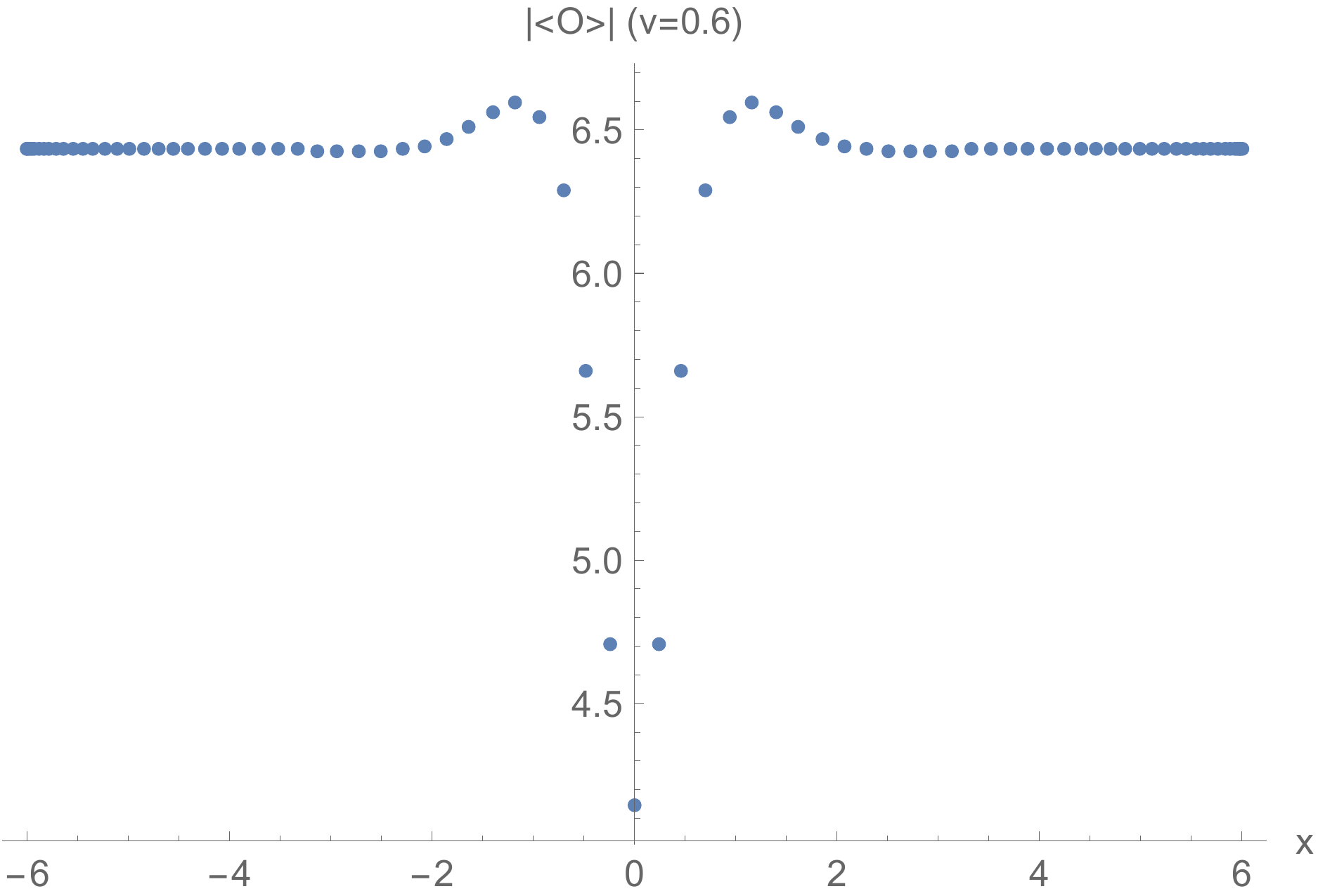}\qquad{}\includegraphics[scale=0.23]{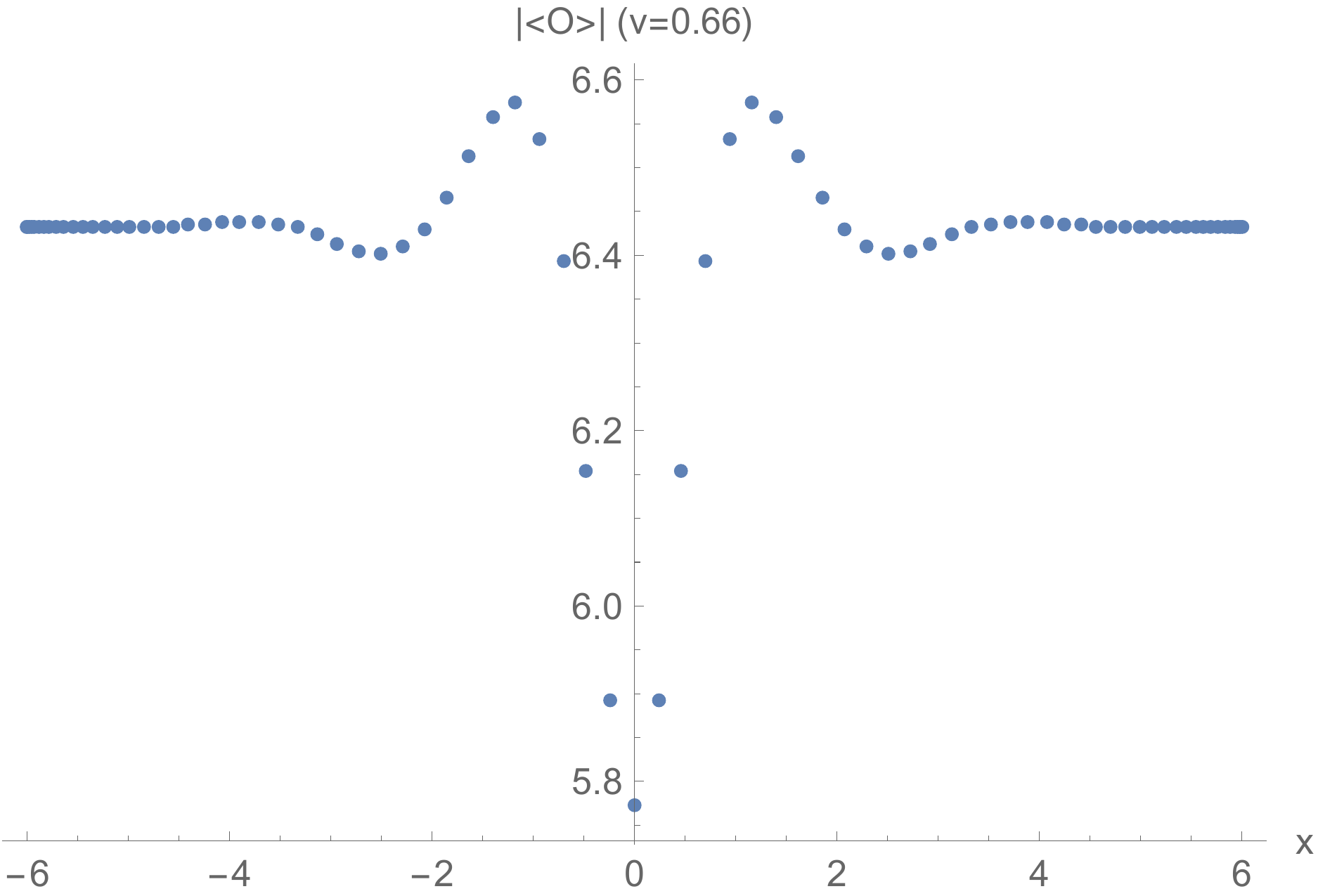}\qquad{}\includegraphics[scale=0.23]{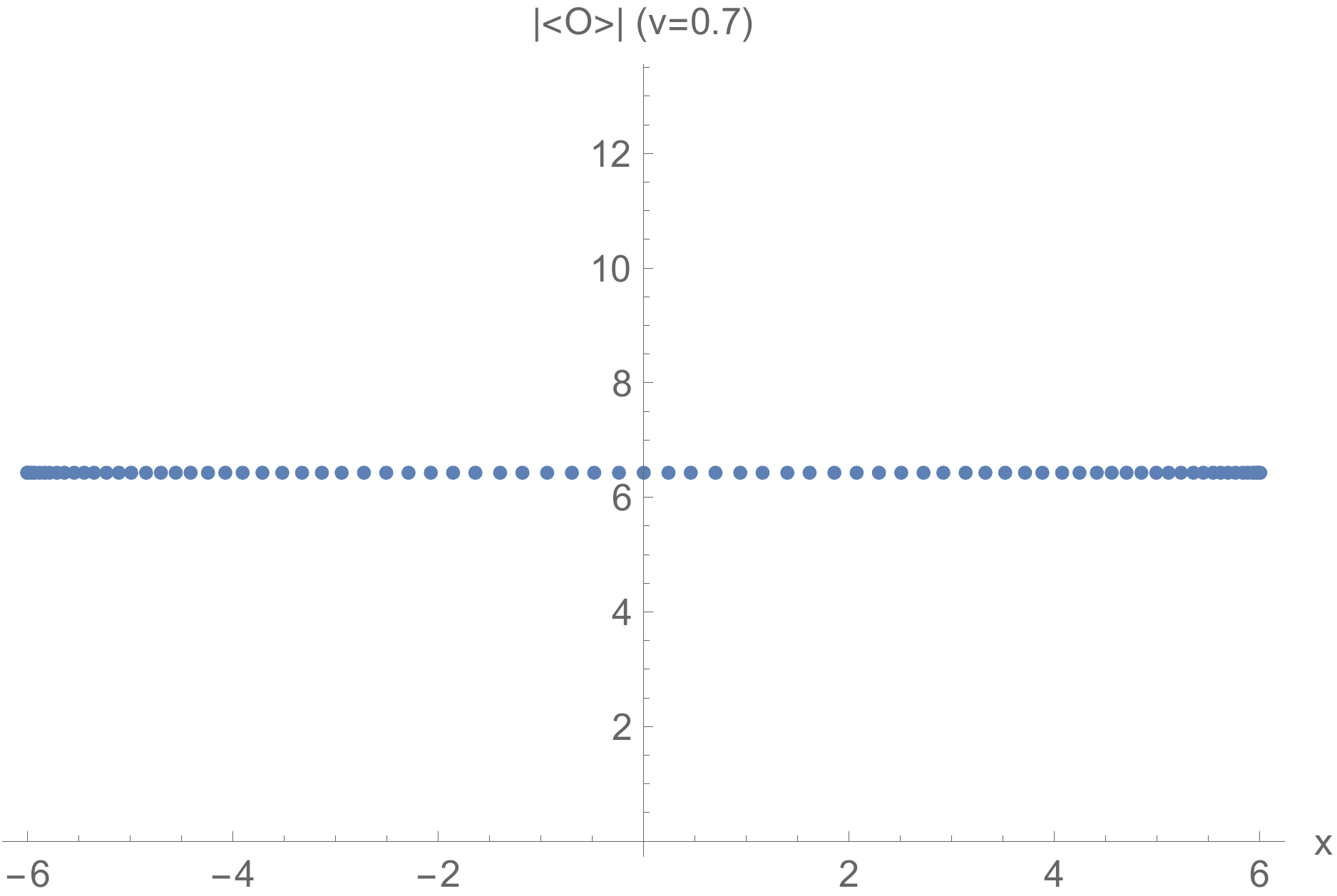}

\caption{The order parameter $\left|\left\langle O\right\rangle \right|$ distributions
at different speeds ($v=0,0.2,0.4,0.6,0.66,0.7$) of gray solitons
with the chemical potential $\mu=5.5$ at standard quantization. The
values of $\frac{\left|\left\langle O\right\rangle \right|_{x=0}}{\left|\left\langle O\right\rangle \right|_{x=6}}$
are in sequence 0, 0.0443334, 0.204758, 0.644198, 0.897435, 1 with
the increasing velocities.\label{standard order parameter gray}}
\end{figure}

\begin{figure}
\includegraphics[scale=0.4]{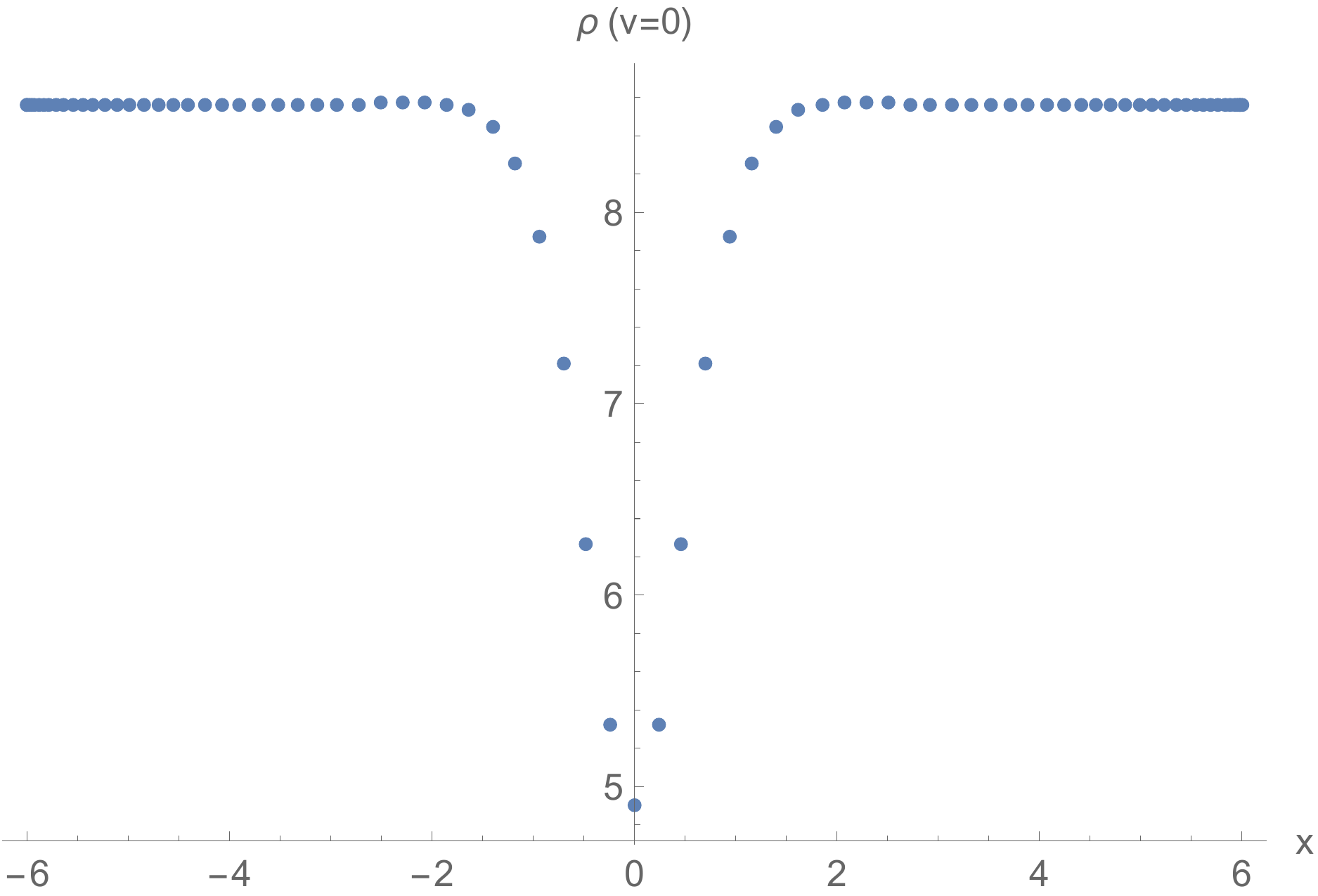}\qquad{}\includegraphics[scale=0.4]{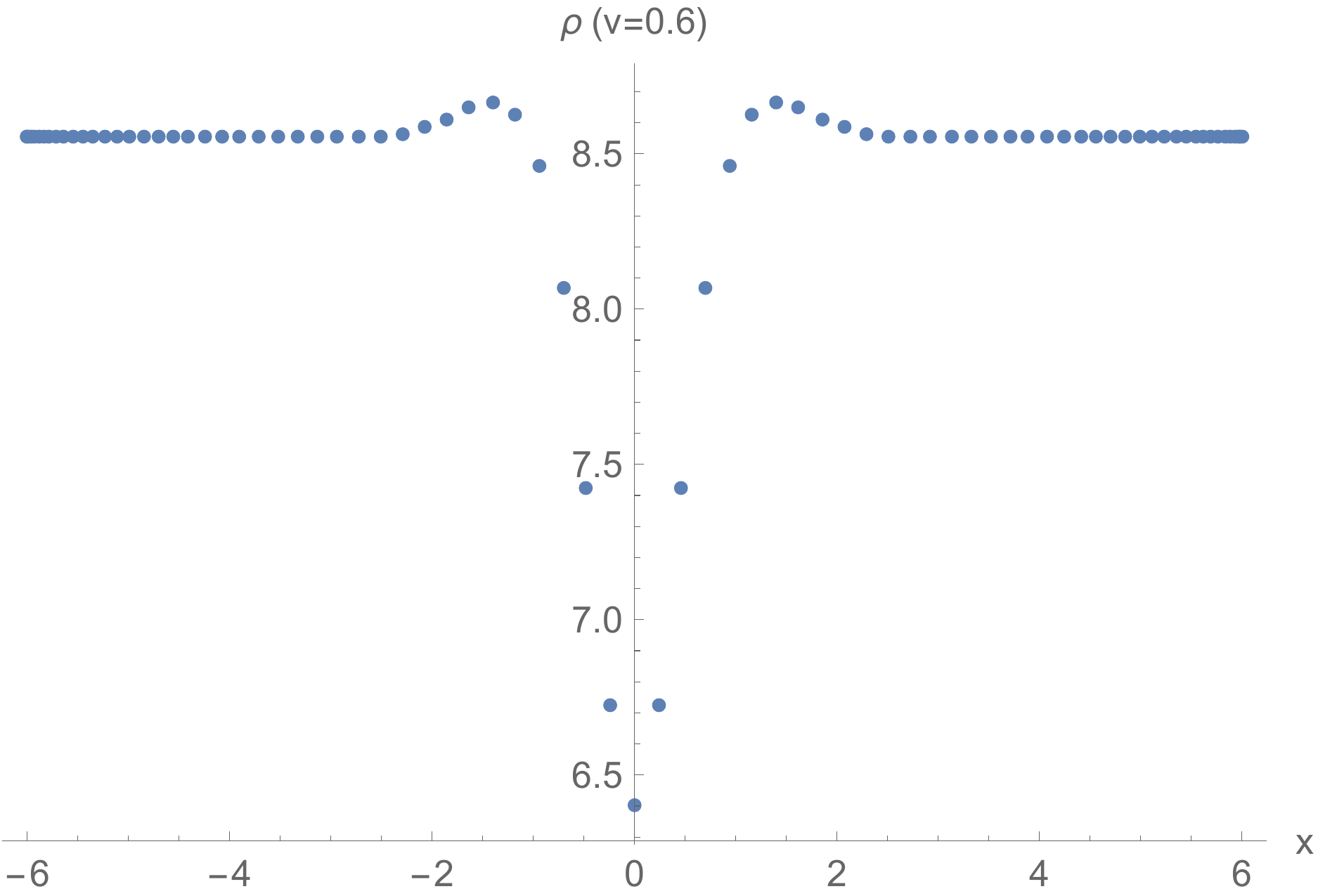}

\includegraphics[scale=0.4]{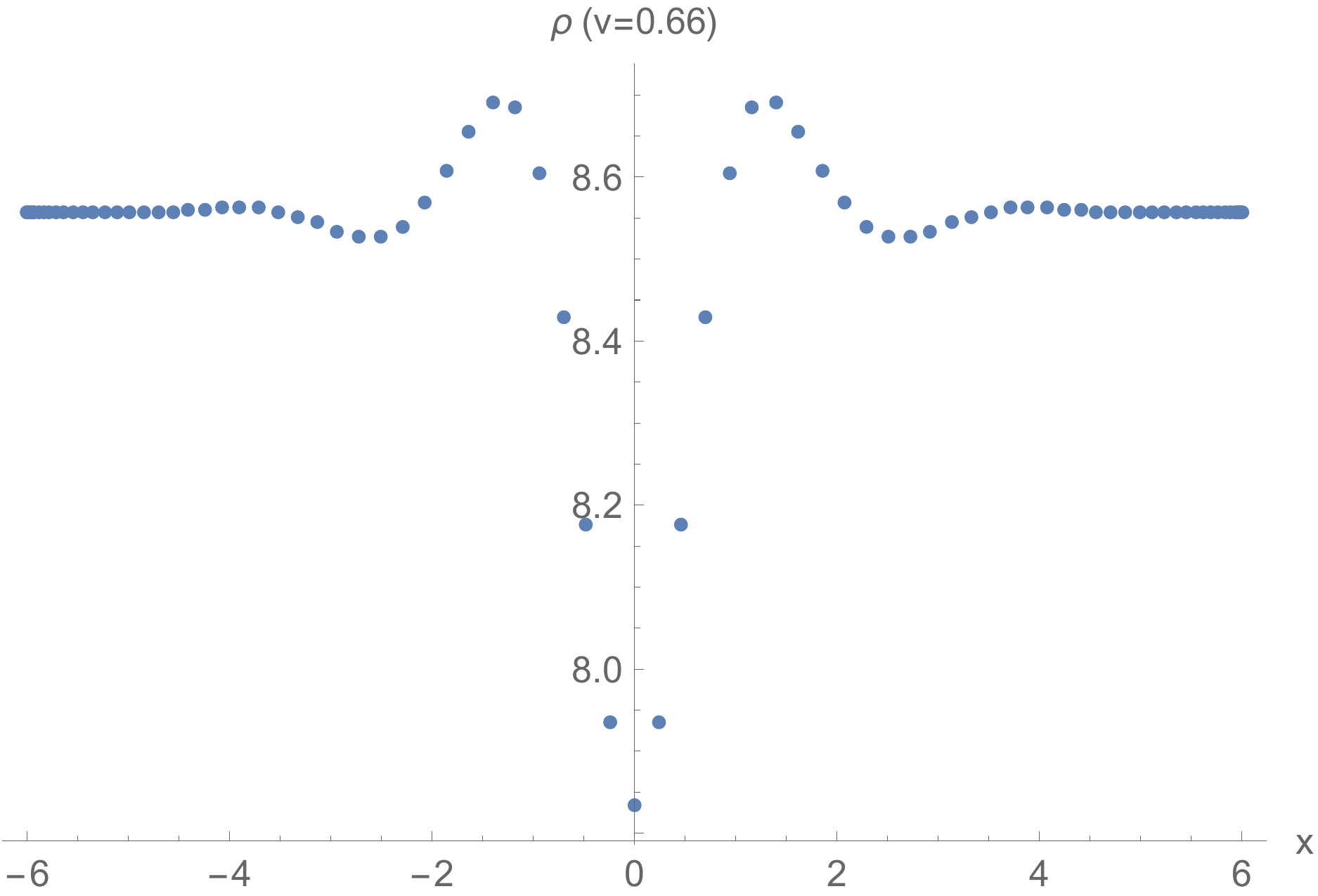}\qquad{}\includegraphics[scale=0.4]{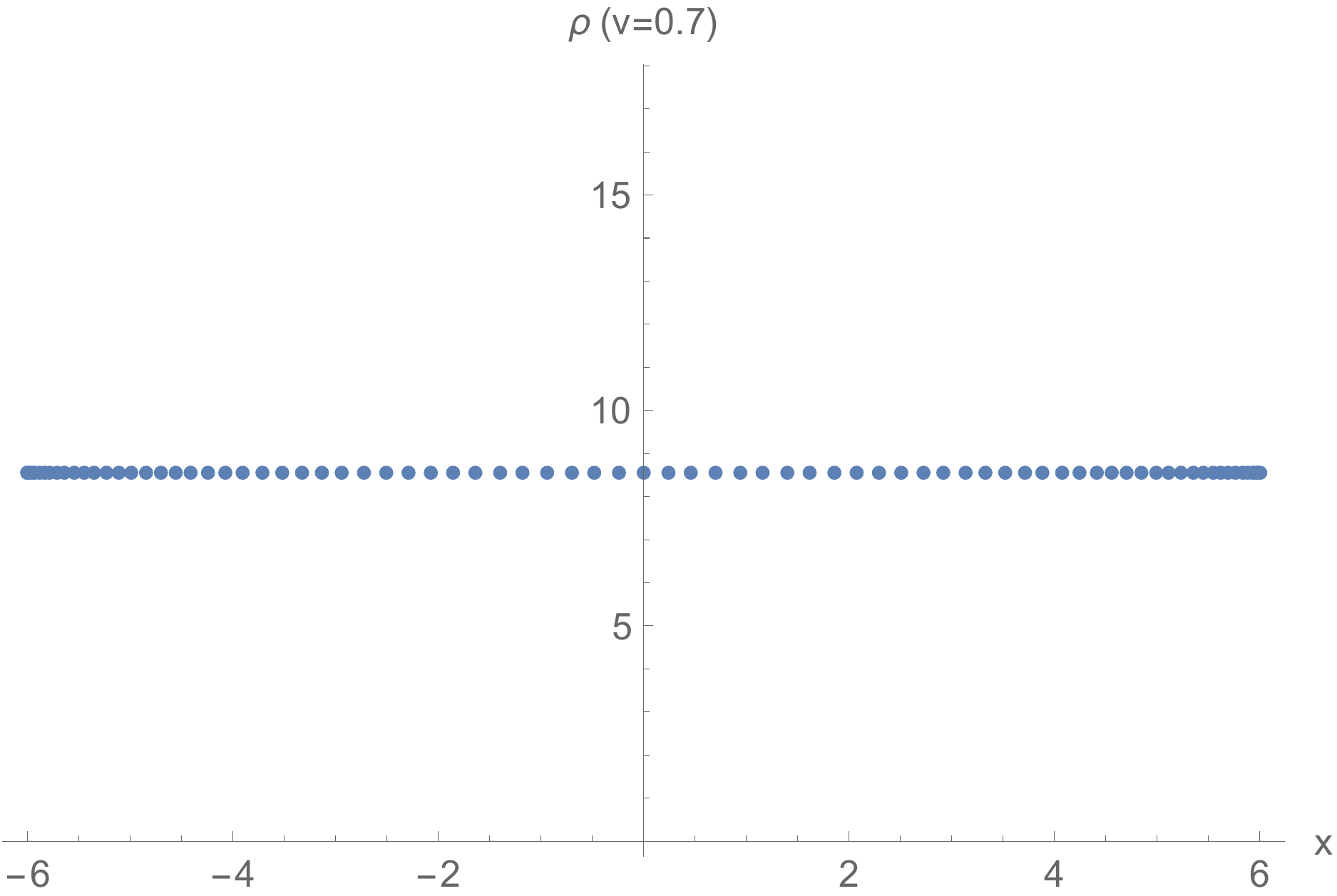}

\caption{The distributions of particle number density at different speeds ($v=0,0.6,0.66,0.7$)
of gray solitons with the chemical potential $\mu=5.5$ at standard
quantization. The values of $\frac{\rho_{x=0}}{\rho_{x=6}}$ are in
sequence 0.572601, 0.74807, 0.915534, 1 with the increasing velocities.\label{standard charge gray}}
\end{figure}

First of all, we present the condensate and particle number density
of gray solitons with different speeds under the standard quantization,
which are shown in Fig.\ref{standard order parameter gray} and Fig.\ref{standard charge gray}.

We can see in Fig.\ref{standard order parameter gray} that the configuration
just becomes the black soliton when the speed is zero, and that the
depletion of the condensate at the soliton core becomes small with
the increase of speed. Eventually, the condensate distribution becomes
homogeneous when the soliton speed approaches the speed of sound ($1/\sqrt{2}\approx0.707$
for the standard quantization\cite{Guo}), since in this case the
moving gray soliton becomes a sound wave, which should actually be
a linear perturbation. Meanwhile, the analogous tendency arises in
the density distribution, as shown in Fig.\ref{standard charge gray}.
Similarly, it becomes homogeneous when the soliton speed reaches the
speed of sound. Most importantly, the peculiar fluctuations similar
to the black soliton case also occur in the general gray soliton case,
and they become even more remarkable when the soliton speed increases.
This enhancement of the Friedel-like oscillation is probably due to
the fact that the oscillation amplitude is less sensitive to the change
of the soliton speed than the depletion of the soliton configurations
are, while in the zero or low speed case the steep variation of the
soliton configuration tends to overwhelm the Friedel-like oscillation.
Additional discussions on these BCS-like configurations can be found
in Appendix \ref{sec:fit_gray}.

\begin{figure}
\includegraphics[scale=0.38]{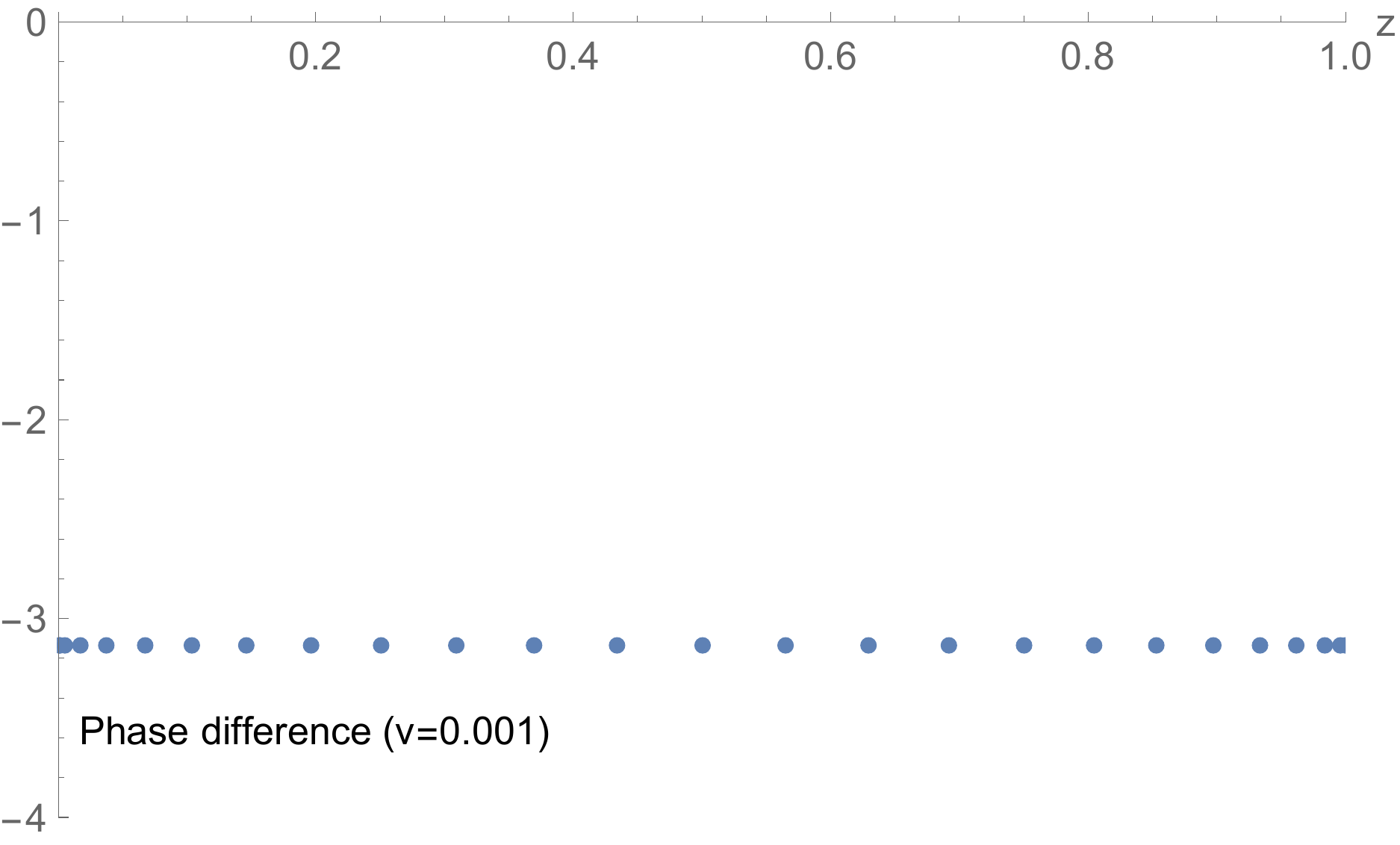}\qquad{}\includegraphics[scale=0.48]{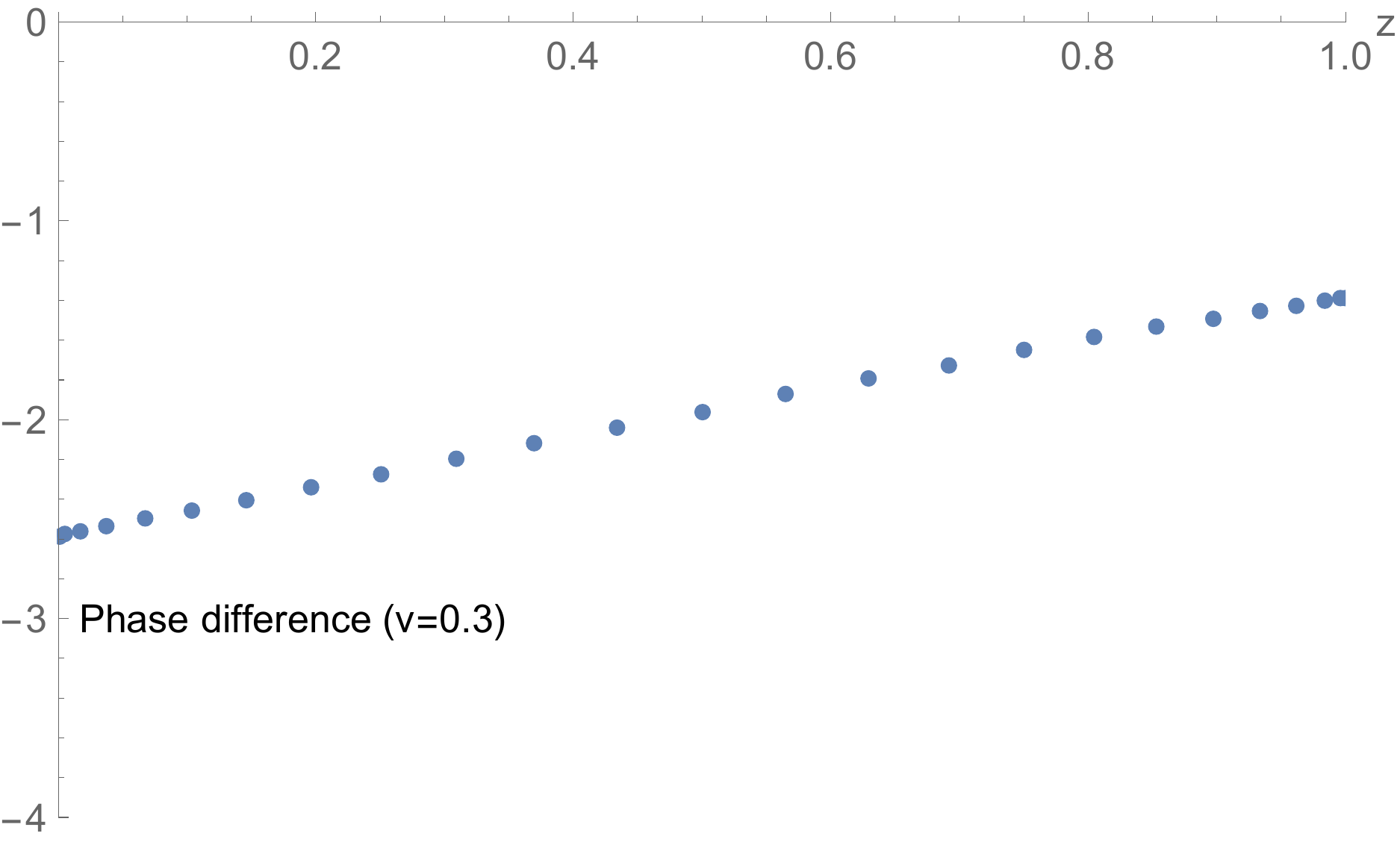}

\includegraphics[scale=0.4]{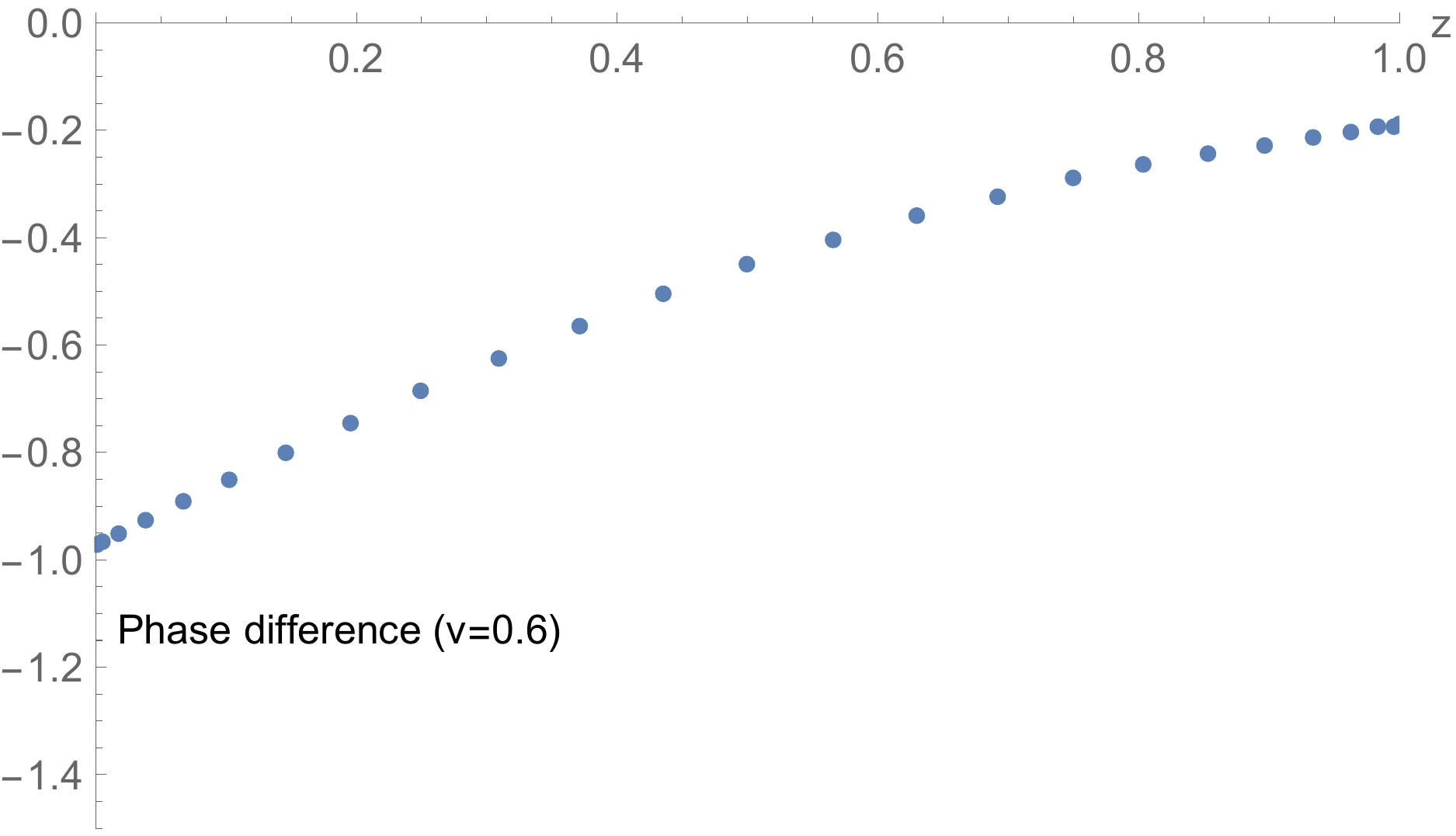}\qquad{}\includegraphics[scale=0.4]{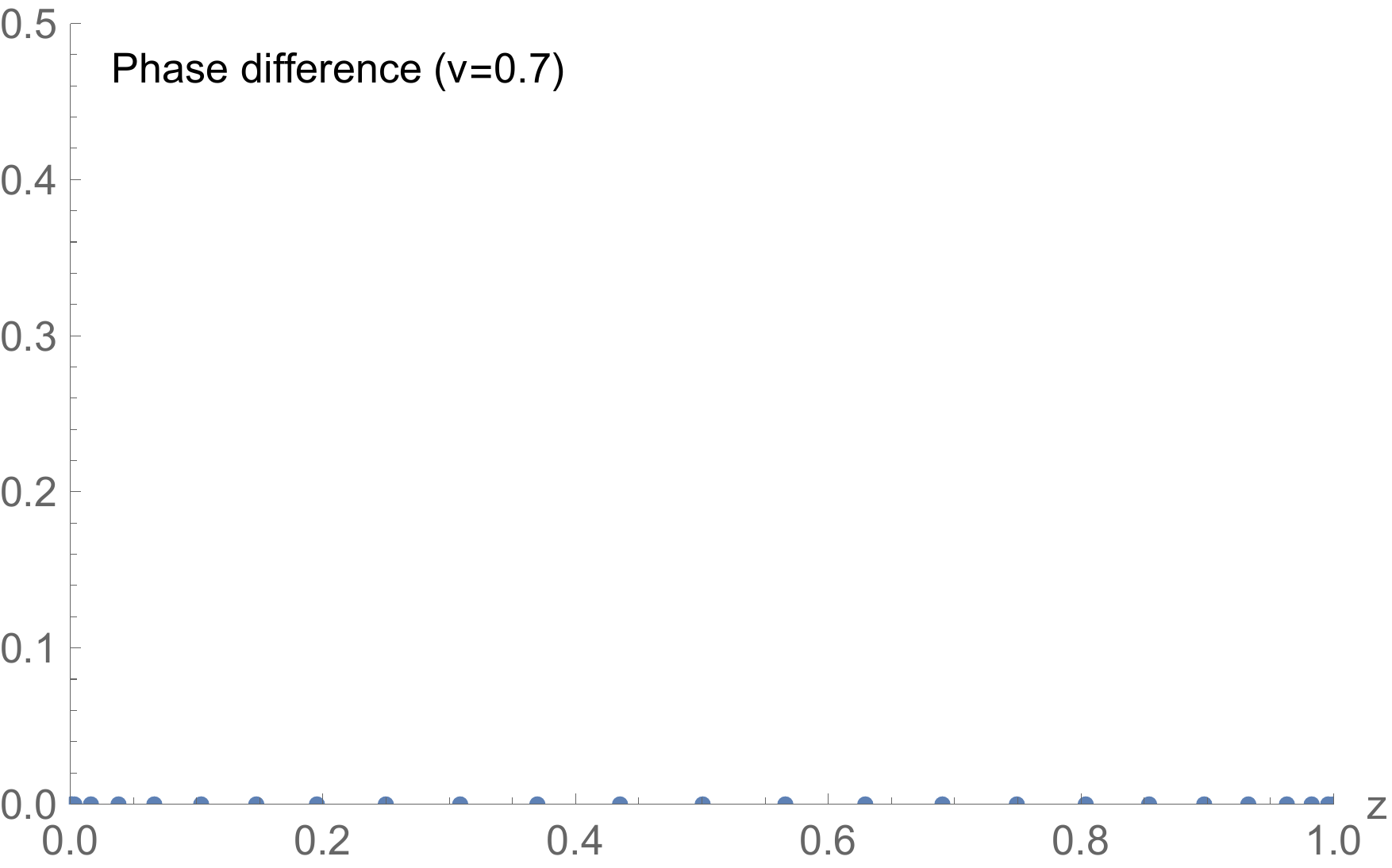}

\caption{The phase differences at different speeds, under the standard quantization
with the chemical potential $\mu=5.5$. \label{standard phase diff gray}}
\end{figure}

\begin{figure}
\includegraphics[scale=0.4]{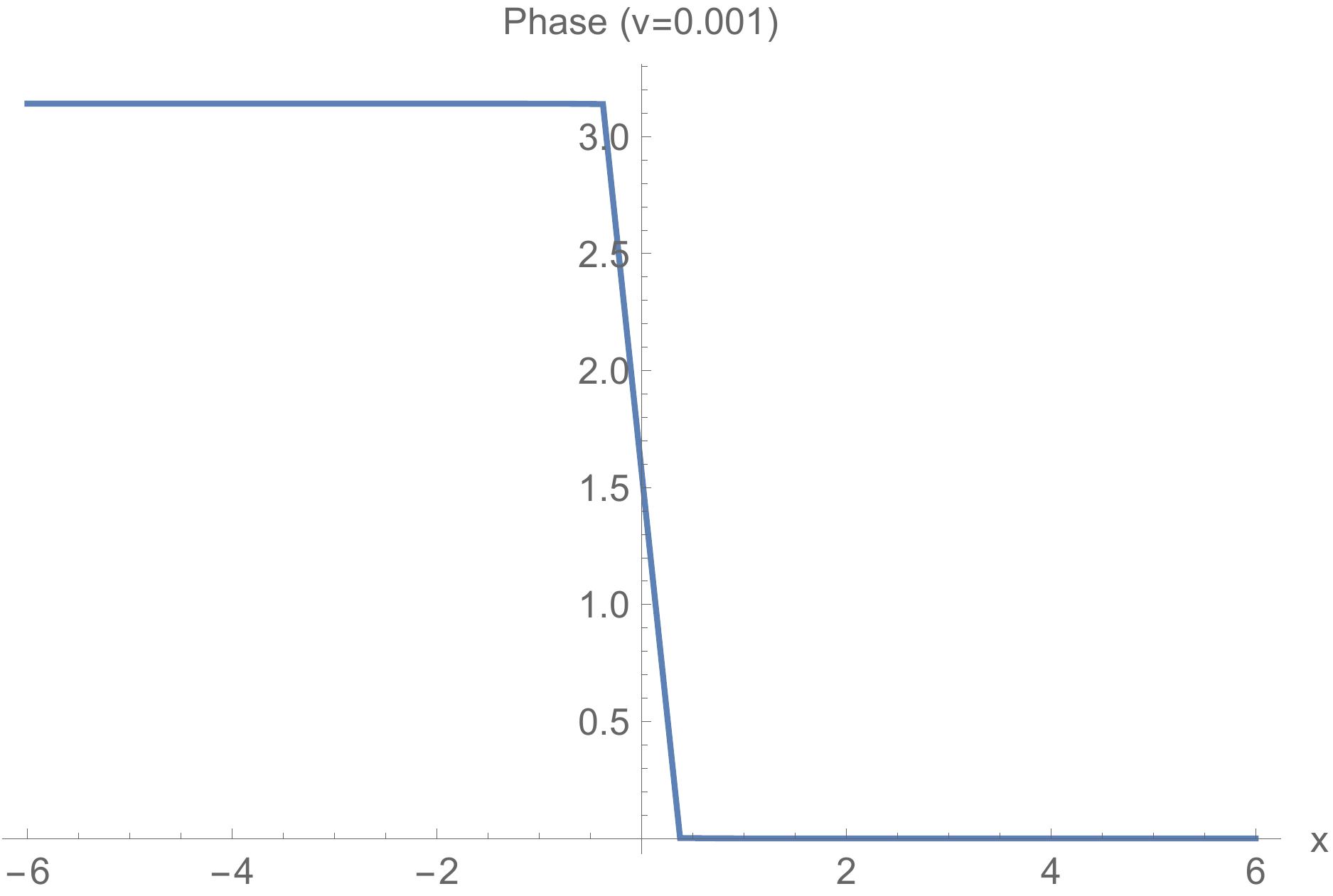}\qquad{}\includegraphics[scale=0.4]{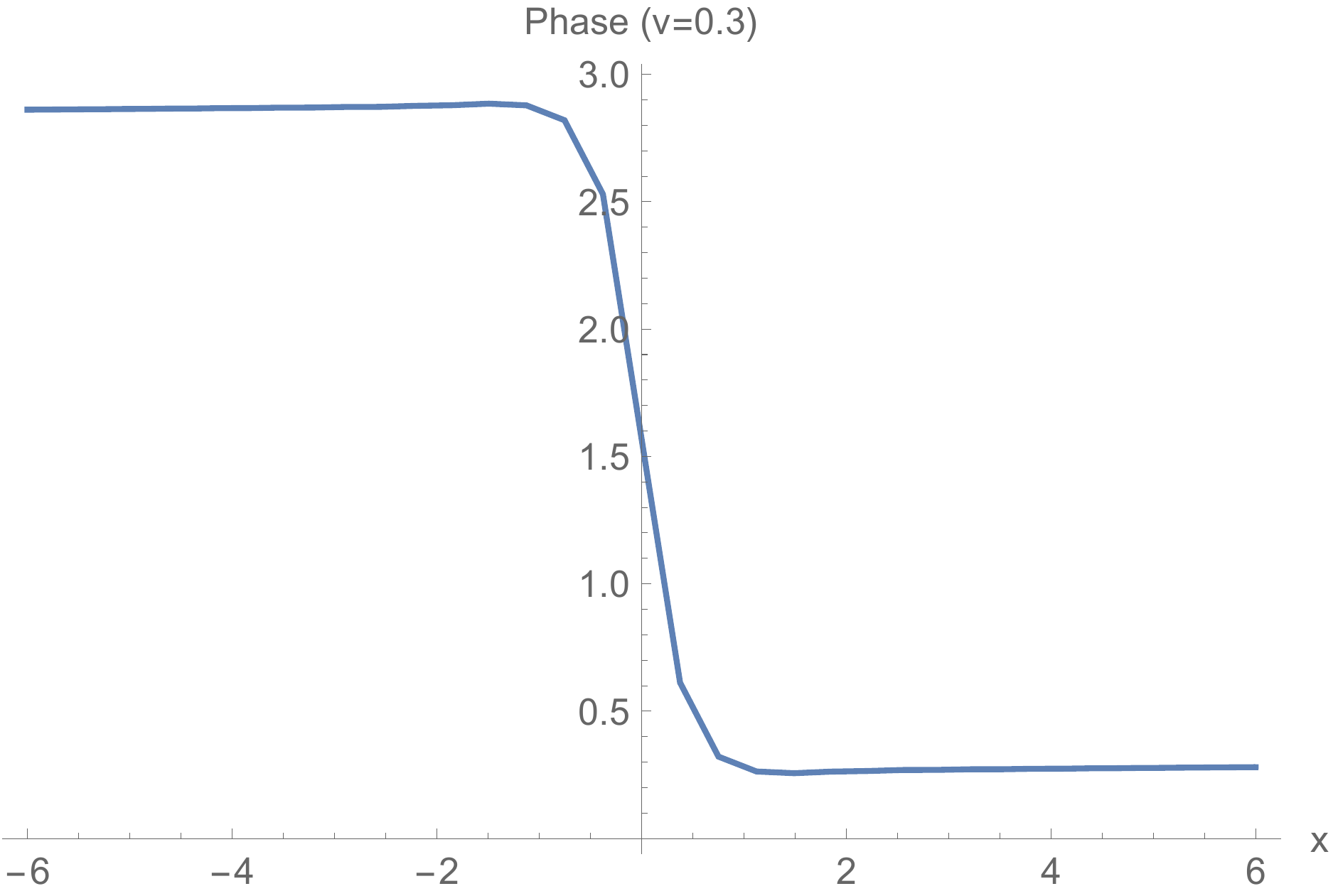}

\includegraphics[scale=0.4]{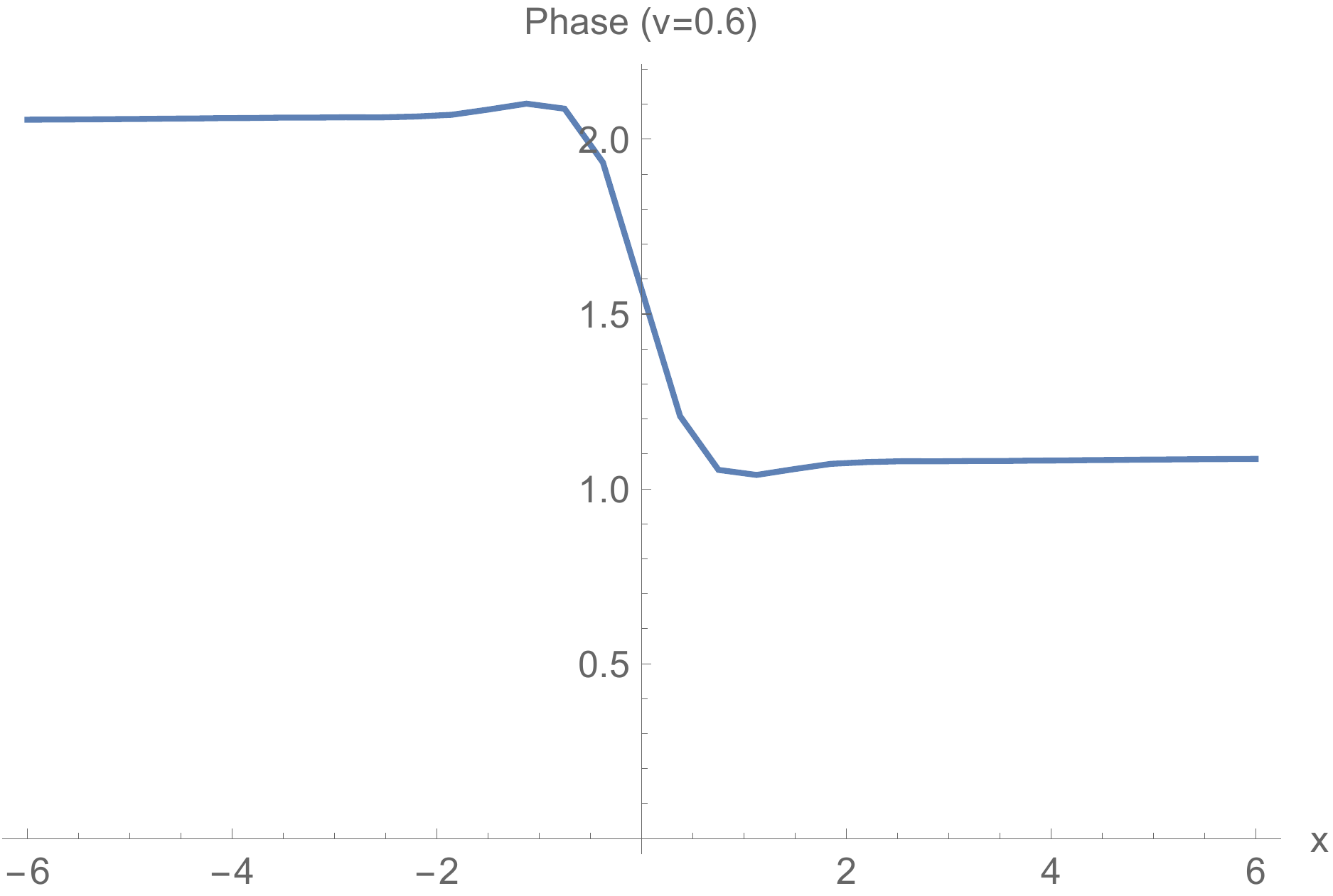}\qquad{}\includegraphics[scale=0.4]{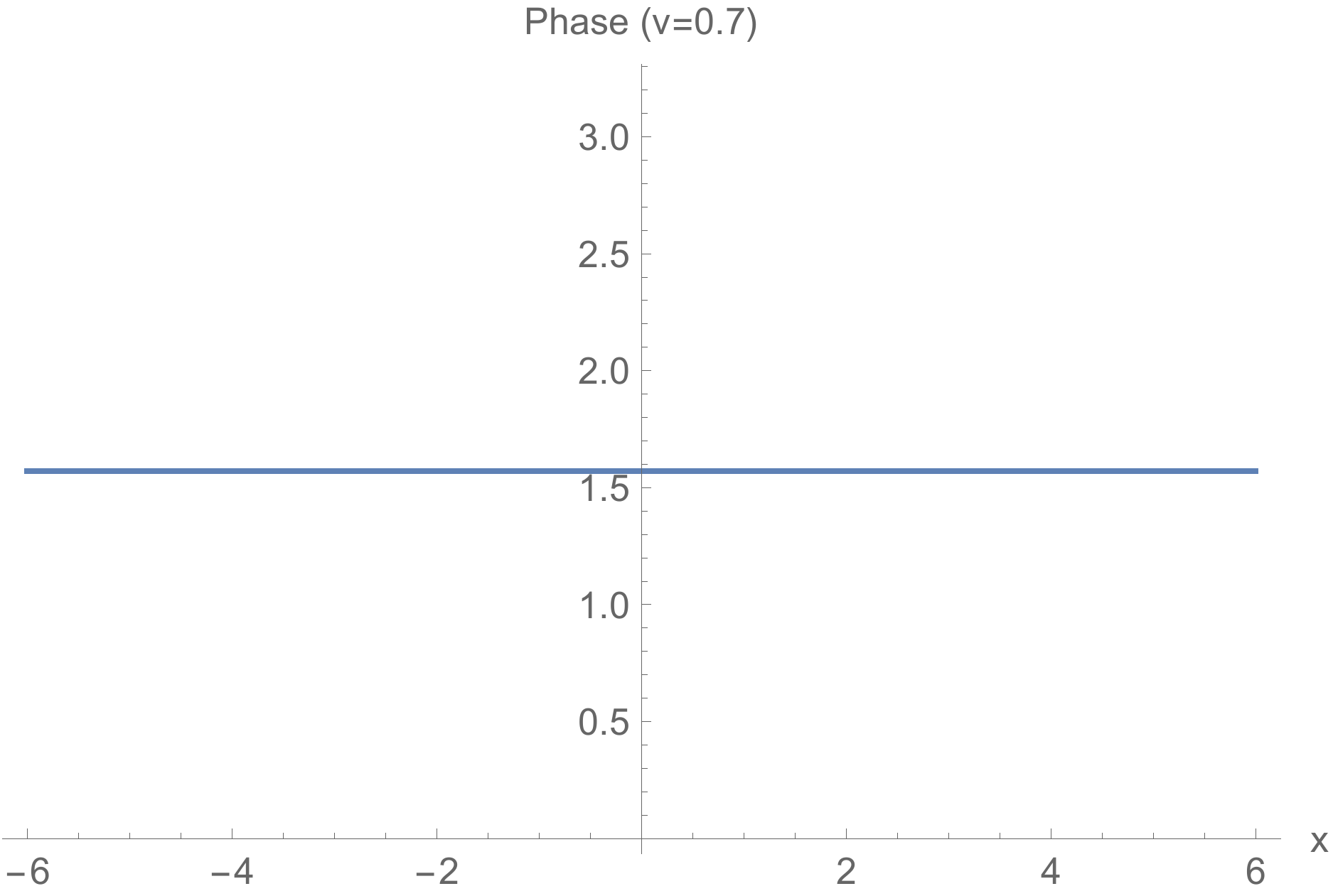}

\caption{The distributions of phase at different soliton speeds, under the
standard quantization with the chemical potential $\mu=5.5$. \label{standard phase gray}}
\end{figure}

It is well known that the phase difference between the condensates
on the two sides of a black soliton is $\pi$, which is not the case
for gray solitons. For the purpose of verifying the gray soliton configurations
above, we investigate the gauge invariant phase differences with respect
to the soliton speeds. This gauge invariant phase difference is written
as $\theta|_{-L}^{L}-\int_{-L}^{L}A_{x}dx$, where $\theta$ is the
argument of the order parameter of a gray soliton. We present the
phase differences and phase distributions in Fig.\ref{standard phase diff gray}
and Fig.\ref{standard phase gray}, respectively.

As can be seen from Fig.\ref{standard phase diff gray} and Fig.\ref{standard phase gray},
the absolute value of the phase difference is approximately $\pi$
when the speed is 0.001, which exactly indicates that the solution
becomes a black soliton when the speed is very small. Then the absolute
value of the phase difference becomes smaller and smaller when the
soliton speed increases. When the soliton speed approaches the speed
of sound, the phase difference tends to zero and the phase of the
condensate is actually a constant. All these features are expected
for gray soliton configurations.

\begin{figure}
\includegraphics[scale=0.4]{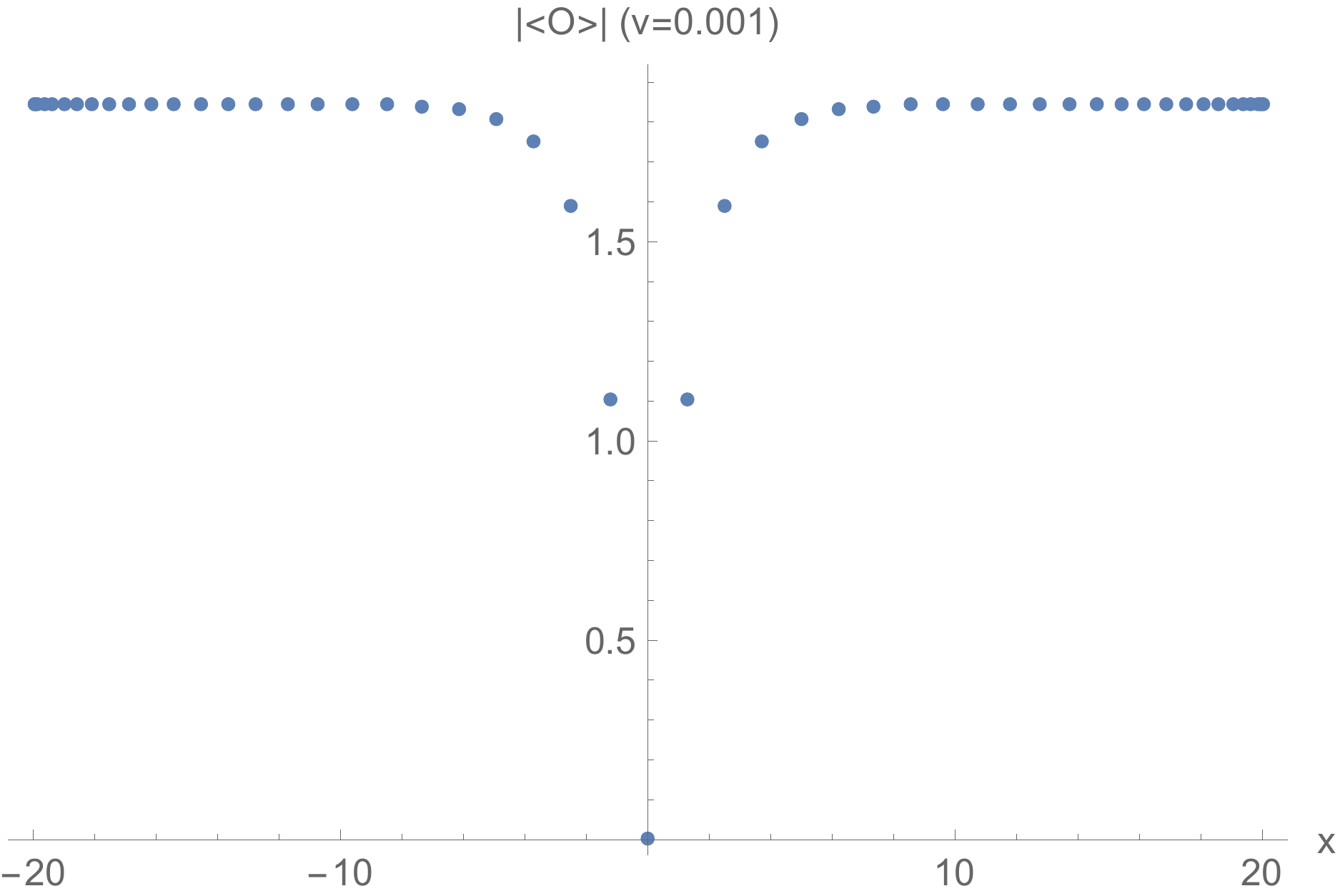}\qquad{}\includegraphics[scale=0.4]{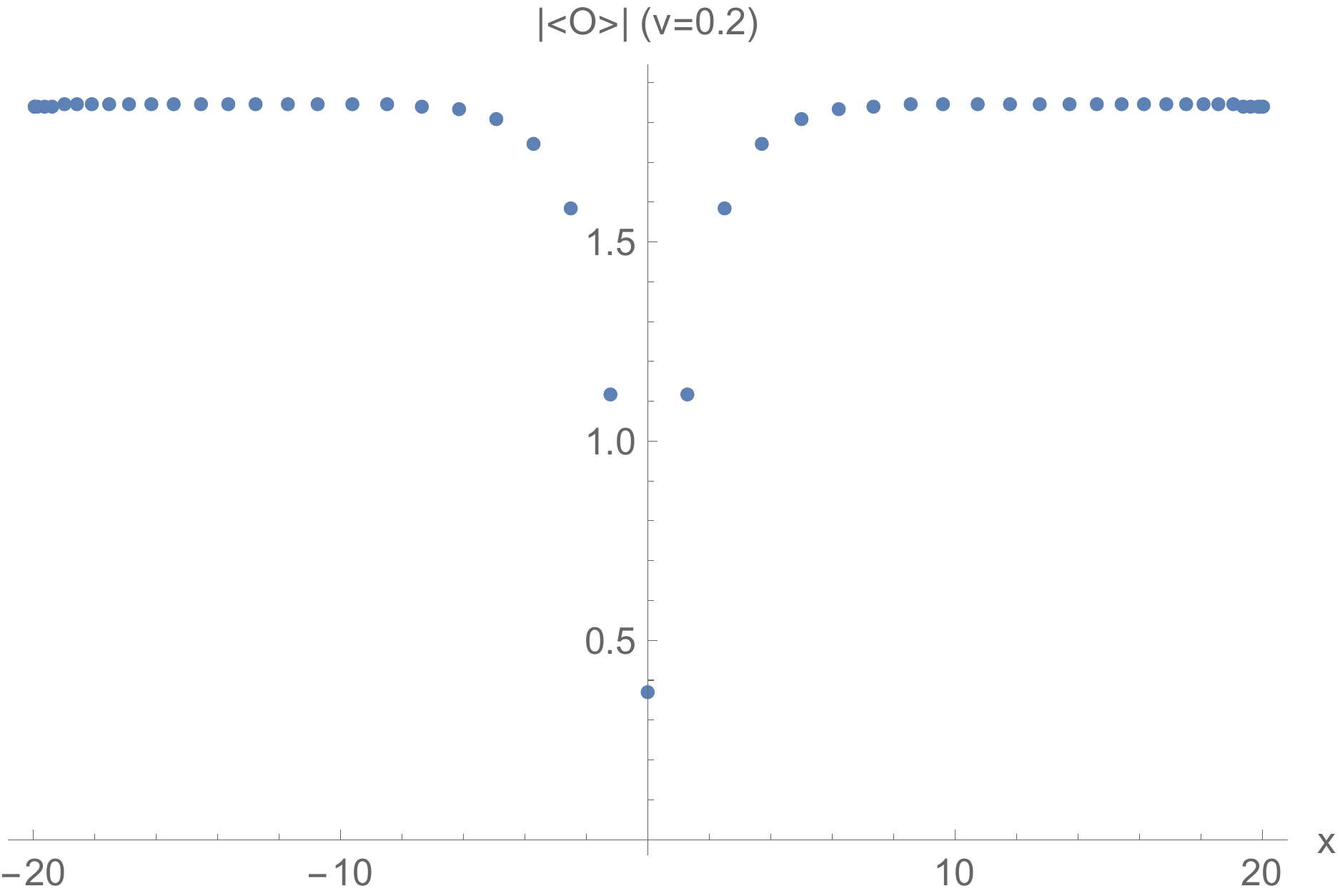}

\includegraphics[scale=0.4]{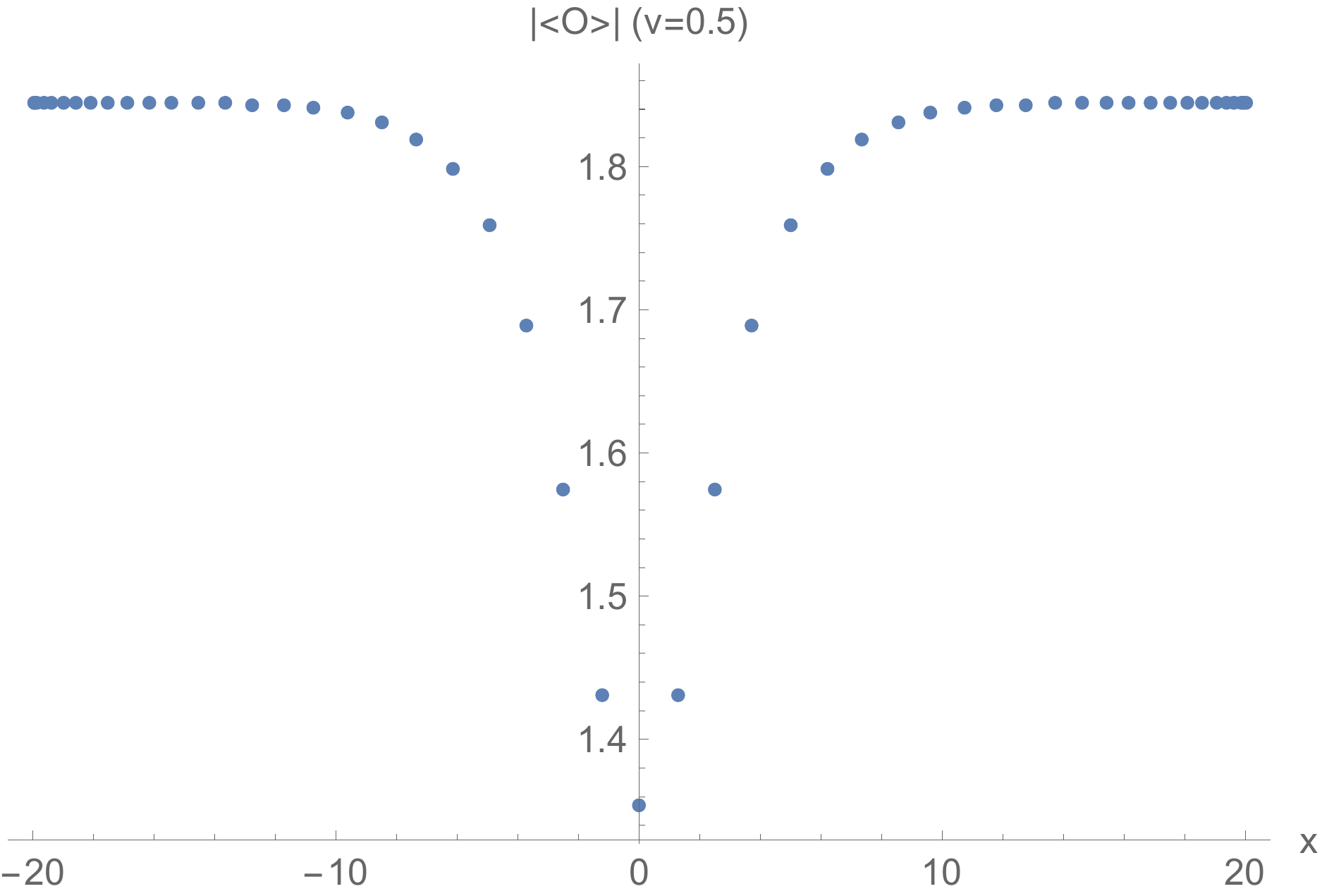}\qquad{}\includegraphics[scale=0.4]{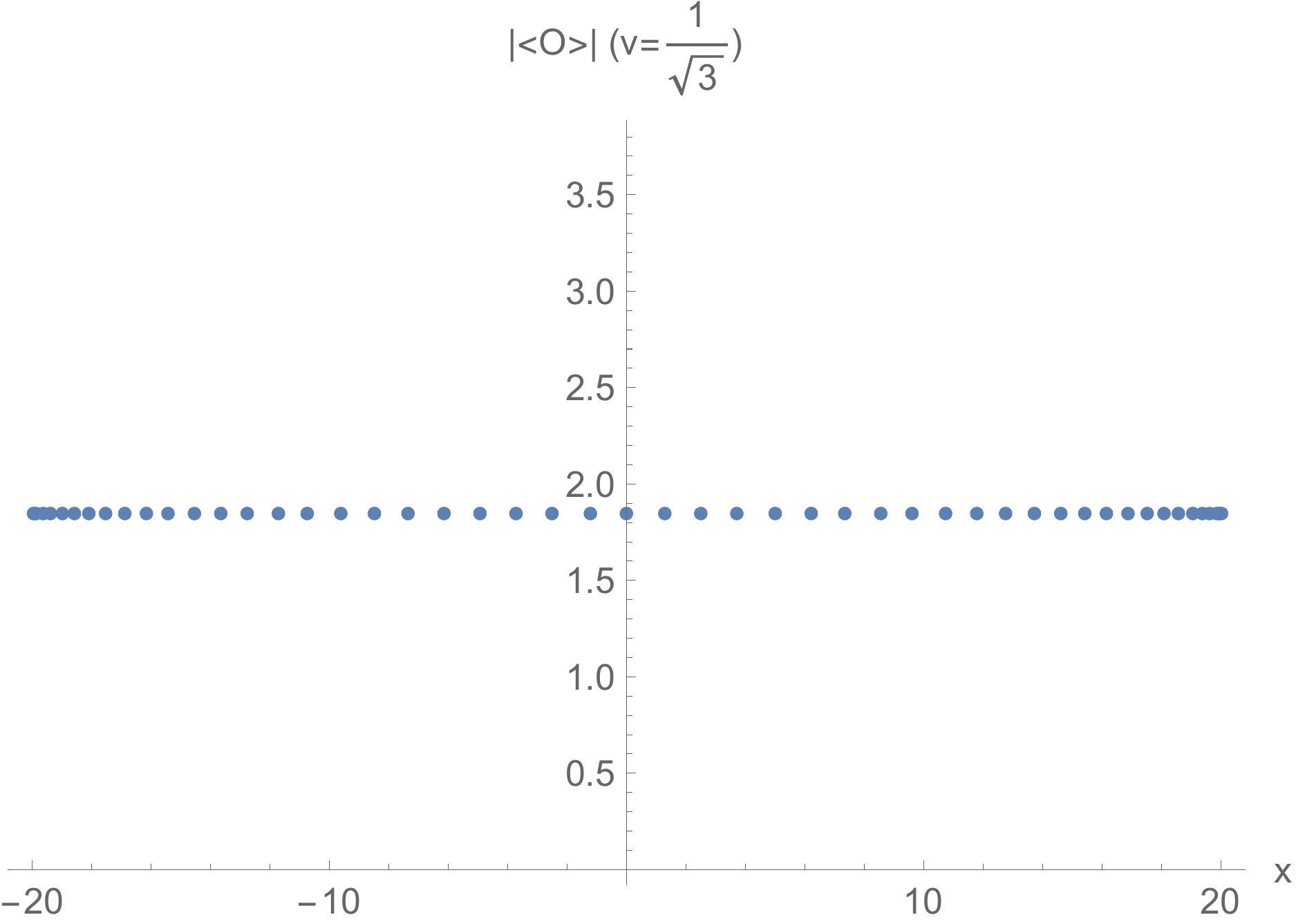}

\caption{The condensate distributions at various soliton speeds ($v=0.001,0.2,0.5,1/\sqrt{3}$)
under the alternative quantization whith the chemical potential $\mu=1.4$.
The values of $\frac{\left|\left\langle O\right\rangle \right|_{x=0}}{\left|\left\langle O\right\rangle \right|_{x=20}}$
are in sequence 0.000959, 0.200329, 0.73415, 1 with the increasing
velocities.\label{alternative order gray}}
\end{figure}

\begin{figure}
\includegraphics[scale=0.4]{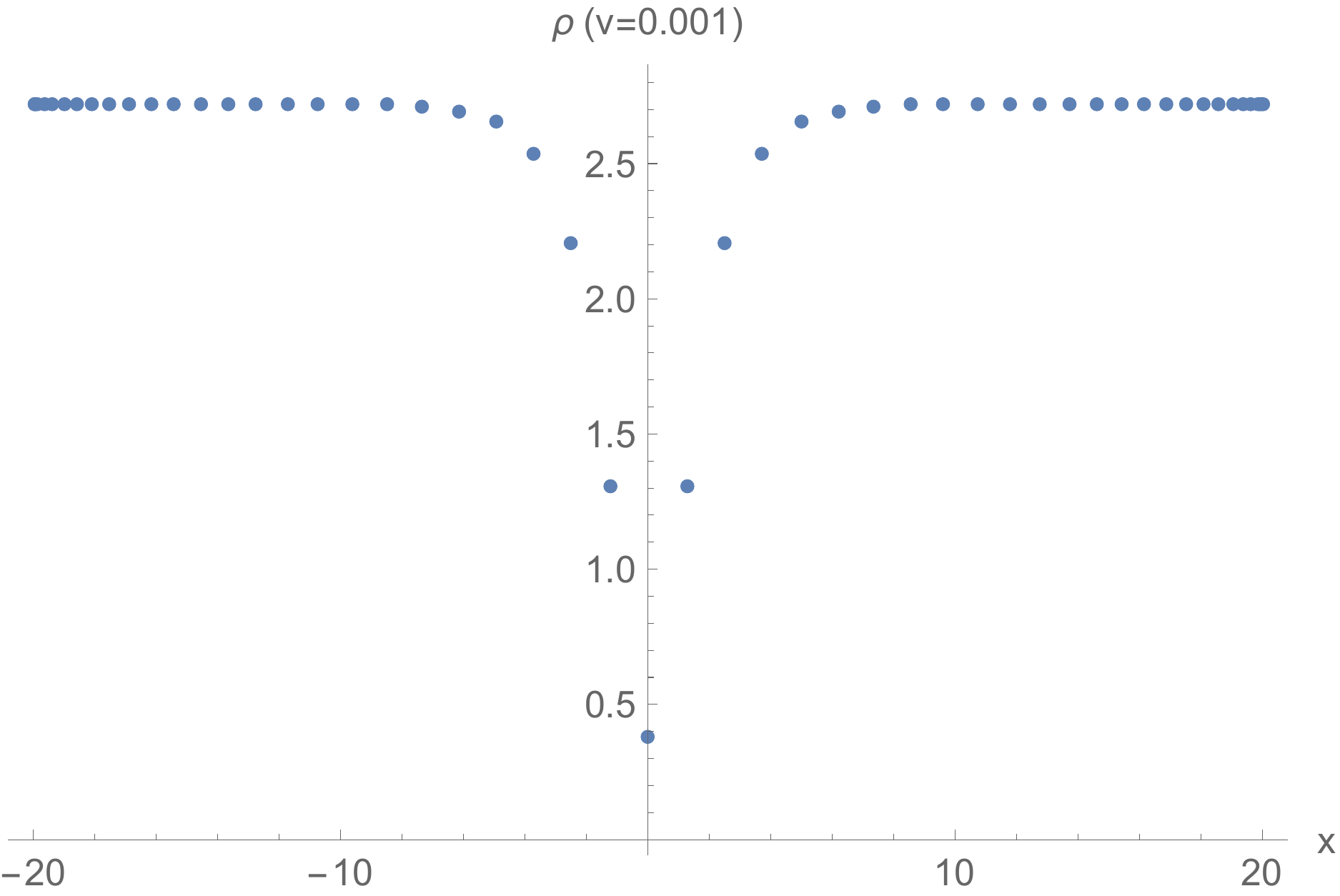}\qquad{}\includegraphics[scale=0.4]{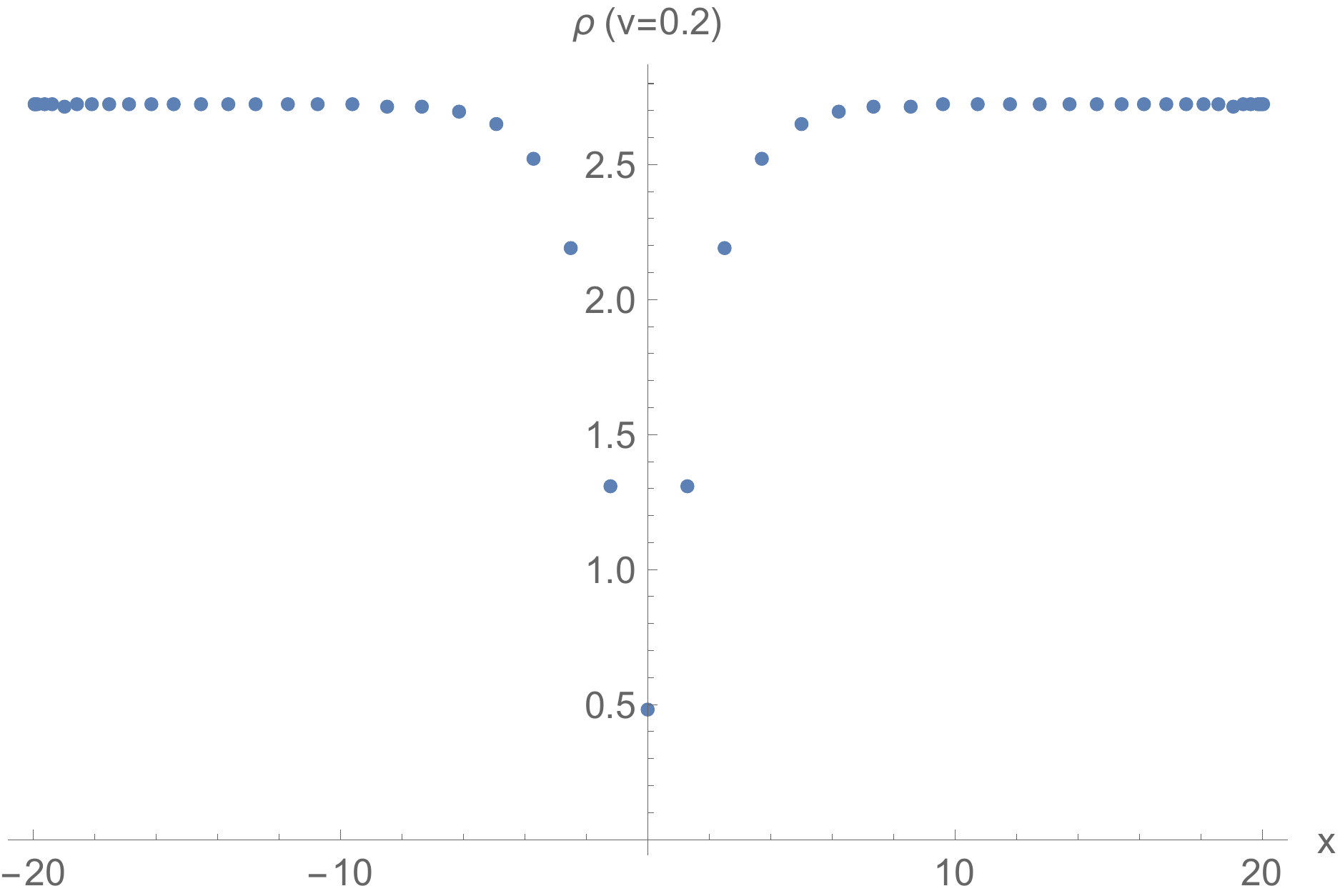}

\includegraphics[scale=0.4]{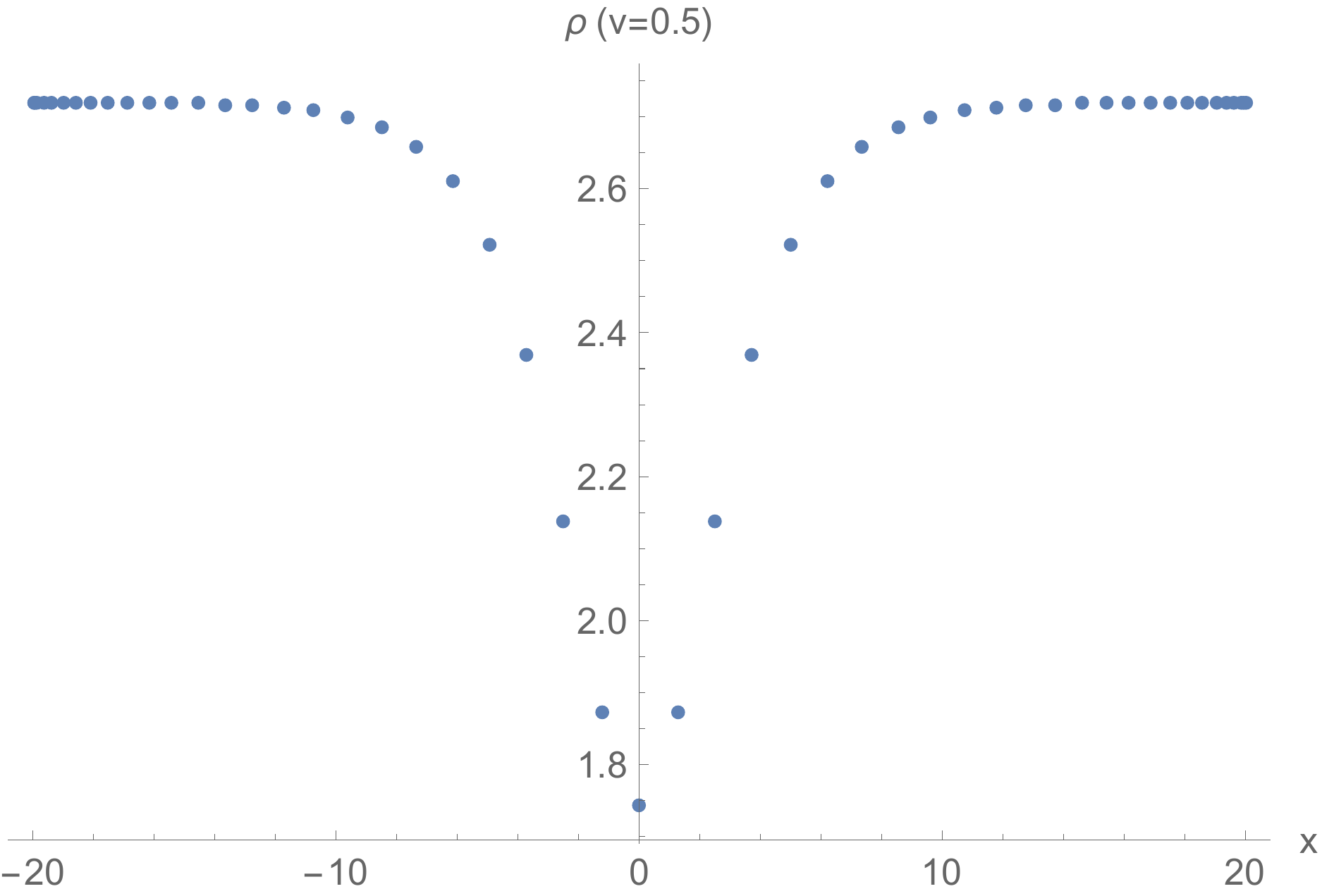}\qquad{}\includegraphics[scale=0.4]{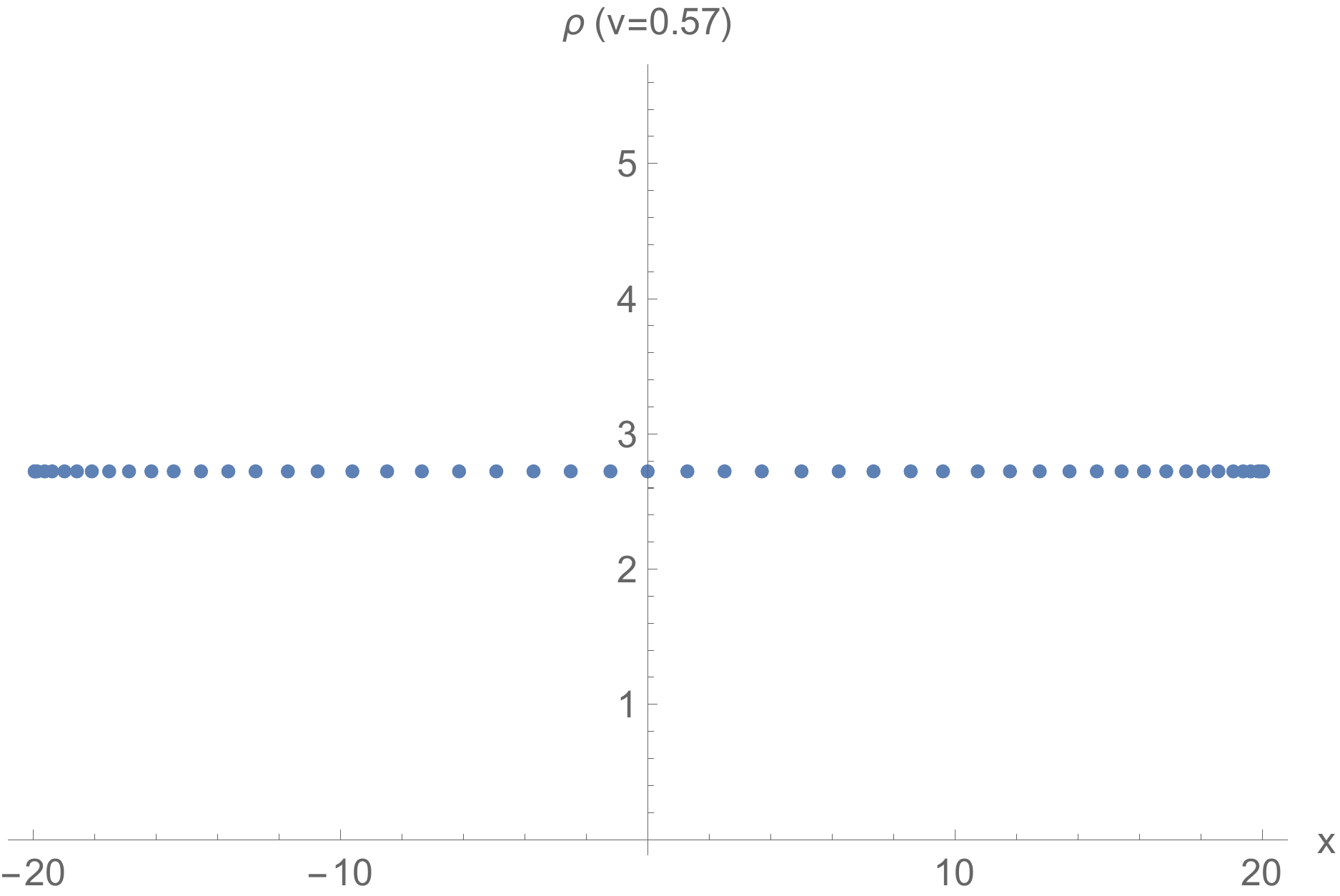}

\caption{The density distributions at various soliton speeds ($v=0.001,0.2,0.5,0.57$)
under the alternative quantization with the chemical potential $\mu=1.3$.
The values of $\frac{\rho{}_{x=0}}{\rho{}_{x=20}}$ are in sequence
0.140459, 0.176112, 0.641066, 1 with the increasing velocities.\label{alternative charge gray}}
\end{figure}

Then we will present the gray soliton solutions under the alternative
quantization. The condensate and particle number density are studied
as in the case of standard quantization and shown in Fig.\ref{alternative order gray}
and Fig.\ref{alternative charge gray}.

It can be seen that there is no fluctuation around the edges of cliffs
of the condensate and density profiles at alternative quantization.
The results are consistent with the proposal that the holographic
superfluid under the alternative quantization is a BEC-like superfluid.
While the speed of sound is $1/\sqrt{3}\approx0.577$\cite{Guo} in
the case of alternative quantization.

\begin{figure}
\includegraphics[scale=0.44]{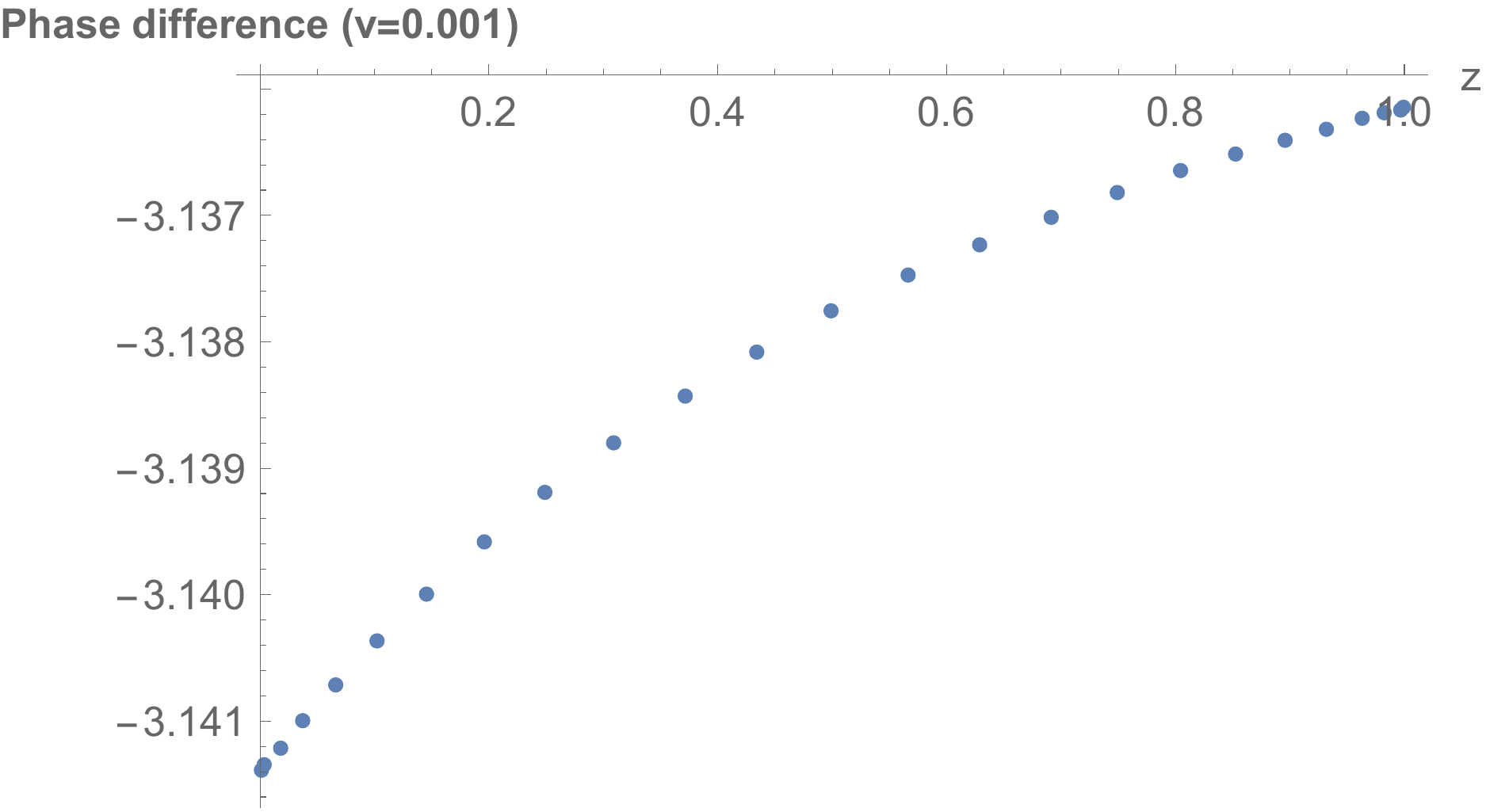} \quad{}\includegraphics[scale=0.38]{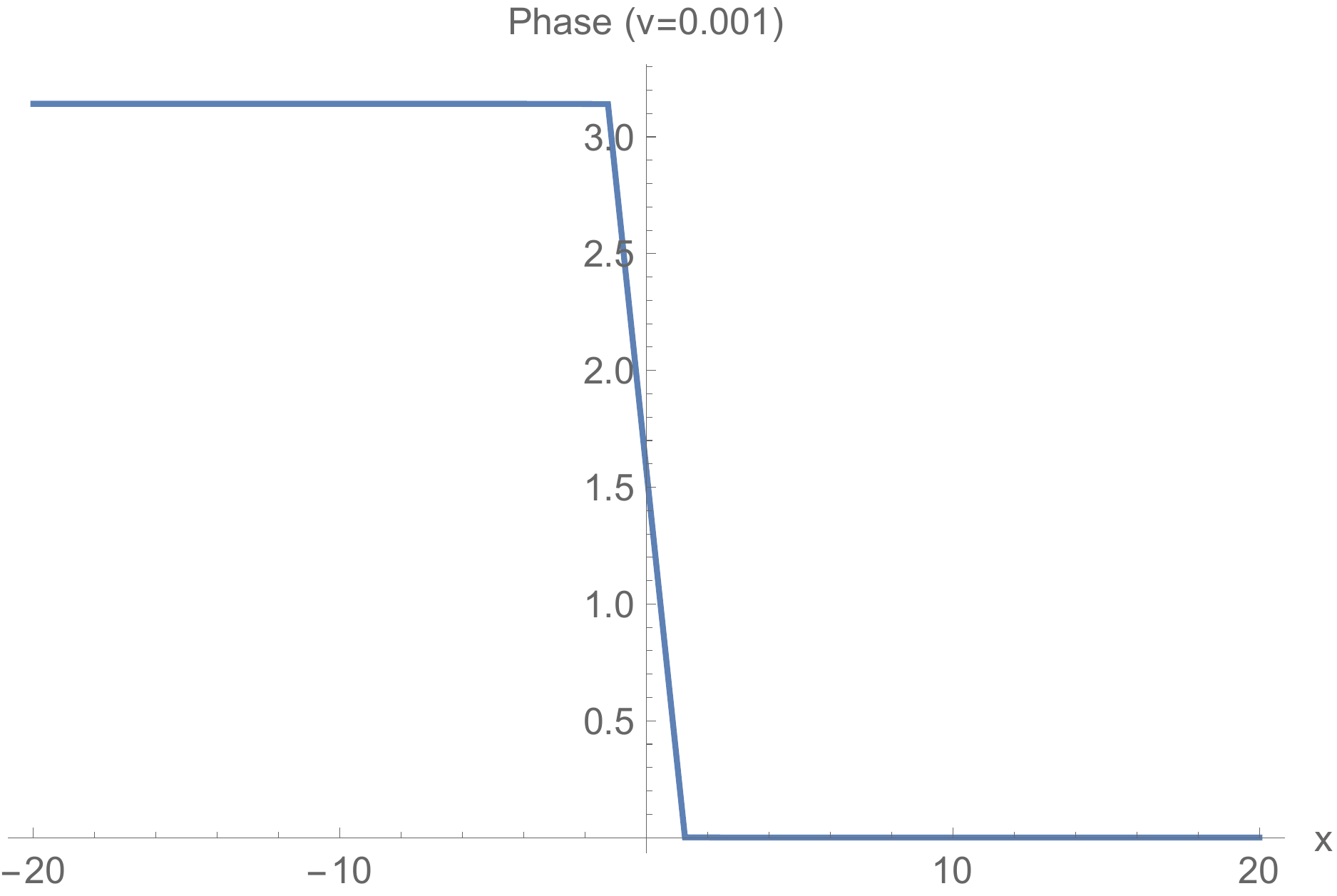}

\includegraphics[scale=0.44]{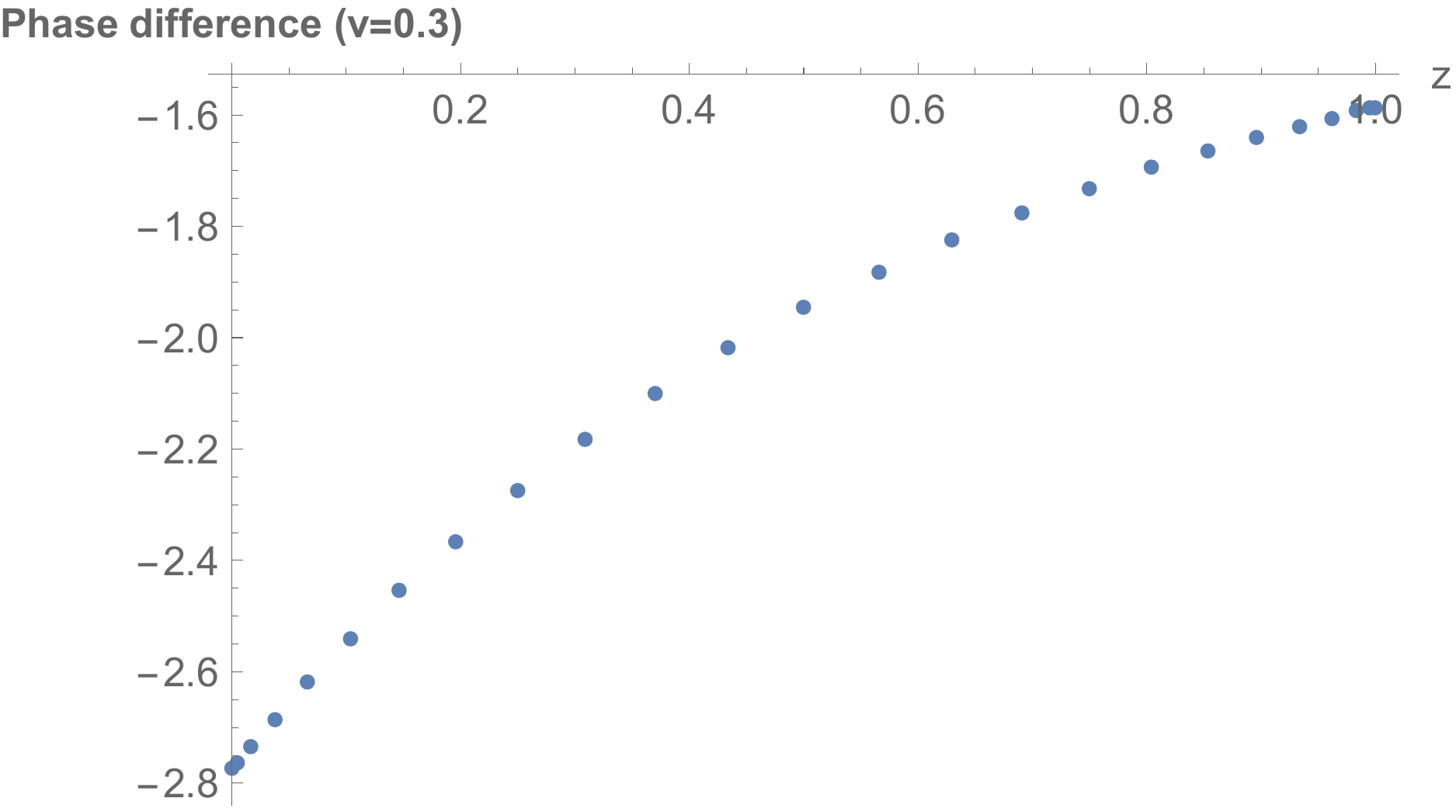} \includegraphics[scale=0.4]{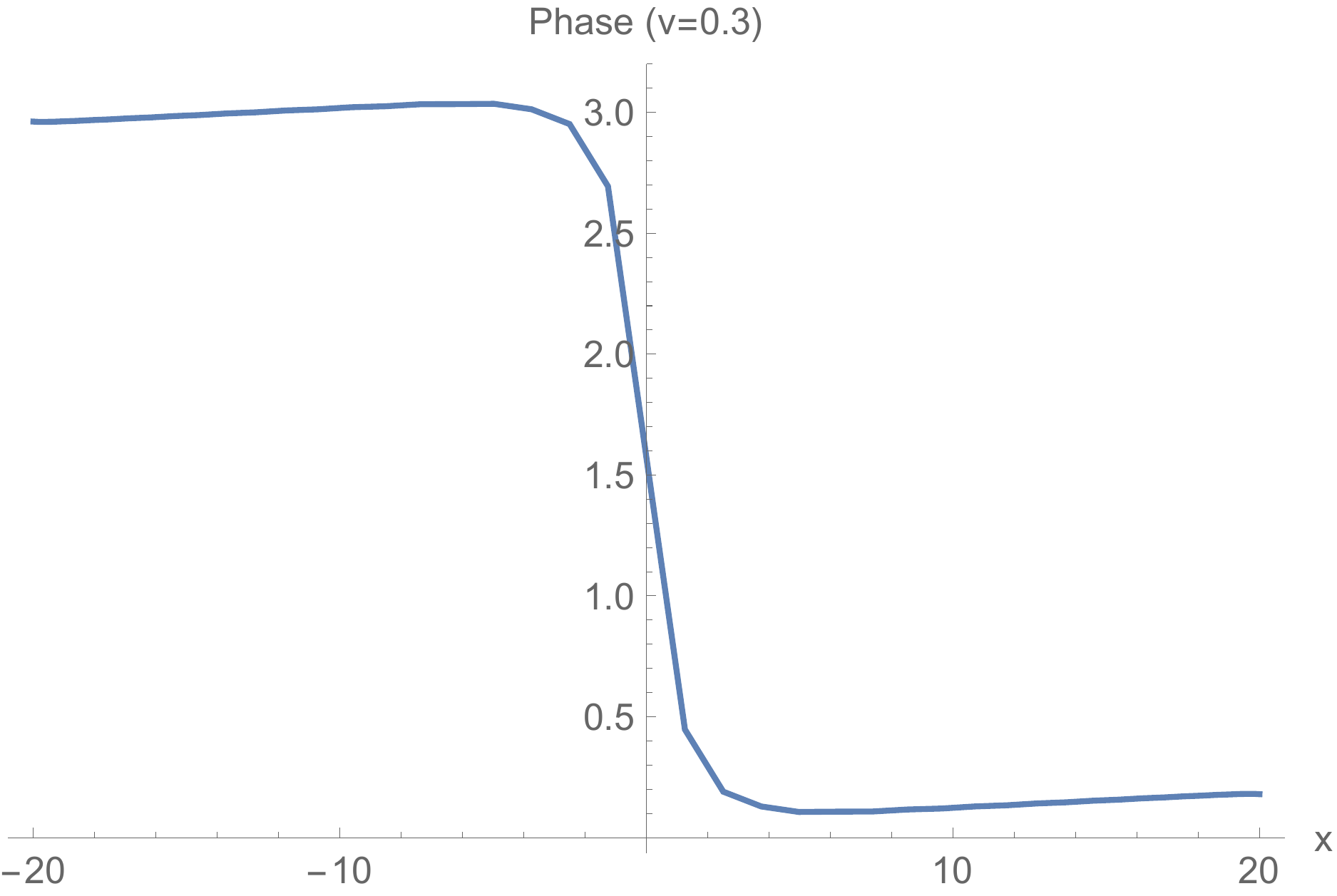}

\includegraphics[scale=0.4]{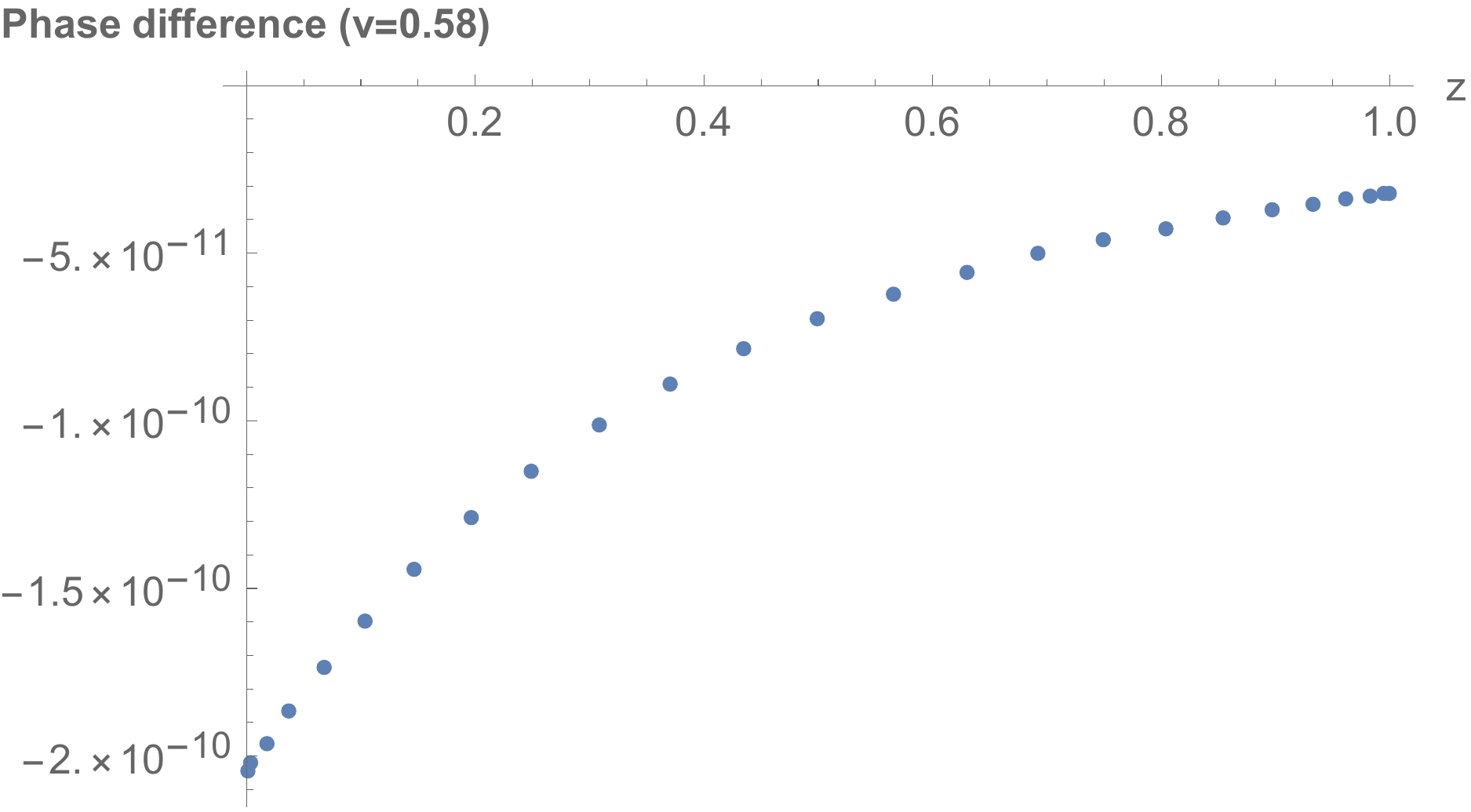} \qquad{}\includegraphics[scale=0.4]{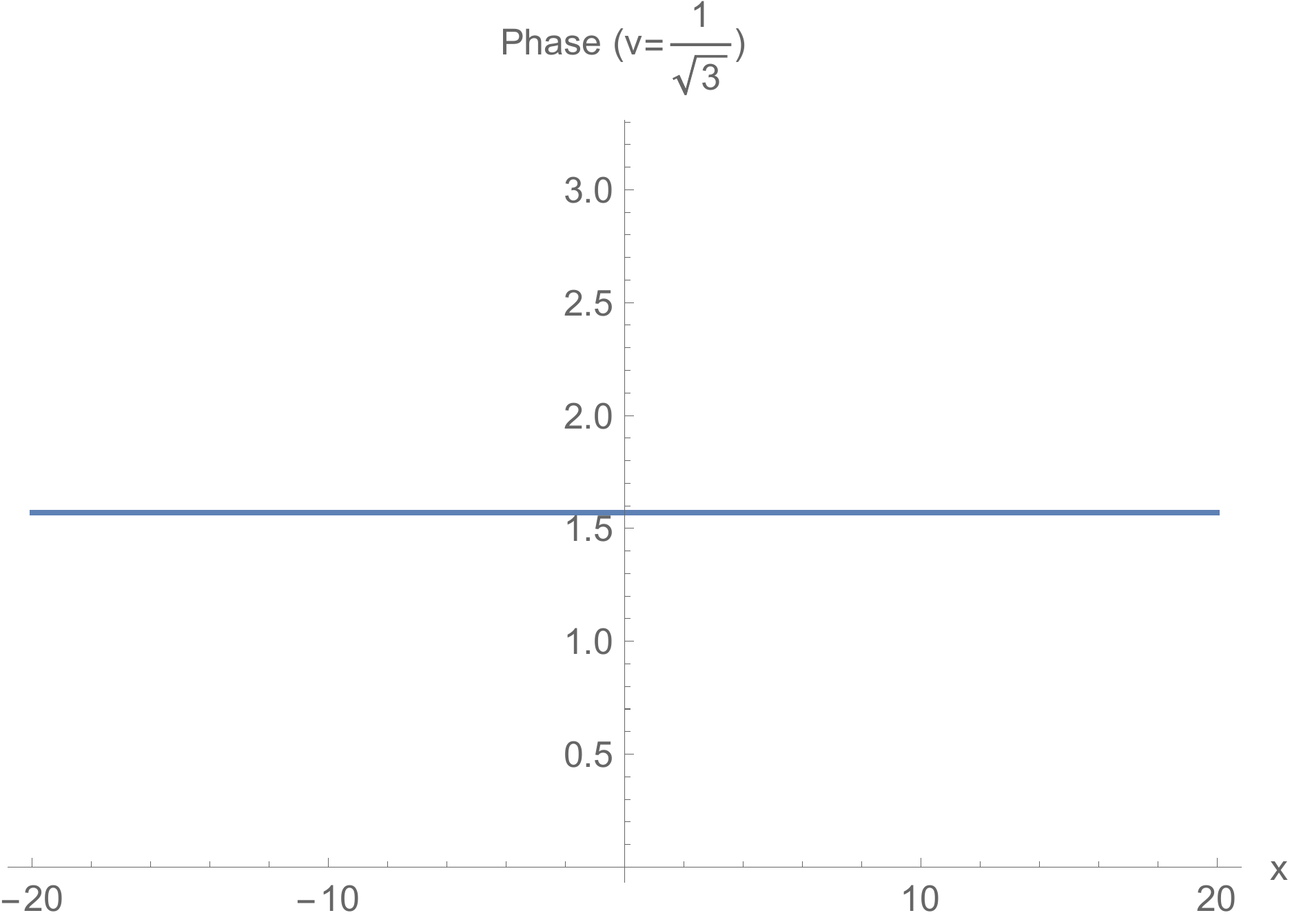}

\caption{The phase differences and phase distributions for gray solitons at
different speeds, under the alternative quantization with the chemical
potential $\mu=1.4$.\label{alternative phase diff gray}}
\end{figure}

Again, we also present the phase differences and phase distributions
for gray solitons under the alternative quantization in Fig.\ref{alternative phase diff gray}.
The qualitative behavior of them is the same as in the case of standard
quantization.

\begin{figure}
\includegraphics[scale=0.4]{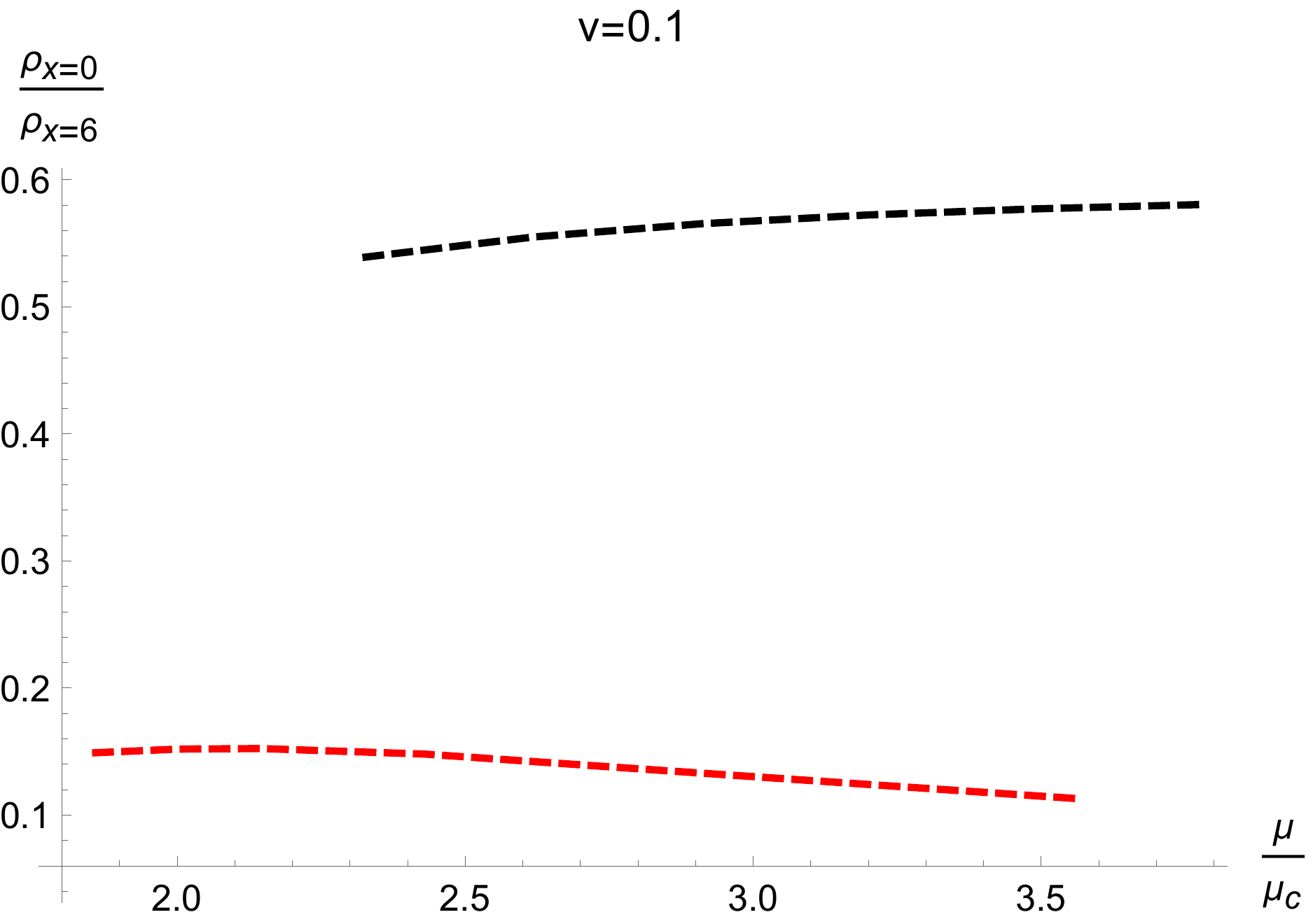}\qquad{}\includegraphics[scale=0.4]{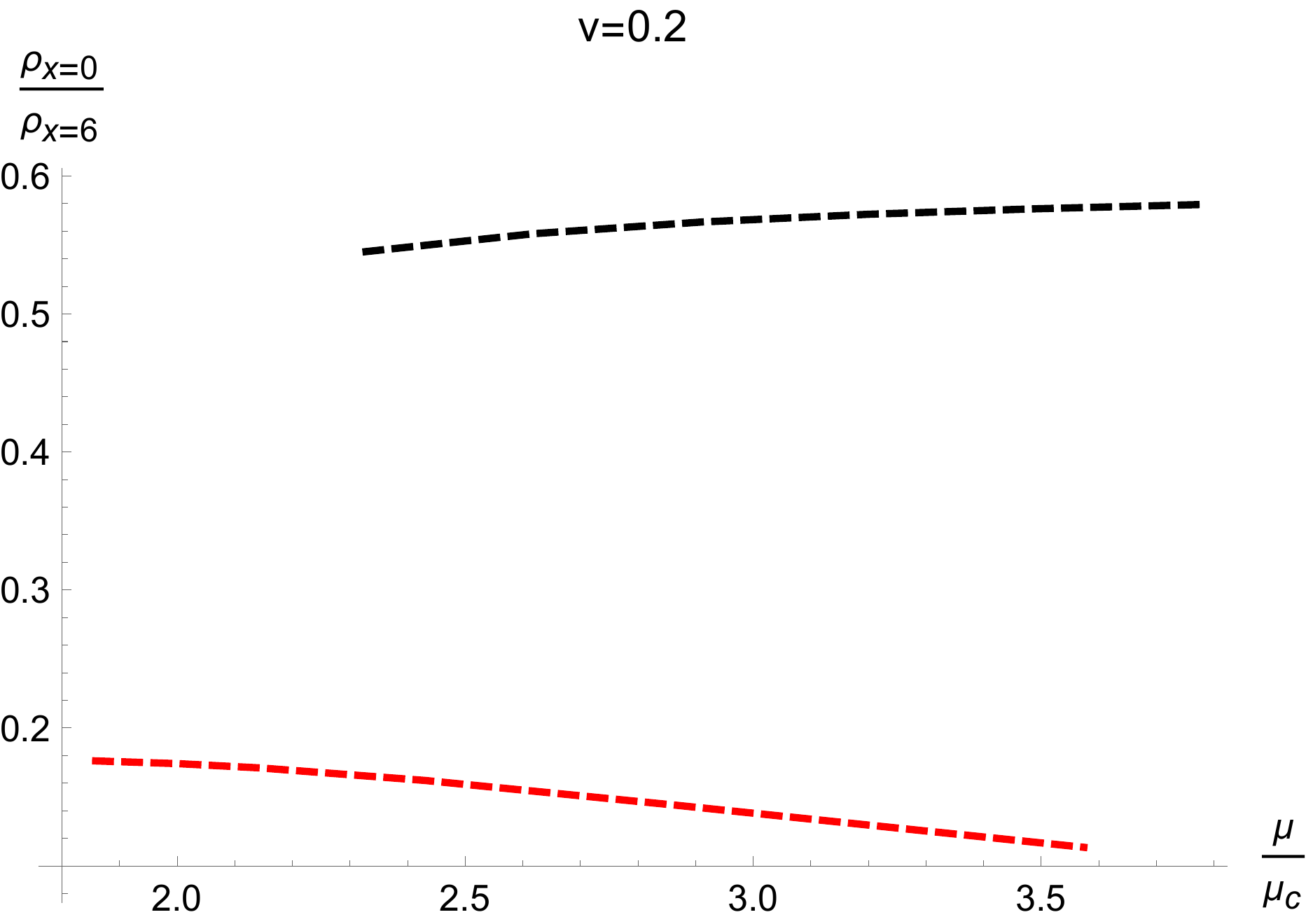}

\includegraphics[scale=0.4]{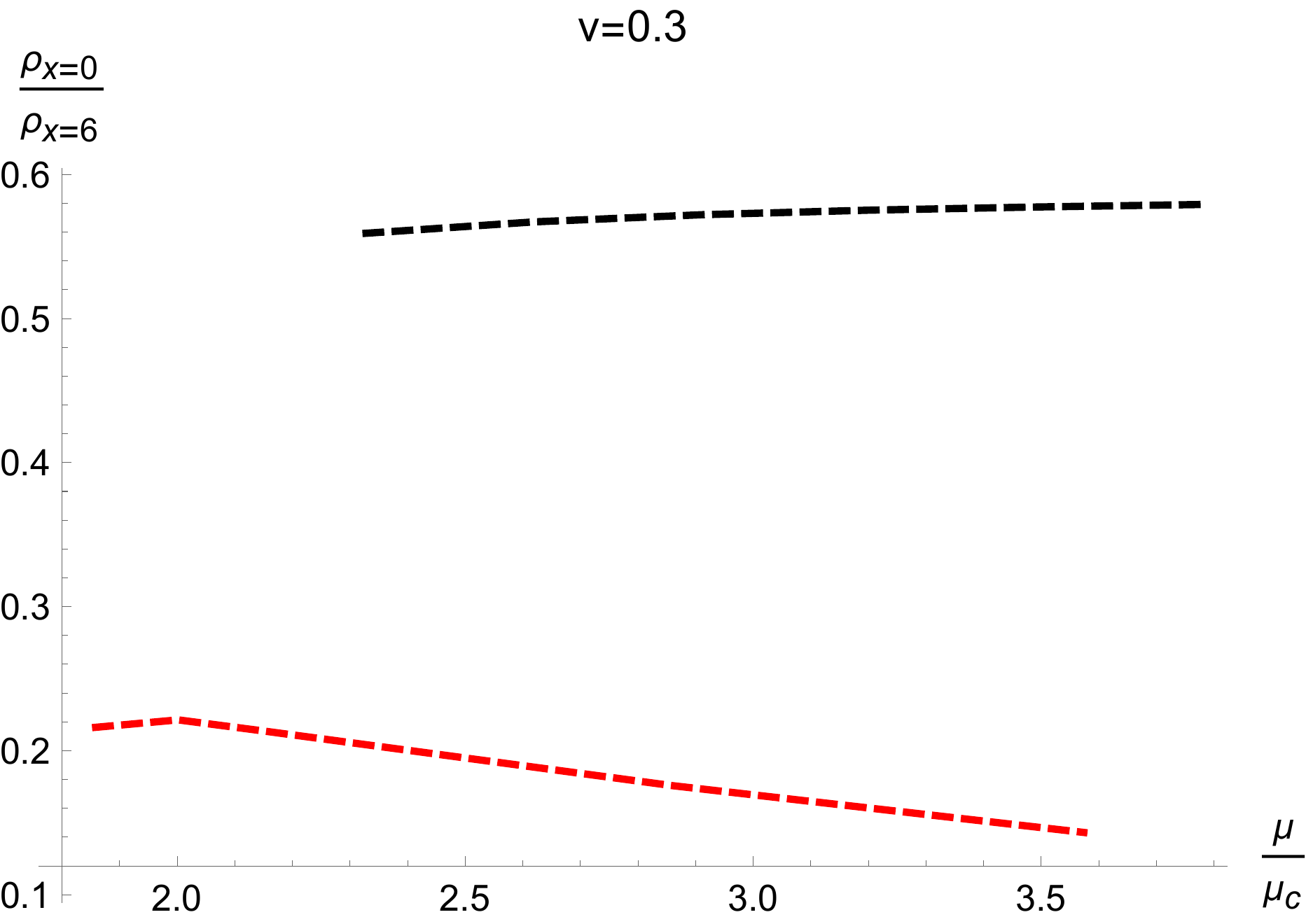}\qquad{}\includegraphics[scale=0.4]{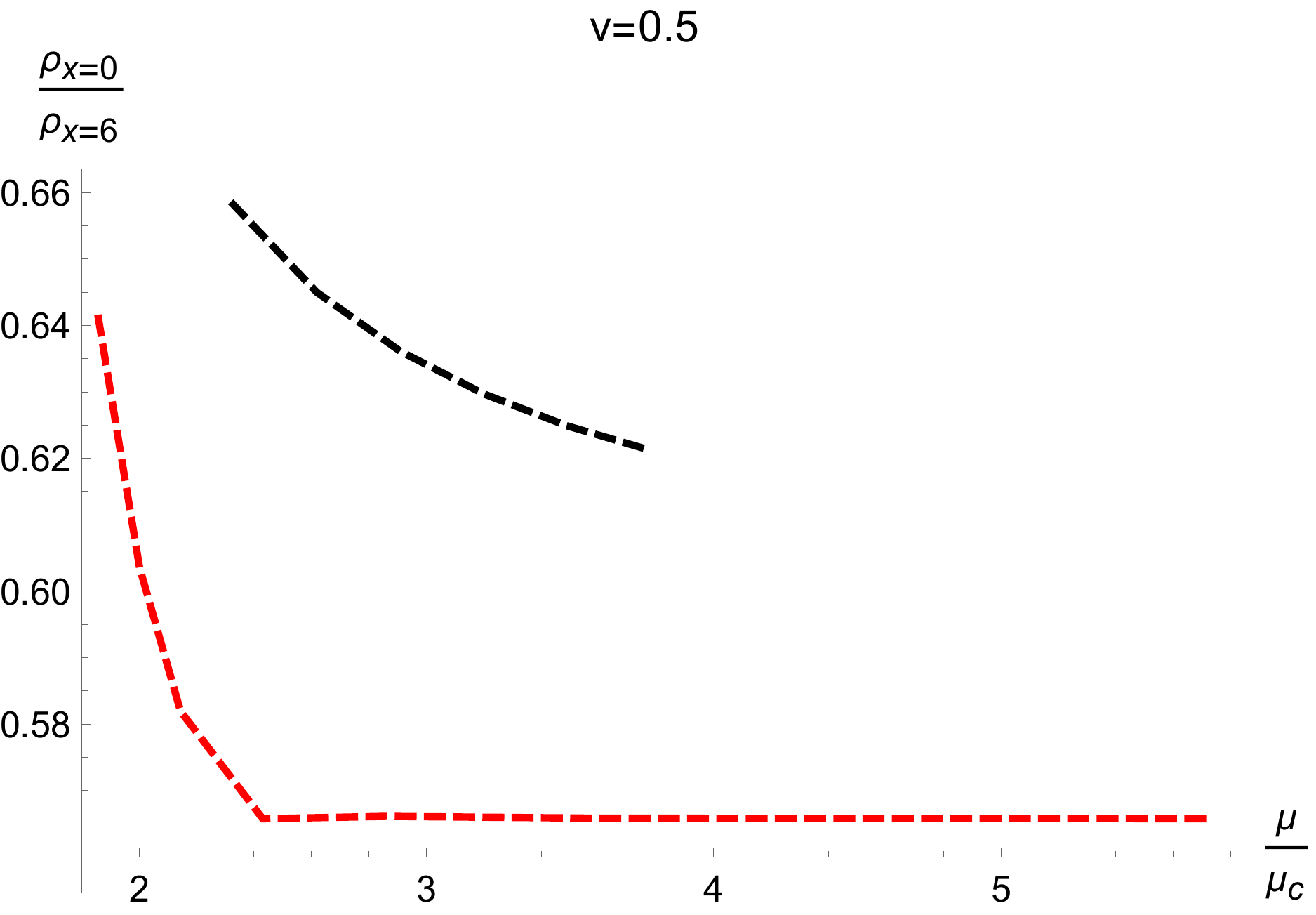}

\caption{The specific values of density $\frac{\rho_{min}}{\rho_{homo}}$ at
different speeds ($v=0.1,0.2,0.3,0.5$) with respect to the relative
chemical potential $\mu/\mu_{c}$ for the standard (black dashed)
and alternative (red dashed) quantizations.\label{depletion difference in gray}}
\end{figure}

For comparison with the black soliton case, we also present the density
depletion of the gray solitons under both quantizations in Fig.\ref{depletion difference in gray}.
It is obvious that the value of $\frac{\rho_{min}}{\rho_{homo}}$
for the standard quantization is in general larger than that for the
alternative one. Moreover, the difference of density depletion for
the two quantizations becomes smaller when the soliton speed increases.

\section{Summary and discussion}

For the zero temperature holographic superfluid model, we have numerically
solved the spatially inhomogeneous equations of motion to obtain the
black and gray soliton solutions under both the standard and alternative
quantizations, and have investigated in detail various aspects of
these holographic superfluid solitons.

The simplest such solitary solutions are the black solitons, which
are static superfluid structures. After having constructed the holographic
black soliton solutions, we focus on the particle number density depletions
at the centers of the solitons for the standard and alternative quantizations
at different chemical potentials, especially the limit of large chemical
potential. As a result, we find that the density depletion is much
larger under the alternative quantization than that under the standard
one. Our holographic model is a genuine zero temperature superfluid
system, so this result acts as a strong support for the correspondence
between different quantizations in holography and different types
of fermionic superfluids proposed in \cite{Keranen}. Specifically,
the alternative quantization corresponds to a BEC-like superfluid,
while the standard quantization corresponds to a BCS-like superfluid.
Moreover, under the standard quantization there are extra fluctuations,
compared to the alternative case, in the black soliton configurations,
which resemble the Friedel oscillation in the case of BCS superfluids
and further support the above proposal.

Then we have investigated the more complicated solitary solutions,
the gray solitons, which keep moving at constant speeds. It is shown
that these solutions perfectly interpolate between the (static) black
solitons and sound waves (moving at the speed of sound). We also find
that the density depletion is evidently larger under the alternative
quantization than that under the standard one at general soliton speeds.
The Friedel-like oscillation under the standard quantization is even
more remarkable in the gray soliton case, probably because the oscillation
amplitude is less suppressed at higher soliton speed than the condensate
variation itself. Finally, we have computed the gauge invariant phase
differences between the condensates on the two sides of our holographic
gray solitons and verified that these configurations do have all the
expected features of gray solitons.

For simplicity, we do not consider the backreaction of the matter
fields onto the bulk geometry in our holographic superfluid model.
But for a full holographic duality, such backreaction should be taken
into account\cite{C.P}. It is known that black solitons at finite
temperature are unstable against the so-called self-acceleration and
snake instabilities, so a natual question is whether it is also the
case for the black and gray solitons at zero temperature\cite{GKLTZ}.
It will be also interesting to investigate the collision dynamics
of holographic gray solitons. In order to do that, one has to transform
the gray soliton solution obtained here in the comoving frame back
to the original frame, manage to combine two or more gray solitons
as an initial configuration, and then perform the numerical time evolution
similar to that in the black hole backgrounds\cite{Chesler,Yiqiang,Y. Tian}.
These topics are left for future exploration. 
\begin{acknowledgments}
This work is partially supported by NSFC with Grant No.11475179 and
No.11675015. YT is also supported by the ``Strategic Priority Research
Program of the Chinese Academy of Sciences\textquotedbl{} with Grant
No.XDB23030000. HZ is supported in part by FWOVlaanderen through the
project G006918N, and by the Vrije Universiteit Brussel through the
Strategic Research Program “High-Energy Physics”. He is also an individual
FWO fellow supported by 12G3515N. 
\end{acknowledgments}

\appendix

\section{Fitting of the holographic BCS-like black soliton configurations}

\label{sec:fit}
\begin{figure}
\includegraphics[scale=0.33]{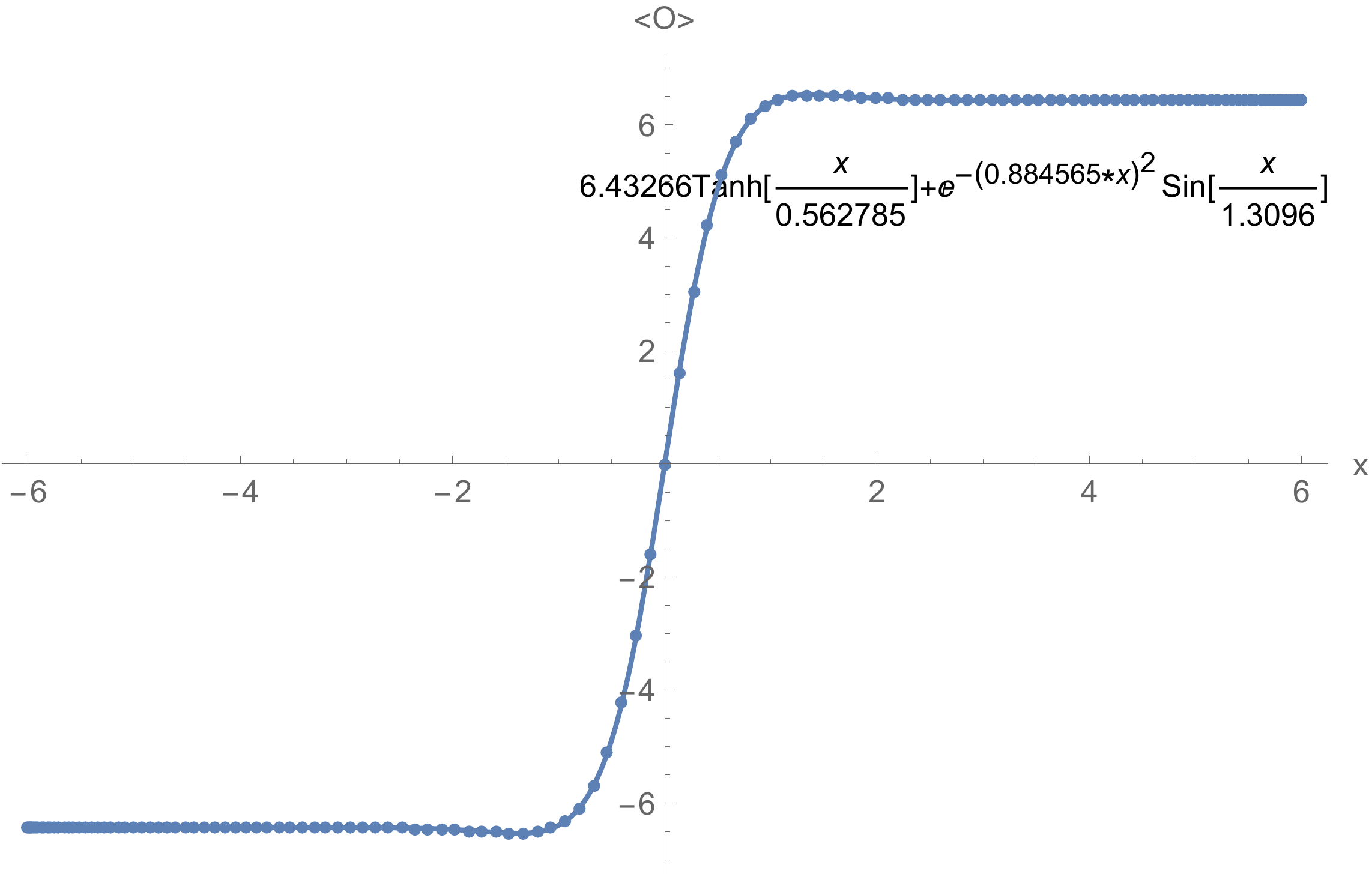}\qquad{}\includegraphics[scale=0.26]{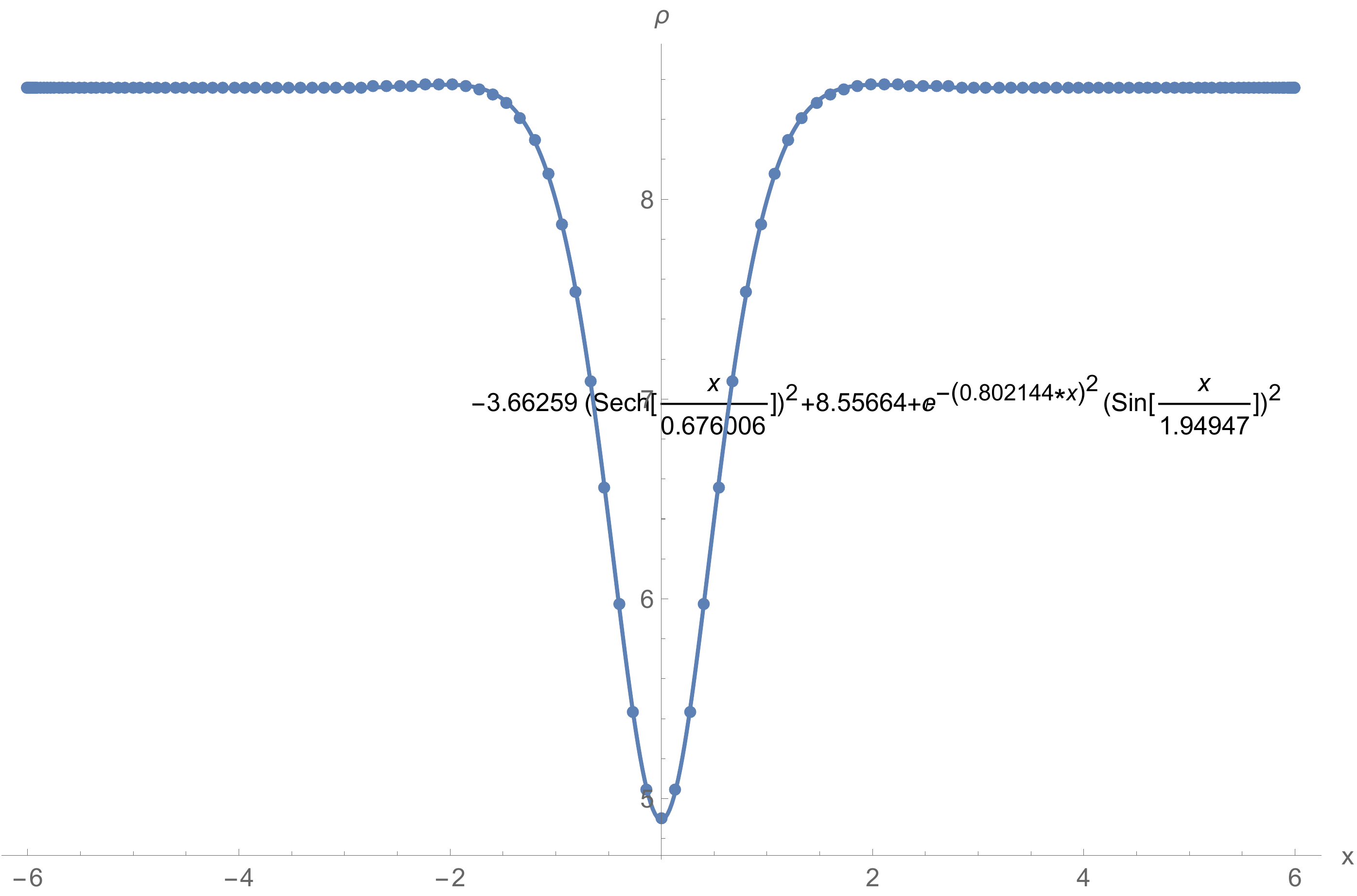}

\caption{The fitting functions of order parameter (left panel) and charge density
(right panel) at standard quantization with the chemical potential
$\mu=5.5$.\label{fit order standard}}
\end{figure}

We have tried to fit the order parameter and density
profiles under the standard quantization with some test functions,
as shown in Fig.\ref{fit order standard}. The fitting function for
the order parameter is $6.43266\tanh\text{\ensuremath{\left(x/0.562785\right)}}+e^{-\left(0.884565x\right)^{2}}\sin\left(x/1.3096\right)$,
where 6.43266 is the condensate for a homogeneous solution with the
same chemical potential, and 0.562785 is the healing length of the
soliton. The fitting function for the charge density is $-3.66259\mathrm{sech}^{2}\left(x/0.676006\right)+8.55664+e^{-\left(0.802144x\right)^{2}}\sin^{2}\left(x/1.94947\right)$,
where 0.676006 is the healing length of the soliton, and 3.66259 is
the charge density depletion. One remarkable difference, compared
to the result in \cite{Lan}, is that both fitting functions have
additional terms, reflecting the fluctuation behaviors.

\section{Fitting of the holographic BCS-like gray soliton configurations}

\label{sec:fit_gray}
\begin{figure}
\includegraphics[scale=0.4]{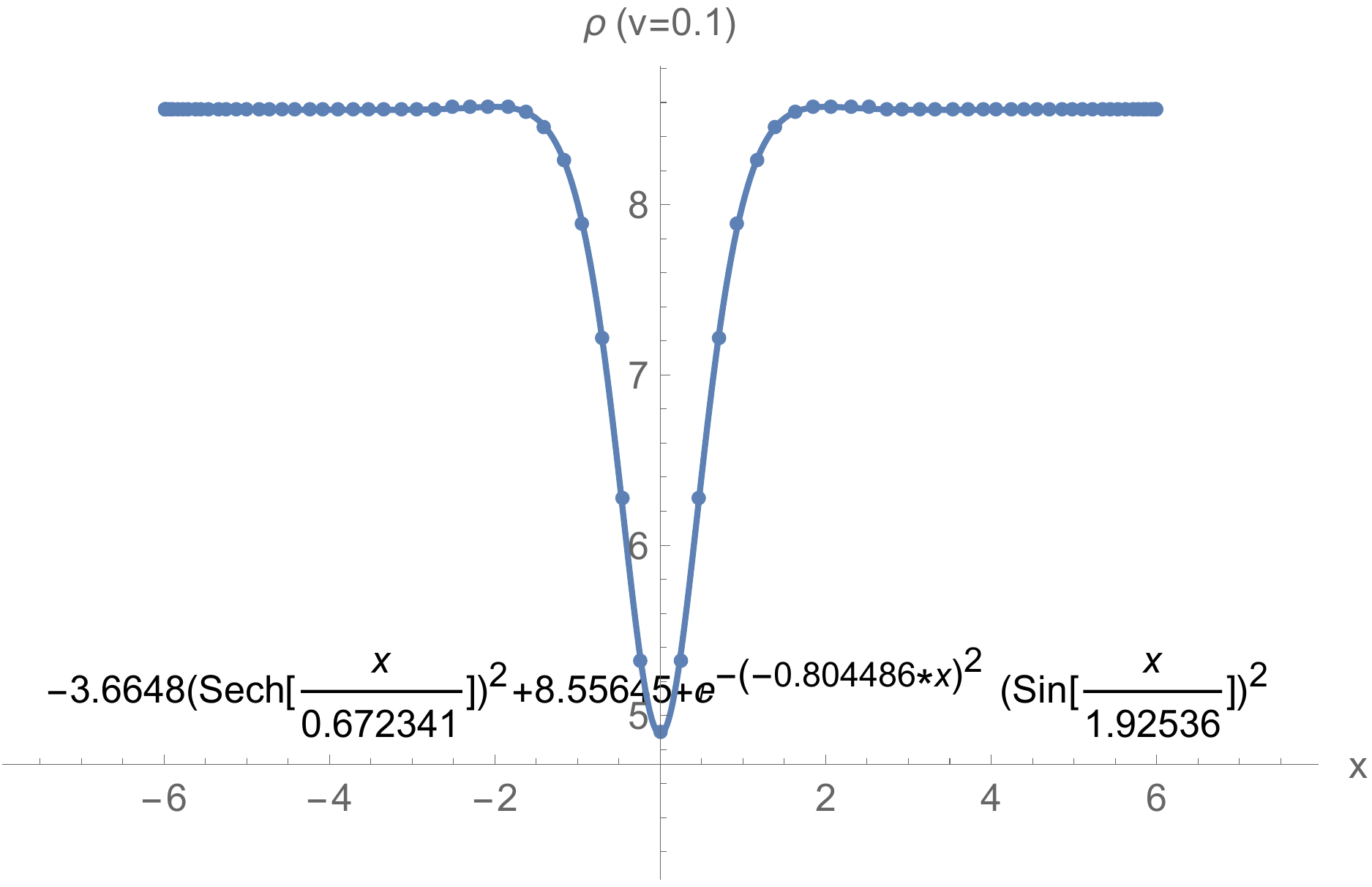}\qquad{}\includegraphics[scale=0.4]{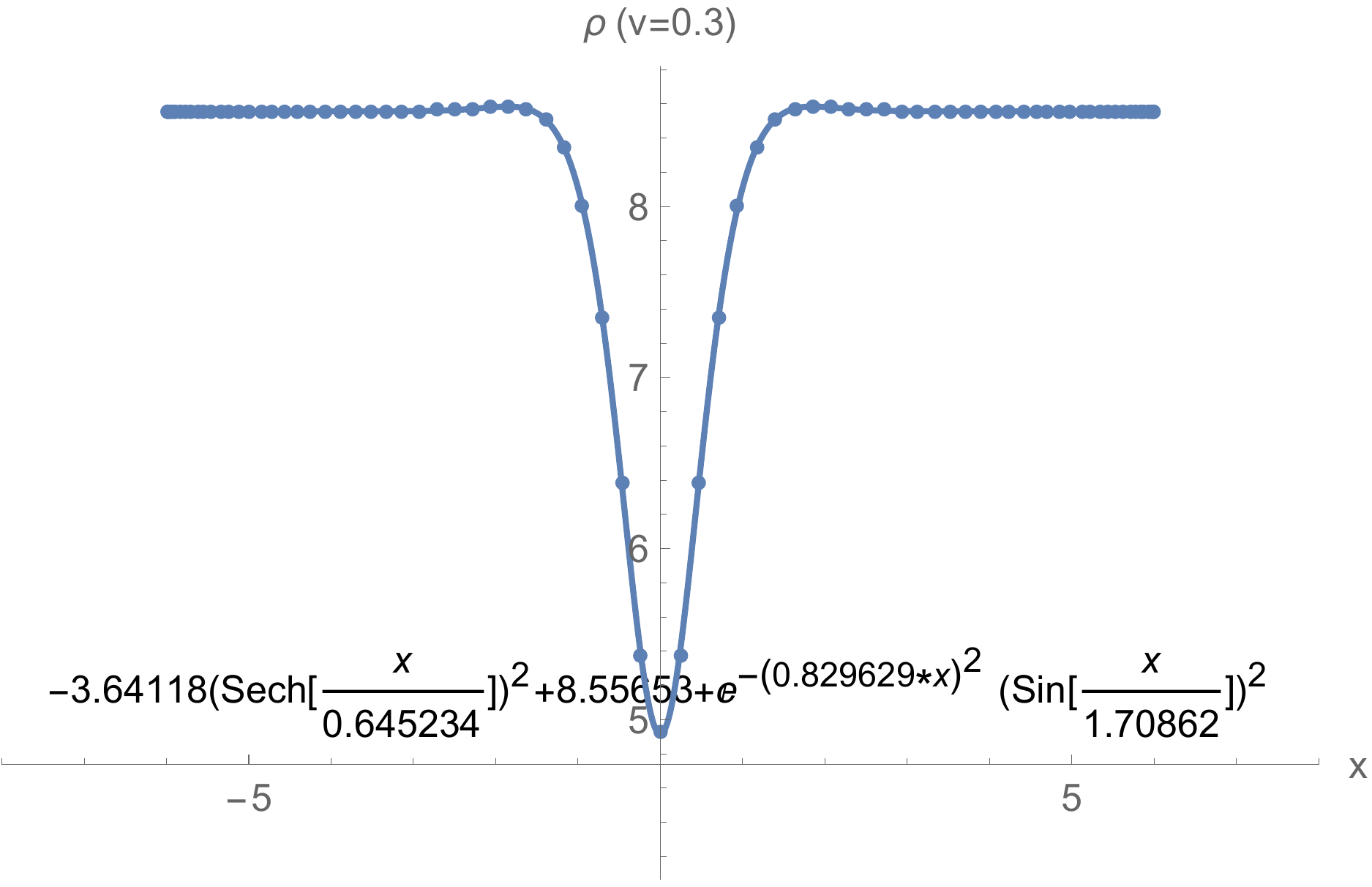}

\caption{The fitting results, at $\mu=5.5$, for the density with different
speeds under the standard quantization.\label{fit density gray}}
\end{figure}

For the sake of presenting the fluctuations in
detail, we have tried to fit the particle number density by functions.
The result is shown in Fig.\ref{fit density gray}.

\end{document}